\documentclass[a4paper,11pt]{article}
\pdfoutput=1 

\usepackage{jheppub} 
\usepackage{amssymb}
\usepackage{mathtools}
\usepackage{slashed}
\usepackage{cleveref}
 
\newcommand{\be}{\begin{equation}}
\newcommand{\ee}{\end{equation}}
\newcommand{\bea}{\begin{eqnarray}}
\newcommand{\eea}{\end{eqnarray}}

\newcommand\ddfrac[2]{\frac{\displaystyle #1}{\displaystyle #2}}
\newcommand{\epm}{e^+e^-}
\newcommand{\softl}{\mathbb{S}}
\newcommand{\soft}[1]{\mathbb{S}_{\mbox{\small {#1}-h}}}
\newcommand{\ftsoft}[1]{\widetilde{\mathbb{S}}_{\mbox{\small {#1}-h}}}
\newcommand{\coll}{\mathbb{C}}
\newcommand{\TheBook}{Ref.~\cite{Collins:2011zzd}}
\newcommand{\hclass}[1]{${#1}$-hadron class}
\newcommand{\lips}[1]{{d^3 \vec{#1}}/{2 E_{#1}}}
\newcommand{\kmeas}[2]{\frac{d^D #1_{#2}}{(2 \pi)^D}}

\title{\boldmath Universality-breaking effects in $e^+e^-$ hadronic production processes}

\author{M. Boglione,}
\author{and A. Simonelli}
 
\affiliation{Dipartimento di Fisica Teorica, Universit\`a di Torino,\\
                Via P.~Giuria 1, I-10125 Torino, Italy}
\affiliation{INFN, Sezione di Torino, Via P.~Giuria 1, I-10125 Torino, Italy}

\emailAdd{elena.boglione@to.infn.it}
\emailAdd{andrea.simonelli@unito.it}

 \abstract{Recent BELLE measurements provide the cross section for single
hadron production in $e^+e^-$ annihilations, differential in thrust and in 
the hadron transverse momentum with respect to the thrust axis.
Universality breaking effects due to process-dependent soft factors make
it very difficult to relate this cross sections to those corresponding to
hadron-pair production in $e^+e^-$ annihilations, where transverse momentum 
dependent (TMD) factorization can be applied. 
The correspondence between these two cross sections is examined in the 
framework of the Collins-Soper-Sterman factorization, in the collinear as well 
as in the TMD approach. 
We propose a scheme that allows to relate the TMD parton densities defined in 
1-hadron and in 2-hadron processes, neatly separating, within the soft and 
collinear parts, the non-perturbative terms from the contributions that can be 
calculated perturbatively.
The regularization of rapidity divergences introduces cut-offs, the
arbitrariness of which will be properly reabsorbed by means of a mechanism 
closely reminiscent of a gauge transformation. 
In this way, we restore the possibility to perform global phenomenological 
studies of TMD physics, simultaneously analyzing data belonging to different 
hadron classes.
}

\begin{document} 

\maketitle

\newpage
 

\section{Introduction\label{sec:intro}}


\bigskip

QCD describes hadronic matter through the dynamics of its elementary 
consituents, quarks and gluons. However, confinement prevents the direct 
observation of partonic degrees of freedom, which are shaded by the   
hadronization mechanism.

Recently, the BELLE Collaboration at KEK has measured the $e^+e^- \to H X$ 
cross section at a c.m. energy of $Q^2\sim 112$ GeV$^2$ as a function of $P_T$, 
the transverse momentum of the observed hadron $h$ relative to the  
thrust axis~\cite{Seidl:2019jei}. The data are binned in $P_T$ and selected in thrust 
in such a way to ensure that $T \sim 1$, which corresponds to a two jet 
configuration.
This is one of the measurements which go closer to being a direct observation 
of a partonic variable, the transverse momentum of the hadron with respect to  
its parent fragmenting parton.

These data have indeed triggered a great interest of the high 
energy physics community, especially among the experts in the 
phenomenological study of TMD phenomena and factorization. 
However, there are difficulties in the analysis of these data as, due to the 
nature of this process, a TMD factorization as that formulated in 
Ref.~\cite{Collins:2011zzd} cannot be directly applied. In this case, in fact, 
collinear factorization would rather be the correct approach. 

In this paper we will follow very closely the formulation proposed by J. 
Collins in Ref.~\cite{Collins:2011zzd} for $e^+e^- \to H_A \, H_B \, X$ 
processes and 
we will give a different definition of TMDs which, by extending their degree of 
universality, becomes suitable to be applied also to the  $e^+e^- \to H \, X$ 
process. We will move along the lines suggested, for instance, in Ref.~\cite{Collins:2007ph}.

In this new definition, the soft factor of the process, which is responsible for potential 
universality breaking effects, is not absorbed in the TMD, 
to prevent it from influencing its genuinely universal nature. 
Instead, it appears explicitly in the cross section which acquires a new 
term, that we will call soft model, $M_S$. After being modelled 
using a suitable parameterization, it can be extracted from 
experimental data. While the TMDs are truly universal and can be extracted 
from any process, $M_S$ is universal only among a restricted number of processes. 
In other words, $M_S$ is universal only within his hadron class. 
Later on in the paper we will define what we mean exactly by ``class''; for the moment being 
we anticipate that, for instance, Drell-Yan, Semi Inclusive Deep Inelastic Scattering (SIDIS) 
and $e^+e^- \to H_A \, H_B\, X$ processes belong to the same hadron class, 
while DIS and 
$e^+e^- \to H\, X$ belong to a different class.

The advantage of this formulation is that a well defined expression relates TMDs 
extracted using different definitions. Consequently, 
all results obtained in past phenomenological analyses can easily be 
reformulated according to this new framework, with no loss of information.

We stress that the factorization procedure itself will not be altered from its 
original form. Rather, we introduce a new methodology to implement the 
phenomenological application of that same scheme, changing the focus on the 
fundamental ingredients of the phenomenological models.

The paper is organized as follows. In Section \ref{subsec:coll_vs_tmd} we will 
outline the basics of TMD and collinear factorization.
Section~\ref{sec:soft_factor} will be devoted to the study of the soft factor 
and its factorization properties, while in Section~\ref{sec:coll_parts} we will 
examine the collinear parts of  hadronic processes. Here we will define the TMDs 
and show how a particular transformation, which we will call ``rapidity 
dilation", allows to consider them invariant with respect to the choice of the 
rapidity cut-offs introduced by the regularization of the rapidity divergences.
In Section~\ref{sec:univ_class} we will briefly outline a new way of classifying 
hadronic processes in terms of their ``hadron class". Section~\ref{sec:2h_class} 
will be dedicated to the study of 2-hadron processes and their cross sections, 
respectively.  Finally, in Section~\ref{sec:epm_1h} we will apply this formalism 
to $e^+e^- \to H\, X$, giving a simple example of how this scheme can be 
applied 
to a phenomenological analysis. Appendices A, B and C will be dedicated to the 
definition of Wilson lines, to  the small $b_T$ behaviour of the soft factor and 
of the TMDs, and to the kinematics of a $e^+e^- \to H\, X$ process, 
respectively.

Throughout the paper we will adopt a pedagogical approach, as we intend to 
provide a review which could be useful to young beginners as well as to 
experienced researchers. 

\bigskip


\subsection{Collinear and TMD Factorization\label{subsec:coll_vs_tmd}}


\bigskip

Modern studies of high energy QCD processes are based on factorization, a  
procedure that allows to separate the cross section of a hadronic process 
involving a hard energy scale $Q$ into a part which is fully computable in 
perturbation theory and a non-perturbative contribution, with an error suppressed by 
powers of $m/Q$, where $m$ is a typical low energy mass scale. 
In general, the 
perturbative part is process dependent but it can be computed, at any given 
order, for any given process. The non-perturbative part, cannot be computed: it 
should rather be inferred from experimental data. However, when 
defined in an appropriate way, it is universal, in the sense that it can be 
extracted from one process and then used in any other. If 
factorization applies and universality is preserved, then the theory can be
predictive.
Nowadays, several different schemes are available to implement factorization. In 
the following, we will adopt the modern version of the Collins-Soper-Sterman 
(CSS) scheme~\cite{Collins:1984kg,Collins:1989gx}, often referred to as CSS2, 
presented in Ref.~\cite{Collins:2011zzd}.

When 
factorization applies, then the cross section of the process will appear 
as a convolution of contributions which can be classified in terms of the 
following three categories:
\begin{enumerate}
\item \textbf{Hard part}. It corresponds to the elementary subprocess and it 
provides the signature of the process, as it identifies the partonic 
scattering uniquely. It is fully computable in perturbation theory in terms of Feynman 
diagrams, up to the desired accuracy.
\item \textbf{Collinear parts}. These contributions are associated to the 
initial and final state hadrons of the process and contain the collinear 
divergences 
related to the massless particles emitted along the hadron direction. 
Each of them corresponds to a bunch of particles strongly boosted along this  
direction, which move almost collinearly, very fast.
Due to their characteristic divergences, collinear parts cannot be 
fully computed in perturbation theory: their non-perturbative content 
has to be extracted from experimental data. 
Among all the particles in the collinear group, two of them deserve special attention:  
the \emph{reference hadron} and the \emph{reference parton}.
If the collinear group refers to the initial state of the process, the 
reference 
hadron coincides with the initial hadron and the reference parton is the parton 
confined inside it that is struck in the hard scattering; if the collinear group 
refers to the final state, the reference hadron is the detected hadron and the 
reference parton is the fragmenting parton, i.e. the particle that initiates the 
hadronization process.
\item \textbf{Soft part}. It embeds the contribution due to the soft 
gluon radiation that connects the collinear parts and that flows through 
the detector. It contains soft divergences and carries 
non-perturbative information, therefore it cannot be computed in perturbation 
theory. 
It cannot be directly extracted from data, either, as the energy of the soft 
radiation is so low that detectors are not sensitive to it. 
Since the collinear parts interact among each 
other \emph{only} through soft gluons, their contribution can affect the 
cross section in a non-trivial way. Moreover, the soft part is 
\emph{always} associated with the collinear terms and there is no way to 
extract them separately. This is sometimes referred to as the soft factor 
problem.
\end{enumerate}
In several cases the contribution of the soft part is trivial. In particular, 
any time in addition to the collinear partons there are real emissions with 
hard transverse momentum, the soft factor fully factorizes 
but its value reduces 
to unity. In these cases, in fact, the soft gluons are kinematically overpowered 
and do not correlate the collinear parts anymore: in this way each collinear 
cluster of partons is totally independent from any other. 
Technically speaking, in such a situation the soft 
factor involves an integration over \emph{all} the components of the total soft 
momentum so that the soft information is washed out in the integral.
Whether there could be a hard real emission or not is determined by kinematics. 
Hence, it is the hard factor that discriminates among 
different cases.
%
\begin{figure}[t]
  \centering 
\begin{tabular}{c@{\hspace*{15mm}}c}  
      \includegraphics[width=4cm]{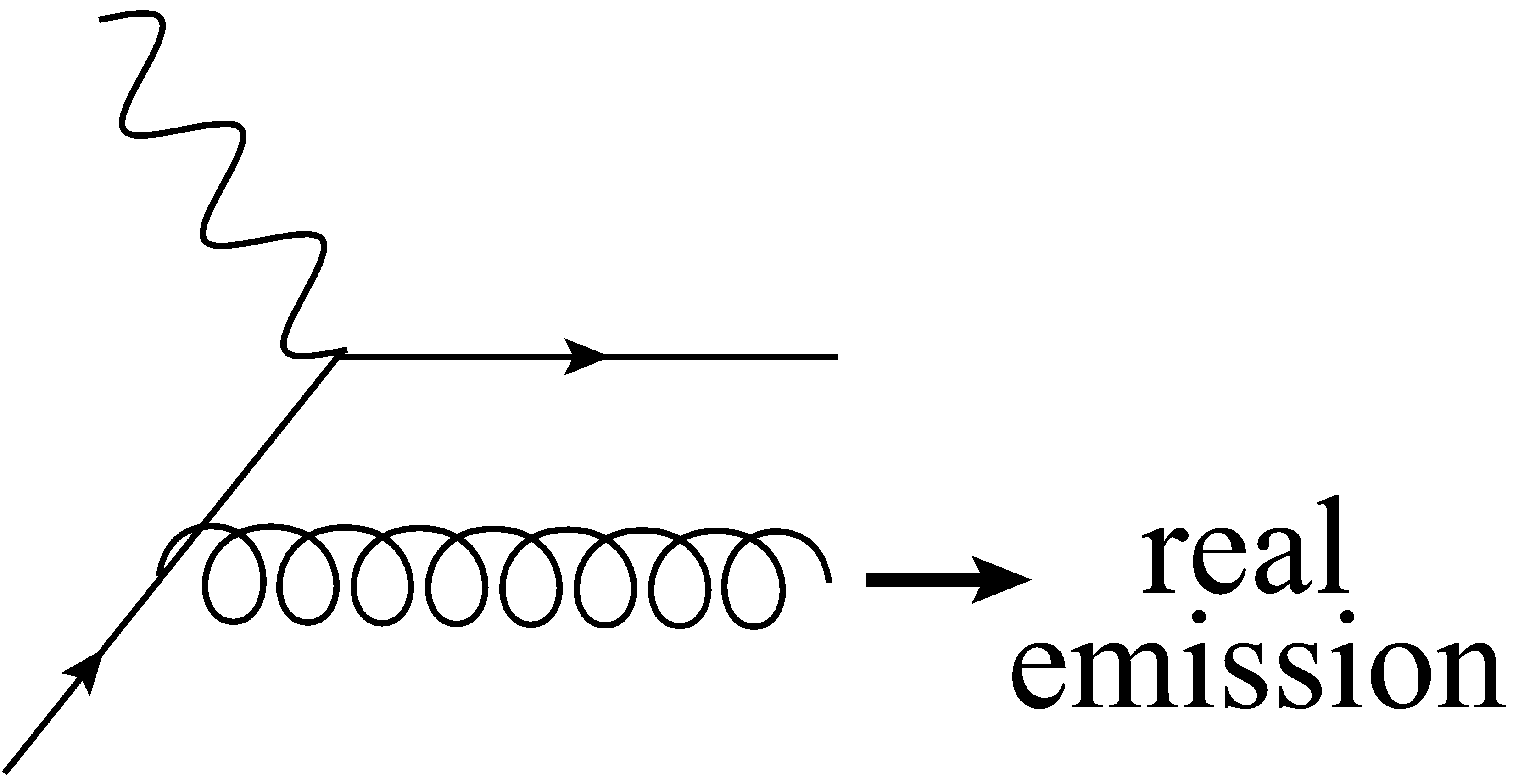}
  &
      \includegraphics[width=4cm]{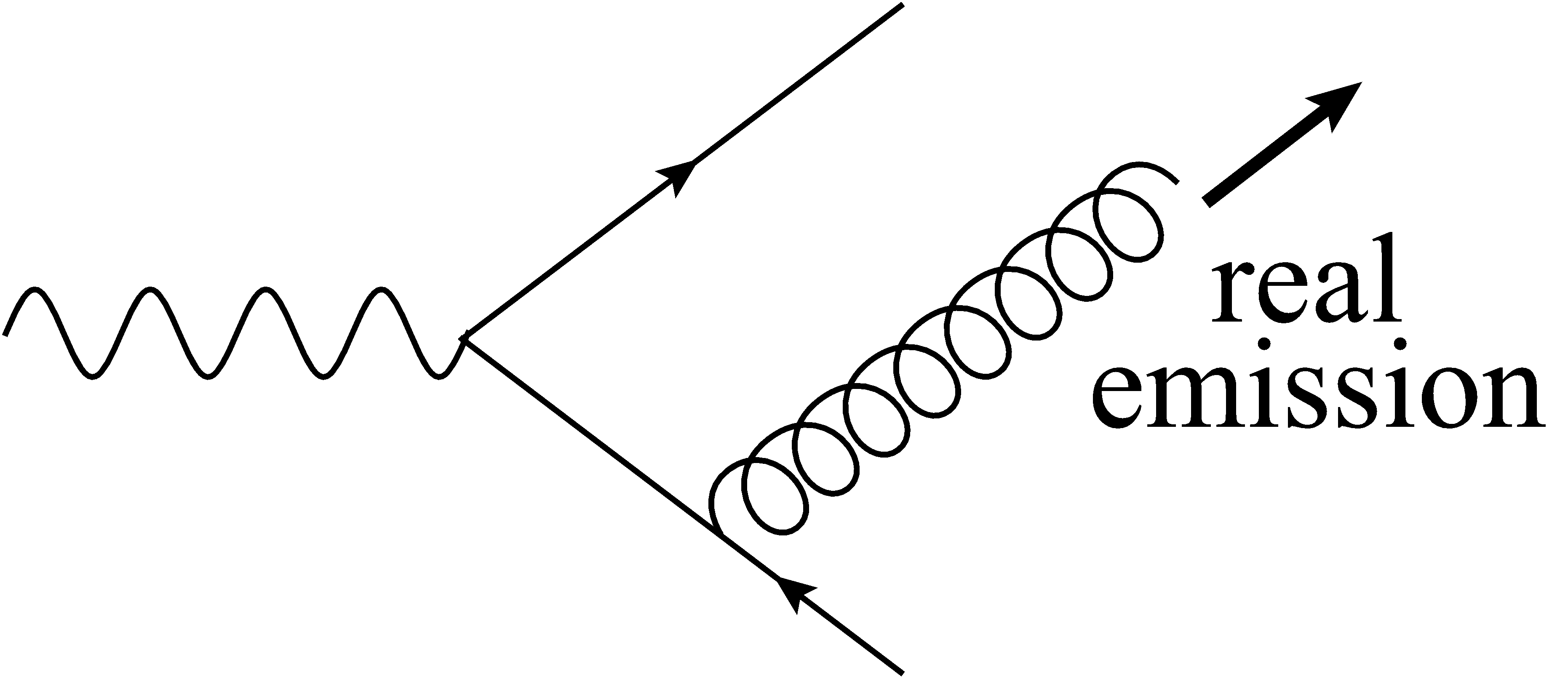}
  \\
  (a) & (b)
\end{tabular}
%
 \caption{(a): Pictorial representation of (the hadronic part of) a DIS process. The struk quark is 
associated to the collinear part relative to the target hadron, while the 
radiated gluon is the hard 
real emission. (b): Pictorial representation of (the hadronic part of) 
$\epm \rightarrow H \, X$. The quark line corresponds to the fragmenting quark 
associated to the 
collinear part representing the final hadron, while 
the radiated gluon is the hard real emission.}
 \label{fig:hard_real_emission}
\end{figure}

Kinematical configurations where hard real emissions are present, see for 
instance 
Fig.~\ref{fig:hard_real_emission},  represent instances in which 
\textbf{collinear factorization} holds. In these cases it is 
possible to relate each collinear part with a Parton Distribution Function 
(PDF) or a Fragmentation Function (FF), depending on whether the associated reference 
hadron is in the initial or in the final state, respectively.
As an  example of a collinearly factorized process, one could consider the 
case of an $e^+e^-$ scattering where two spinless hadrons $H_A$ and $H_B$ 
are produced in the final state, in a configuration far from being 
back-to-back in the center of mass frame (which, in this case, 
corresponds to the lab frame). The resulting cross section is given by (see Eq. 
(12.84) in Ref.~\cite{Collins:2011zzd}):
\begin{equation} \label{eq:coll_fact}
\frac{d \sigma}{(\frac{d^3 \vec{p}_A} {E_A}) \, (\frac{d^3 \vec{p}_B}{E_B})} = 
\sum_{j_A,\,j_B} 
\int \, \frac{d \widehat{z}_A}{\widehat{z}_A^2} \, d_{H_A / j_A} (\widehat{z}_A) 
\,
\int \, \frac{d \widehat{z}_B}{\widehat{z}_B^2} \, d_{H_B / j_B} (\widehat{z}_B) 
\,
\frac{d \widehat{\sigma}}{(\frac{d^3 \vec{k}_A}{\epsilon_A}) \,
(\frac{d^3\vec{k}_B} {\epsilon_B})}\,,
\end{equation}
where $d \widehat{\sigma}$ is the partonic cross section, i.e. the hard part, 
while $d_{H_i / j_i} (\widehat{z}_i)$, for $i = A, \,B$, are the usual FFs 
associated to the outgoing hadrons, with momenta $\vec{p}_A$ and $\vec{p}_B$, 
and to the fragmenting partons of flavor $j_A$ and $j_B$,
corresponding to the two collinear contributions to the cross 
section of the process.

Configurations in which kinematics forbid hard real emissions, instead,  
are extremely complex, but still very interesting. 
Here, the soft factor does not reduce to unity, and soft gluons 
have a non-trivial impact on the cross section as they correlate the  
collinear parts.
This correlation originates from momentum conservation laws in the transverse 
direction. In fact, with no hard real emissions and consequently no large 
transverse momentum entering into the game, 
the low transverse momentum components of soft and collinear particles cannot 
be neglected anymore: the information regarding the 
(total) soft transverse momentum survives and the soft factor results in an 
integration over the plus and minus (but not over the transverse) components.
In these cases it is not possible to associate a PDF or a FF to 
the collinear contributions: parton densities are now related to different and 
more general objects, known as Transverse Momentum Dependent (TMD) parton 
functions, either TMD PDFs or TMD FFs depending on whether they refer to an 
initial or a final state hadron.
In this cases collinear factorization breaks and a different, more 
involved, factorization scheme has to be applied, commonly referred to as 
\textbf{TMD factorization}.
As an example of a TMD factorized process, we can once again consider the 
production of two spinless hadrons from an $e^+e-$ scattering where, this 
time, the two hadrons are almost back-to-back in the $e^+e-$ center of mass 
frame. In this case, there are no hard real emissions and the hadronic part of 
the 
cross section is given by (see Eq. (13.31) in Ref.~\cite{Collins:2011zzd}):
\bea
& & W^{\mu\,\nu}(Q,\,p_A,\,p_B) = \frac{8 \pi^3 z_A z_B}{Q^2} \, \sum_f 
H^{\mu\,\nu}_{f,\,\overline{f}}(Q) \,\times \notag \\
& & \times \,\int d^2\vec{k}_{A, \,h \,T} \,d^2\vec{k}_{B, \,h \,T} \, 
\softl(\vec{q}_{h \,T} - \vec{k}_{A, \,h \,T} - \vec{k}_{B, \,h \,T}) \,
D_{H_A / f} (\vec{k}_{A, \,h \,T}) \, D_{H_B / \overline{f}} (\vec{k}_{A, \,h 
\,T}), \nonumber \\ 
\label{eq:tmd_fact}
\eea
where $H^{\mu\,\nu}_{f,\,\overline{f}}(Q)$ is the hard part, $\mathbb{S}$ 
represents the soft factor and the functions $D_{H_i / f}$, for $i = A, \,B$, 
are the TMD FFs associated to the outgoing hadrons and to the fragmenting partons 
of flavor $f$ and $\bar{f}$.

Once again, it is kinematics that determines which fatorization scheme has 
to be used: if the two hadrons are back-to-back then TMD factorization, 
Eq.~\eqref{eq:tmd_fact}, must be applied, otherwise Collinear 
factorization, Eq.~\eqref{eq:coll_fact}, will be appropriate.

\bigskip


\section{Soft Factor}\label{sec:soft_factor} 


\bigskip

In this Section we focus on those kinematical configurations in which there 
is no hard radiation, where TMD factorization has to be applied and the soft 
factor plays a non trivial role. 

From the point of view of the soft gluons, each collinear group is 
simply a bunch of particles strongly boosted in a certain direction.
The boost is so 
strong that the soft gluons are only sensitive to the color charge and to the 
direction of the collinear particles. As a consequence, the propagation of 
collinear particles is well approximated by a Wilson line, see Appendix~\ref{app:WL},  
in the direction of the corresponding collinear group, usually represented by double lines,  
see Eq.~\eqref{eq:soft_kspace}.
In the massless limit, the versor which identifies this 
direction is light-like. However, a light-like Wilson line brings unregulated rapidity 
divergences. In order to cancel them, it is common to introduce a rapidity 
cut-off $y_i$ which tilts the 
corresponding Wilson line away from its original 
light-like direction.
Obviously, the final result for the cross section should not depend on these 
rapidity cut-offs, which then have to be removed in the final stage of the 
computation.
As explained in Ref.~\cite{Collins:2011zzd}, the self-interactions of these 
Wilson lines should not be included into the 
definition of the soft factor.
If $\vec{k}_{S,\,T}$ is the total transverse momentum of the real soft radiation flowing through the 
detector, then the soft factor of a generic process 
is defined as~\cite{Collins:2011zzd}:
\begin{align}
&\softl(\vec{k}_{S,\,T};\, \mu,\,y_i,\,j_k)=
Z_S(\mu,\,\,y_i,\,j_k) 
\, \times \notag \\
&\quad \times \,
\int \frac{d k_{S}^+\,d k_{S}^-}{(2 \pi)^D} \quad 
\begin{gathered}
\includegraphics[width=5cm]{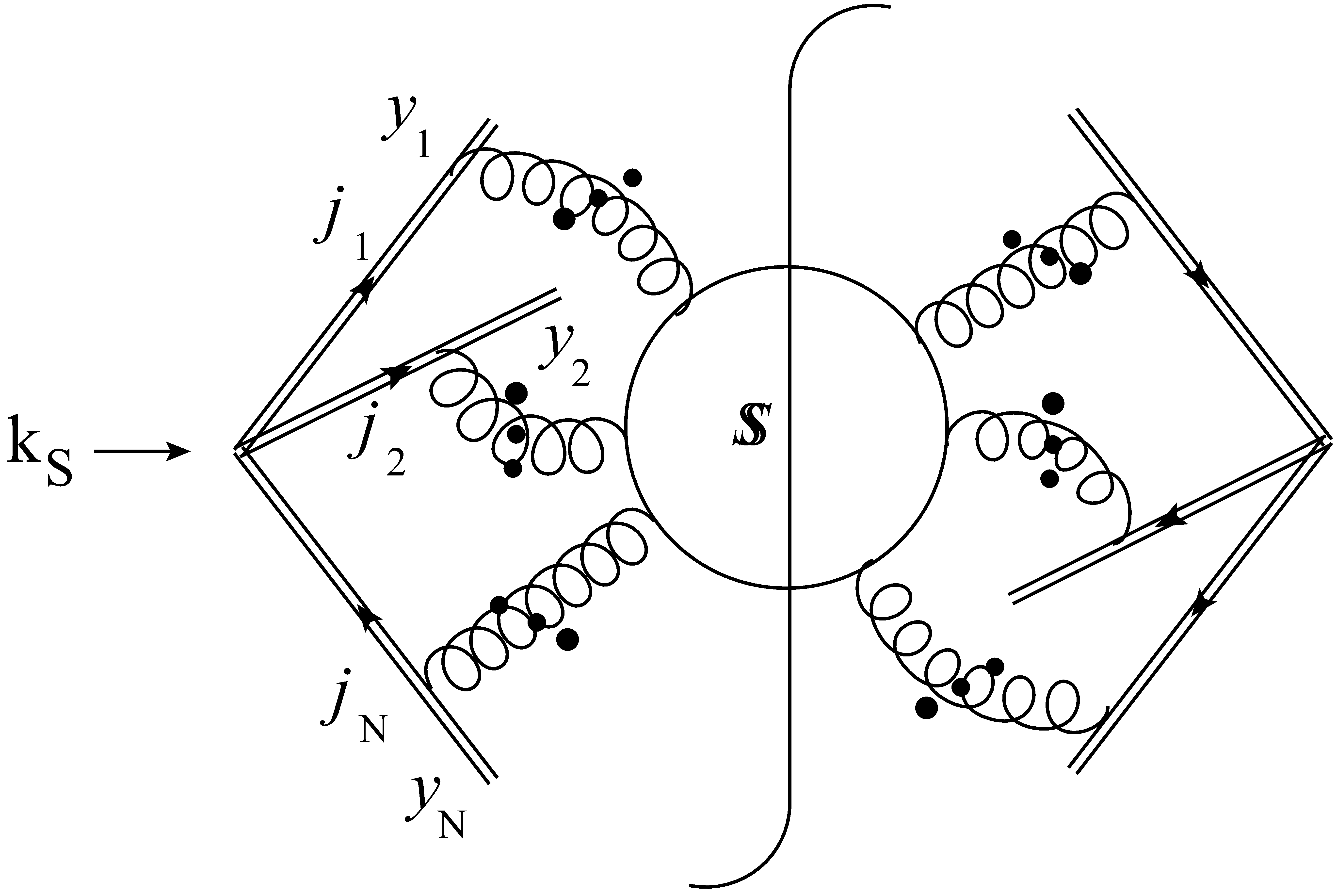}
\end{gathered} 
\Bigg \vert_{\mbox{NO S.I.}},\label{eq:soft_kspace}
\end{align}
where 
$D$ is the dimension of space 
time ($D = 4 - 2 \epsilon$ in dimensional regularization), 
$\mu$ is the renormalization scale and 
$\{y_i\}$ are Lorentz invariant combinations of the rapidity cut-offs ($i = 1, 
\dots N$ where 
$N$ is the number of collinear parts in the process).
The dependence on the 
parton-types $j_1 \dots j_N$ of the partons associated to the 
collinear parts is only on their Wilson line approximation, which changes 
according to their color representation (fermions or gluons).
The label ``NO S.I.'' reminds us not to consider the Wilson lines self energies~\TheBook. 
This implies that $N = 1$ is excluded, since it would correspond only to a 
Wilson line self energy-like contribution.
Finally, the factor $Z_S$ is a UV-renormalization factor that cancels, order by order, the
UV divergences 
generated when 
the integration region stretches outside of the soft region. The role of $Z_S$ will become clear later on, 
when the soft factor will be defined in the Fourier conjugate space, Eq.~\eqref{eq:ft_soft}.

It is important to stress that, with this definition, the soft factor is 
sensitive to the number $N \geq 2$ of collinear groups, each one associated to a reference hadron $h$. 
Therefore it is not totally blind to the rest of the process, but carries some residual information about the overall process.
For this reason, in what follows we will always add a label ``N-h'' to the soft factor $\softl$ in order to take into account this dependence. 

It is usually more convenient to define the soft factor in the 
Fourier conjugate $\vec{b}_T$ space of $\vec{k}_{S,\,T}$, where the quantities involved in the cross section can be identified through an 
operator definition.
In the following, the Fourier transformed quantities will be labeled by a 
tilde. 
In particular, the Fourier transform of the soft factor, $\ftsoft{N}$, is
a matrix in color space, given by the vacuum expectation value of a product of Wilson lines:
\begin{align}
& \ftsoft{N}(\vec{b}_{T};\, \mu,\,y_i,\,j_k)= 
\int d^{D-2} \vec{k}_{S,\,T} \, 
e^{i \, \vec{k}_{S,\,T} \cdot \vec{b}_{T}} 
\,\soft{N}(\vec{k}_{S,\,T},\, \mu,\,y_i,\,j_k) = \notag \\
&\quad = Z_S(\mu,\,\,y_i,\,j_k)
\langle 0 | 
\prod_{i = 1}^N \, 
W_{j_i}(\infty,\,-{\vec{b}_T}/{2}; \, n_i(y_i)\,)^\dagger \, 
\times \notag \\
&\quad\hspace{1.5cm} \times \, 
\prod_{k = 1}^N \, 
W_{j_k}(\infty, \,{\vec{b}_T}/{2};\, n_j(y_j)\,) 
| 0 \rangle \, \vert_{\mbox{NO S.I.}}.
\label{eq:ft_soft} 
\end{align}
The Wilson line $W_{j_i}( \infty,\,{\vec{b}_T}/{2};\, n_i)$ goes from 
${\vec{b}_T}/{2}$ towards infinity in the direction of $n_i$, which is not 
light-like thanks to the rapidity cut-off $y_i$, and has the color 
representation given by the 
parton type $j_i$. 

We can obtain more information about the soft factor by studying its structure in detail. 
Since all the collinear information is replaced by spinless eikonal propagators, $\vec{k}_{S,\,T}$ is the only vector appearing in the soft factor.
Therefore, 
$\softl$ is always rotational invariant 
and depends only on the 
modulus $|\vec{k}_{S,\,T}| = 
k_{S,\,T}$.
This reflects 
on the Fourier conjugate space, 
where the dependence on $\vec{b}_T$ is only through its modulus $|\vec{b}_T| = b_T$.
Moreover 
the natural leading momentum region of $\softl$ is where  
all the momenta are soft, with 
components of size $\lambda_S = {\lambda^2}/{Q}$, where $\lambda << Q$ 
is a very low energy scale.
When the soft factor is Fourier transformed, the total transverse soft 
momentum $\vec{k}_{S,\,T}$ is integrated out and its dependence is replaced by 
$\vec{b}_T$.
At fixed $b_T$ we can roughly access all momenta with $k_{S,\,T} \leq 
\frac{1}{b_T}$, hence this operation can be regarded as a sort of analytic 
continuation of the function $\soft{2}(k_{S,\,T})$ outside of its natural momentum region, since
when $b_T$ is small $k_{S,\,T}$ can be very large.
This generates UV divergences which will have to be canceled order by order by the UV counterterm $Z_S$.

\begin{figure}[t]
\centering
\includegraphics[width=10cm]{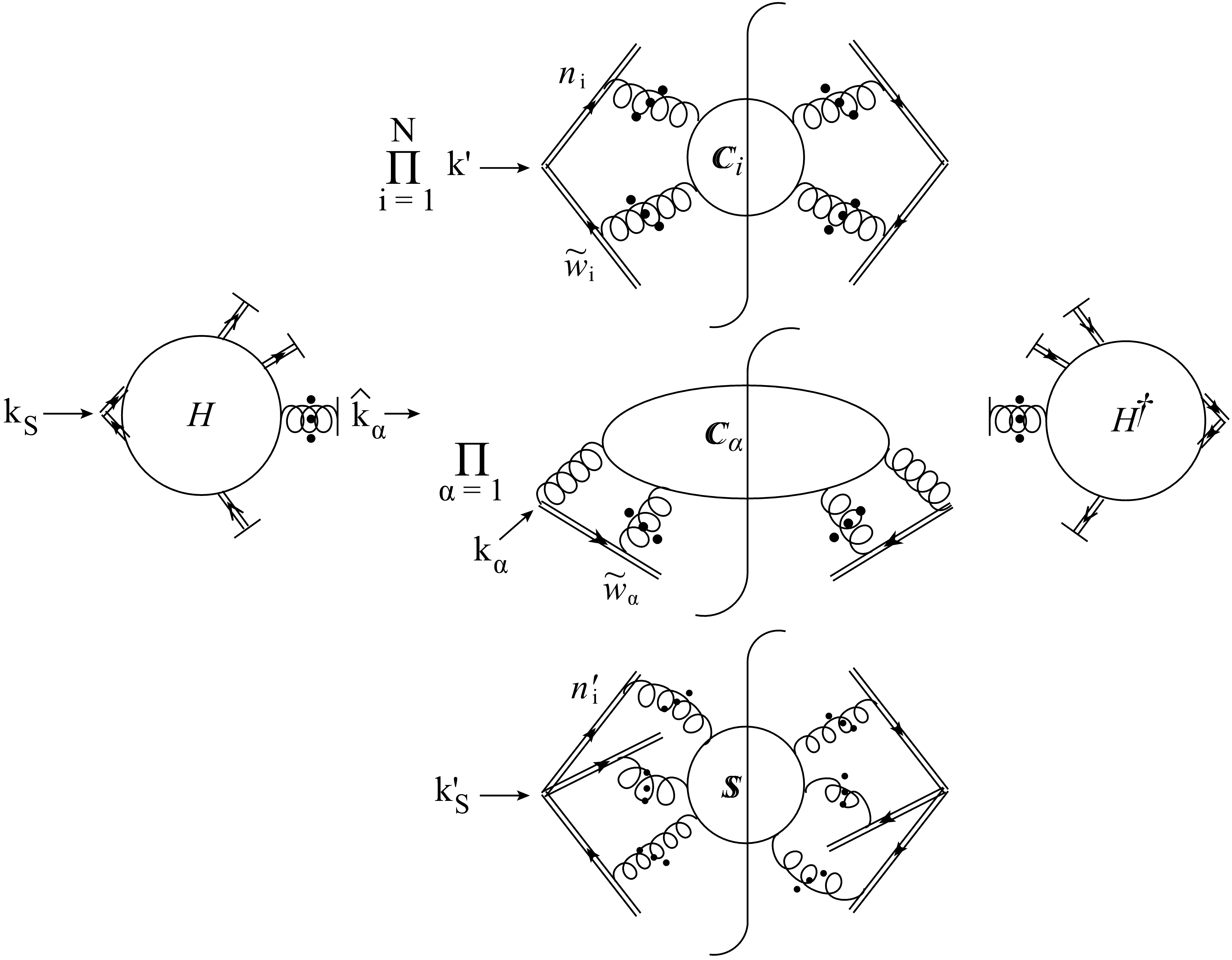}
\caption{Leading regions for the soft factor, $\soft{N}$, at small $b_T$.}
\label{fig:FactSoft}
\end{figure}

The application of the factorization procedure to the soft factor itself gives us the possibility 
to express $\ftsoft{N}$ in terms of perturbative and non-perturbative parts (see~\TheBook).
Leading regions involve hard, collinear and soft subgraphs, as represented in Fig.~\ref{fig:FactSoft}.
%
%
The hard factor is associated to the external Wilson line vertices and contains hard subgraphs with highly virtual loops.
There is a collinear subgraph corresponding to each Wilson line and all of them are connected by the soft subgraph.
Furthermore, if the entering transverse momentum $k_s$ is large enough, there 
can be more hard subgraphs $\coll_{\alpha}$ 
with production of final-state jets of high transverse momentum (i.e. hard gluon 
emissions which cross the cut).
In each hard jet there is a fully inclusive sum/integral over final states, 
hence 
the sum-over-cuts argument presented in \TheBook ~allows us to consider them as 
being far off-shell and part of the hard factor.
In this case collinear factorization holds and the soft factor is unity. 
Furthermore, there is no convolution between the hard part and the collinear 
factors $\coll_i$, since the cut eikonal propagators that exit from the hard 
subgraphs do not carry momentum. 
As a consequence, all the collinear parts are integrated 2-h soft factors and are unity as well.
Therefore, the only remaining effective region is the hard factor with all the 
extra hard jets. 
It has the same structure of $\soft{N}$ but now it is fully computable in 
perturbation theory.
In particular, it is a standard result that general soft functions exponentiate 
and that the exponent can be computed by using web technology, see for example 
Ref.~\cite{White:2015wha,White:2018vfr,Vladimirov:2015fea,Falcioni:2014pka}.
Hence at small $b_T$ the soft factor can be written schematically as:
\begin{align} \label{eq:soft_webs}
\soft{N}
(b_T;\, \mu,\,y_i,\,j_k)
&\overset{\mbox{low }b_T}{\sim}
\int d^{D-2} \vec{k}_{S,\,T} \, 
e^{i \, \vec{k}_{S,\,T} \cdot \vec{b}_{T}} \,
\mbox{exp} \left[ 
\sum_{\mathcal{W}} \mathcal{W}(k_{S,\,T};\, \mu,\,y_i,\,j_k)
\right] ,
\end{align}
where the sum is extended to all the (multiparton) webs and the sums over the diagrams in each web, corresponding to a certain color mixing matrix, has not been shown for simplicity.
In order to separate the small and large $b_T$ behavior of $\ftsoft{N}$, we can modify its functional dependence on $\vec{b}_T$ by 
introducing a function $\vec{b}_T^\star(\vec{b}_T)$ such that 
it coincides with $\vec{b}_T$ at small $b_T$, while at large $b_T$ it is no larger 
than a certain $b_{\rm{max}}$. A possible choice, according to 
Ref.~\cite{Collins:2011zzd,Collins:1984kg,Collins:1989gx}, is given by:
\begin{equation} \label{eq:bstar}
\vec{b}_T^\star \left(b_T\right) = \dfrac{\vec{b}_T}{\sqrt{1 + {b_T^2}/{b_{\rm{max}}^2}}}
\end{equation}
Then, 
by dividing and multiplying $\ftsoft{N}$ in Eq.~\eqref{eq:soft_webs} by its 
small $b_T$ behavior, we easily obtain a factorized expression which holds 
valid 
at any value of $b_T$:
\begin{align}
&\ftsoft{N}(b_T; \, \mu,\,y_i,\,j_k) =
\ftsoft{N}(b_T^\star; \, \mu,\,y_i,\,j_k) \, \times \,
\frac{\ftsoft{N}(b_T; \, \mu,\,y_i,\,j_k)}
{\ftsoft{N}(b_T^\star; \, \mu,\,y_i,\,j_k)}
= \notag\\
&\quad= \,
\int d^{D-2} \vec{k}_{S,\,T} \, 
e^{i \, \vec{k}_{S,\,T} \cdot \vec{b}_T^\star} \,
\mbox{exp} \left[ 
\sum_{\mathcal{W}} \mathcal{W}(k_{S,\,T};\, \mu,\,y_i,\,j_k)
\right] \times 
M_S (b_T; \, \mu,\,y_i,\,j_k),
\label{eq:soft_full}
\end{align}
where $M_S (b_T; \, \mu,\,
\{y_i\}_{i = 1 \dots N}, \,\{j_i\}_{i = 1 \dots N})$ is the fully  
non-perturbative function that models the $N$-h soft factor at large 
$b_T$, while the whole perturbative content is gathered in the webs.

In the t'Hooft limit\footnote{$N_C \rightarrow \infty$ and $\alpha_S \, N_C$ is fixed.}, the soft factor is strongly simplified.
Regarding the perturbative part, the only surviving diagrams are planar and the exponentiation becomes trivial.
Furthermore, we can also make some guess on the non-perturbative part which is, in principle, a fully arbitrary function, since there is no way to extract it independently from experiments.
In this limit the non-perturbative contribution of $\ftsoft{N}$ 
only regards the incoherent emission of  free glueballs, of every possible kind 
\footnote{In order to preserve unitarity the sum must run over \emph{all} the 
possible final states.}.
The function that models 
this kind of emission is a Poisson distribution, similarly to what happens for photons in QED.

\bigskip


\subsection{2-h Soft Factor}\label{subsec:2h_S} 


\bigskip

In the $2$-h class, there are two directions for the collinear parts which can be identified  
to the plus and the minus direction in the c.m.   frame.
The Wilson lines are tilted with respect to these light-like directions by introducing two 
rapidity cut-offs $y_1$ and $y_2$.
The original plus and minus directions are restored if the cut-offs are removed, i.e. by 
taking the limits $y_1 \rightarrow +\infty$ and $y_2 \rightarrow -\infty$.
In total, there are four Wilson lines, two on each side of the final state cut.
The only relevant case for applications involves Wilson lines that replace fermionic collinear partons.
Hence, in the following we will drop the dependence on the 
parton types for simplicity.
Furthermore, the $2$-h soft factor is color singlet, proportional to the identity matrix in color space, i.e. $( \ftsoft{2} )^i_{\;j} \propto \delta^i_{\;j}$.
Then, $\ftsoft{2}$ is defined as the coefficient in front of the delta function. By using the definition in Eq.~\eqref{eq:ft_soft} we have:
\begin{align}
& \ftsoft{2}(\vec{b}_{T};\, \mu, \,y_1 - y_2) =
Z_S(\mu,\,y_1 - y_2)
\times \notag \\
&\quad \times \,  
\frac{\mbox{Tr}_C}{N_C} \,
\langle 0 | 
W(-{\vec{b}_T}/{2}, \,\infty; \, n_1(y_1)\,)^\dagger \, 
W({\vec{b}_T}/{2}, \,\infty; \, n_1(y_1)\,) \,
\times \notag \\
&\quad \times \,  
W({\vec{b}_T}/{2}, \,\infty; \, n_2(y_2)\,)^\dagger \, 
W(-{\vec{b}_T}/{2}, \,\infty; \, n_2(y_2)\,) 
| 0 \rangle \, \vert_{\mbox{NO S.I.}},
\label{eq:ft_soft2} 
\end{align}
where $N_C$ is the number of colors available for quarks and antiquarks (3 in QCD).
The Eq.~\eqref{eq:ft_soft2} describes a loop for the full path outlined 
by the Wilson lines. It starts (e.g.) from $-{\vec{b}_T}/{2}$ and goes to 
${\vec{b}_T}/{2}$, passing through $\infty$, along the almost plus direction 
$n_1$. Then it comes back, again passing through $\infty$, along the almost 
minus direction $n_2$.
Notice that the only Lorentz invariant combination for a function depending on 
two rapidities (e.g. $y_1$ and $y_2$) is their difference (e.g. $y_1 - y_2$).
It is possible to write the evolution 
equation for $\soft{2}$ in the $b_T$-space with respect to both  
rapidity cut-offs, $y_1$ and $y_2$, using a single 
rapidity-independent kernel $\widetilde{K}(\vec{b}_T;\,\mu)$ defined 
as~\cite{Collins:2011zzd}:
\begin{align}
&\lim_{y_2 \to -\infty} \frac{\partial \log{\ftsoft{2}(\vec{b}_{T}; \, \mu, \, 
y_1 - y_2)}}{\partial y_1} = 
\frac{1}{2}\, \widetilde{K}(\vec{b}_T;\,\mu) \label{eq:soft2evo_1}\\
&\lim_{y_1 \to +\infty} \frac{\partial \log{\ftsoft{2}(\vec{b}_{T}; \, \mu, \, 
y_1 - y_2)}}{\partial y_2} = 
- \frac{1}{2}\, \widetilde{K}(\vec{b}_T;\,\mu) \,, \label{eq:soft2evo_2}
\end{align}
It has an anomalous dimension $\gamma_K$:
\begin{equation} \label{eq:gammaK_evo}
\frac{d  \widetilde{K}(\vec{b}_T;\,\mu)}{d \log{\mu}} = - 
\gamma_K(\alpha_s(\mu)),
\end{equation}
where $\gamma_K$ depends on $\mu$ through the strong coupling $\alpha_S$ and is independent of $b_T$.
Then, $\widetilde{K}$ can be written as:
\begin{equation} \label{eq:K_evo_sol}
\widetilde{K}(\vec{b}_T;\,\mu) = \widetilde{K}(\vec{b}_T;\,\mu_0) -
\int_{\mu_0}^\mu \, \frac{d \mu'}{\mu'} \, \gamma_K(\alpha_s(\mu')) .
\end{equation}
For large values of $(y_1 - y_2)$, the solution to the evolution equations for $\ftsoft{2}$ is given by:
\begin{align}
\ftsoft{2}&(\vec{b}_T; \, \mu, \, y_1 - y_2) = \ftsoft{2}(\vec{b}_{T}; \, 
\mu_0, 
\, 0) \, 
\mbox{exp} \Big \{ \frac{y_1 - y_2}{2} \, \widetilde{K}(\vec{b}_T;\,\mu)\Big \} 
+ \mathcal{O}\Big( e^{-(y_1 - y_2)} \Big) = \notag \\
&= \ftsoft{2}(\vec{b}_{T}; \, \mu_0, \, 0) \, 
\mbox{exp} \Big \{ - \frac{y_1 - y_2}{2} \Big[ \int_{\mu_0}^{\mu}\, \frac{d 
\mu'}{\mu'}\, \gamma_K(\mu) \; - 
\widetilde{K}(\vec{b}_T;\,\mu_0)\Big]\Big \} \,
+ \mathcal{O}\Big( e^{-(y_1 - y_2)} \Big), \label{eq:soft_evo_sol}
\end{align}
where the reference values of the RG scale and of the rapidities are 
chosen to be $\mu_0$ and $y_{1,\,0} = y_{2,\,0}$, respectively.
In the solution of the evolution equation, 
two functions appear: the fixed scale 
soft factor $\ftsoft{2}(b_T; \, \mu_0, \, 0)$ and the soft 
kernel $\widetilde{K}(b_T;\,\mu)$.
Both of them can be separated in terms of their perturbative and 
non-perturbative contents 
by using the $b^\star$ prescription, similarly to what was done in 
Eq.~\eqref{eq:soft_full}:
\begin{align}
&\ftsoft{2}(b_T; \, \mu_0, \, 0) = \ftsoft{2}(b_{T}^\star; \, \mu_0, \, 0)  \, 
M_S^{(0)}(b_T) \label{eq:soft_ref_star} \,; \\
&\widetilde{K}(b_T;\,\mu)  = \widetilde{K}(b_T^\star;\,\mu) - g_K(b_T) 
\,.\label{eq:K_star}
\end{align}
Finally, consistency between 
Eqs.~\eqref{eq:soft_full} and \eqref{eq:soft_evo_sol} requires that:
\begin{align}
&\lim \limits_{\substack{y_1 \to +\infty\\y_2 \to -\infty}} \, 
\int d^{D-2} \vec{k}_{S,\,T} \, 
e^{i \, \vec{k}_{S,\,T} \cdot \vec{b}_T^\star} \,
\mbox{exp} \left[ 
\sum_{\mathcal{W}} \mathcal{W}(k_{S,\,T};\, \mu,\,y_1-y_2)
\right] = \notag \\
&\quad = \frac{y_1 - y_2}{2}\widetilde{K}(b_T^\star;\,\mu) = 
\frac{y_1 - y_2}{2} \left[\widetilde{K}(b_T^\star;\,\mu_0)  - 
\int_{\mu_0}^{\mu}\, \frac{d \mu'}{\mu'}\, \gamma_K(\mu') \right]; 
\label{eq_webs_comp} \\
&\lim \limits_{\substack{y_1 \to +\infty\\y_2 \to -\infty}} \, M_S(b_T; \, 
\mu,\,y_1 - y_2) = M_S^{(0)}(b_T) \, e^{-\frac{y_1 - y_2}{2} \, g_K(b_T)} \,; 
\notag \\
&\;\;\;\;\ftsoft{2}(b_T^\star; \, \mu_0, \, 0) = 1 \,. \label{eq_softref_comp} 
\end{align}
Notice that the non-perturbative function $M_S(b_T; \, \mu,\,y_1 - y_2)$ loses 
its dependence on $\mu$ in the large rapidity limit, as $g_K$ does not depend on the RG scale.
Since we are only interested in the asymptotic behaviour of $\ftsoft{2}$, we will drop 
the label $(0)$ from $M_S(b_T)$ 
and we will refer to it as the \textbf{soft model}, i.e. the 
non-perturbative part which will have to be parametrized and treated 
phenomenologically, possibly taking inspiration from the properties of the soft factor in the t'Hooft limit. 
The two non-perturbative functions $M_S$ and $g_K$ should not contribute at 
small $b_T$ by definition, 
hence we require that $g_K(b_T)\rightarrow 0$ and $M_S(b_T) \rightarrow 1$ when 
$b_T \rightarrow 0$. 
Furthermore, since the Fourier transform of $\ftsoft{2}$ has to be well behaved, 
the contribution 
of $g_K$ and $M_S$ should be suppressed at large $b_T$. Notice that the factor 
in front of $g_K$, being proportional to the difference of the rapidity 
cut-offs, is always large and negative in the large rapidity cut-off limit.
In conclusion, the $2$-h soft factor in $b_T$ space can be written as:
\begin{equation} \label{eq:soft_asy}
\ftsoft{2}(b_T; \, \mu,\,y_1 - y_2) = e^{\frac{y_1 - 
y_2}{2}\widetilde{K}(b_T^\star;\,\mu)} \, 
M_S(b_T) \, e^{-\frac{y_1 - y_2}{2} \, g_K(b_T)} + 
\mathcal{O}\Big( e^{-(y_1 - y_2)} \Big) \,.
\end{equation}
This result shows that the soft factor itself 
can be factorized in a purely perturbative part, process dependent 
but calculable within pQCD, and a part which is genuinely 
non perturbative and, inevitably, will have to be committed to a 
phenomenological model, in this case embedded in the functions $M_S(b_T)$ and 
$g_K(b_T)$.

\bigskip

Although the definition of Eq.~\eqref{eq:ft_soft2} implies that $\ftsoft{2} = 1$ at $b_T = 0$, a direct fixed order perturbative computation of $\widetilde{K}$ does not reproduce the correct behavior  in this region.
In this regard, since the soft factor is unity at $b_T = 0$, then $\widetilde{K}$ goes to zero at small $b_T$, but an explicit calculation gives instead a larger and larger value as $b_T$ decreases, forcing $\ftsoft{2}$ to vanish in $b_T = 0$.
This kind of problems arise because the integrated soft factor can be defined through perturbative QCD only as a bare quantity.
A solution can be found by applying some regularization procedure, for instance one can modify the $b^\star$ prescription of Eq.~\eqref{eq:bstar} allowing for the introduction of a new parameter $b_{\mbox{\tiny MIN}} \neq 0$ that provides a minimum value for $b_T$ (see Appendix~\ref{app:smallbT}).

\bigskip


\section{Collinear Parts and TMDs}\label{sec:coll_parts} 


\bigskip

Let's now consider a generic collinear part. 
If kinematics forbid hard real emissions, the information about 
the transverse 
momentum $ \vec{k}_{T}$ of the reference parton survives. 
All the collinear particles are boosted very strongly in the collinear group
direction, that we can identify with the plus direction without loss of generality. To them, everything outside of the collinear group is 
moving very fast in the opposite direction, so fast that the only surviving information is the color 
charge and the direction.
In other words, as seen from the collinear factor,  the rest of the process is well approximated  by a light-like Wilson line flowing in the direction opposite to that of the collinear group.

Assuming for simplicity that the reference parton is a quark, if $\vec{k}_T$ is the total transverse momentum of the collinear group, then the collinear factor (along the plus direction) is defined as in \TheBook:
\begin{align}
&\coll_{j,\,H}(\xi,\, \vec{k}_{T}; \, \mu, \, y_P ,\,-\infty) = 
Z_C(\mu,\, y_P ,\,-\infty) \times \notag \\
&\quad\times \frac{\mbox{Tr}_{\mbox{\small 
C}}}{N_C}\,  \int \frac{d k^-}{(2 \pi)^D}\,
\begin{cases}
 \begin{gathered} \hspace{.5cm}
 \includegraphics[width=3.5cm]{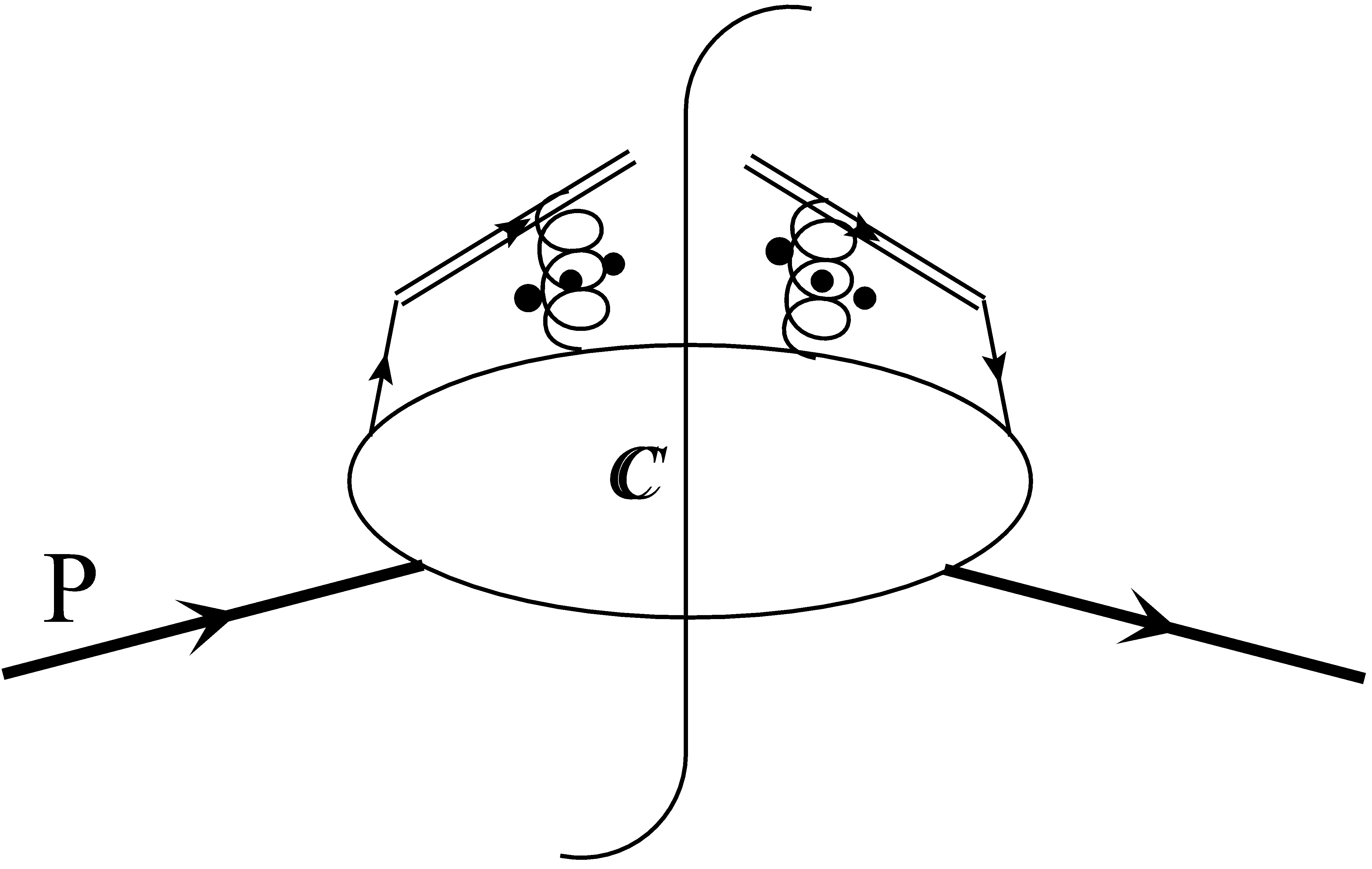}
 \end{gathered} \hspace{.2cm}\Bigg \vert_{\mbox{NO S.I.}}\, &\mbox{ initial state,} \\
\dfrac{1}{\xi} \times \,
 \begin{gathered}
 \includegraphics[width=3.0cm]{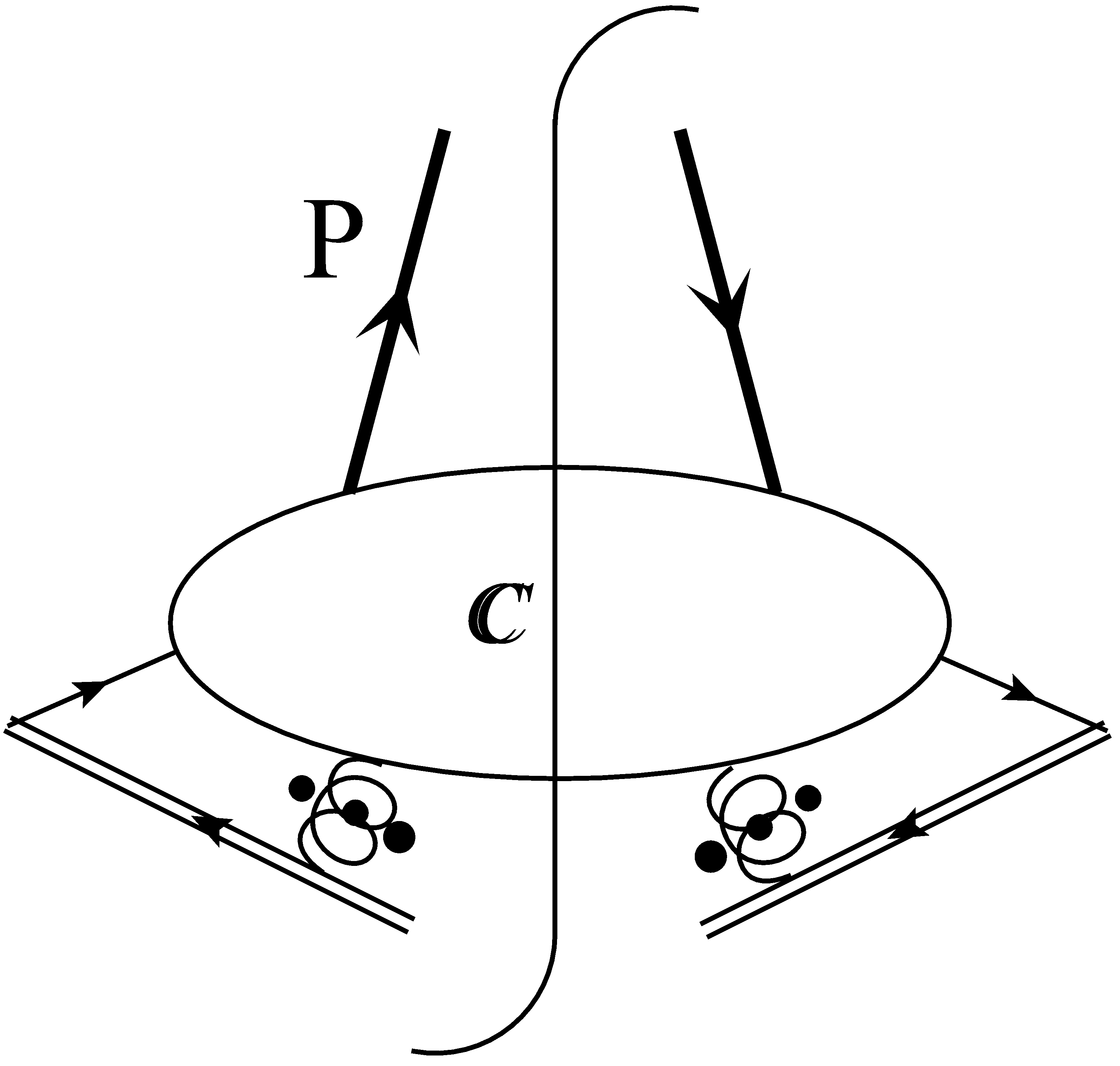}
 \end{gathered} \hspace{.4cm}\Bigg \vert_{\mbox{NO S.I.}}\, &\mbox{ final state,}
\end{cases}
\label{eq:coll_kspace}
\end{align}
where the color average ${\mbox{Tr}_{\mbox{\small C}}}/{N_C}$ is due to the fact that collinear factors are color singlets and, analogously to the $2$-h soft factor, they are defined as the coefficient in front of the delta in color space.
The variable $\xi$ is the light-cone fraction of the 
momentum $k$ of the reference parton, quark of flavor $j$, with respect to the momentum $P$ of the reference hadron $H$, $\mu$ is the renormalization 
scale at 
which $\coll$ is evaluated and $y_P$ is the (very large) rapidity of the reference 
hadron. 
The definition of $\xi$ in the initial and final state is given by:
\begin{align}
\xi = 
\begin{cases}
x = \frac{k^+}{P^+}
&\mbox{ initial state hadron;} \\[10pt]
z = \frac{P^+}{k^+}
&\mbox{ final state hadron,}
\end{cases}
\label{eq:xi_def}
\end{align}
To the Wilson line, instead, we can associate a very large and negative rapidity. 
Similarly to the definition in Eq.~\eqref{eq:soft_kspace}, $Z_C$ is the UV-counterterm of $\coll$, while the label ``NO S.I.'' reminds us not to consider the Wilson lines self interactions.
It is important to stress that the collinear factor $\coll$ is totally blind 
to the rest of the process, it only depends on its 
\emph{intrinsic} variables.
As for the soft factor, the operator definition is simpler 
in the Fourier conjugate space.
Here, we have:
\begin{align}
&\widetilde{\coll}_{j,\,H}(\xi,\, \vec{b}_{T}; \, \mu, \, y_P ,\,-\infty)  = 
\int d^{D-2} \vec{k}_T \, e^{i \, \vec{k}_T \cdot \vec{b}_T} \,\coll_{j,\,H}(\xi,\, 
\vec{k}_{T}; \, \mu, \, y_P,\,-\infty) = \notag \\
&\quad= Z_C(\mu,\, y_P ,\,-\infty) \, \frac{\mbox{Tr}_{\mbox{\small C}}}{N_C} \, \int \frac{d x^-}{2 \pi} \, 
e^{i k^+ \, x^-} \times \notag \\
&\quad\times \begin{cases}
\langle P \,(H) | \overline{\psi}(-{x}/{2}) \, W_j(-{x}/{2}, \, {x}/{2};\, w_{-}\,) 
\, \psi({x}/{2})| P \,(H) \rangle \, \vert_{\mbox{NO S.I.}} 
\, &\mbox{ initial state,}  \\
\begin{aligned}
\dfrac{1}{\xi} \, &\times \, \sum\limits_{X} \, 
\langle P\,(H),\,X;\,\mbox{out} | \overline{\psi}(-{x}/{2}) \, W_j(-{x}/{2}, \infty;\, 
n_1(y_1)\,)^\dagger | 0 \rangle \times \\
&\quad\times \, \langle 0 | W_j({x}/{2}, \infty;\, w_{-}\,) \, \psi({x}/{2})|  
P\,(H),\,X;\,\mbox{out} \rangle \, \vert_{\mbox{NO S.I.}}
\end{aligned}
\, &\mbox{ final state,}
\end{cases}
\label{eq:ft_coll}
\end{align}
where $x = (0, x^-, \vec{b}_T)$ and $w_{-}$ is the light-like minus direction of the Wilson line.

The presence of a light-like Wilson line allows for 
particles with a low, or even a very large negative rapidity, to be considered as part of the collinear group. 
This contradiction reflects in the computation by inducing the presence of unregulated rapidity divergences.
This problem can be solved by subtracting out these unphysical contributions from the collinear factor 
by using the subtraction method described in \TheBook. 
Since all the non truly collinear contributions are due to the overlapping with the soft region, they can be  rearranged in one global term which 
turns out to be a $2$-hadron soft factor.
Hence we can use the definitions given in Eqs.~\eqref{eq:soft_kspace} and \eqref{eq:ft_soft} to subtract them out and obtain: 
\begin{align}
\widetilde{\coll}_{j,\,H}^{\mbox{\small sub}} (\xi,\, \vec{b}_{T}; \, \mu, \, y_P - y_1) &= 
Z_C^{\mbox{\small sub}}(\mu,\, y_P - y_1) \,
Z_2 \left(\alpha_S(\mu)\right) \times \notag \\
&\quad \times
\lim_{y_{u_2} \to -\infty}
\dfrac{\widetilde{\coll}_{j,\,H}^{(0)} (\xi, \,\vec{b}_{T}; \, \mu, \, 
y_P - y_{u_2})}
{\ftsoft{2}^{(0)} (b_{T}; \, \mu, \,y_1 - y_{u_2})} \,,
\label{eq:sub_coll}
\end{align}
where $y_1$ is the rapidity cut-off carried by $\soft{2}$ that, similarly to the case of the soft factor, should be removed in
the final result for the cross section.
After subtraction, the particles in $\coll$ can only have a rapidity $y$ such that $y_1 < y < y_P \sim +\infty$. 
Hence if $y_1$ is chosen to be sufficiently large, only strongly boosted particles in the plus direction contribute to $\coll$, according to the naive physical intuition.
The subtracted collinear factor has its own UV-counterterm $Z_C^{\mbox{\small sub}}$, for this reason in the previous definition the quantities inside the limit are bare, in the sense that they have to be considered without their UV-renormalization factors. 
Since the unsubtracted collinear part is defined with renormalized quark fields, see Eq.~\eqref{eq:ft_coll}, then if $Z_C^{\mbox{\small sub}}$ is the ratio of the renormalized collinear part to the unrenormalized collinear part, we have to multiply explicitly by the wave-function renormalization factor of the quark field, $Z_2$.

Having given the general definition of $\coll$, TMDs can be obtained straightforwardly.
In fact, as $\coll$  is an operator acting onto the space of Dirac spinors, it belongs to the Clifford algebra built from the Dirac matrices $\left \{ \gamma^\mu \right \}$. Therefore, we simply expand $\coll$ on the basis of this 
algebra. Neglecting all the dependences on partonic and hadronic variables, 
we have:
\begin{equation} \label{eq:FF_Clifford}
\coll^{\mbox{\small sub}} = \mathcal{S} \, \mathbb{I} +  \mathcal{V}^\mu \, \gamma_\mu + 
\mathcal{A}^\mu \, \gamma^5 \, \gamma_\mu + i \, \mathcal{P} \, \gamma^5 + i 
\mathcal{T}^{\mu \, \nu} \, \sigma_{\mu \, \nu} \gamma^5.
\end{equation}
Then, the TMDs are related to the coefficients $\mathcal{S},\mathcal{V}, 
\dots, \mathcal{T}^{\mu \, \nu}$ of the Clifford Algebra expansion and
the definition in Eq.~\eqref{eq:sub_coll} naturally extends to TMDs.
Such coefficients can be further expanded in terms of all the Lorentz tensors 
contributing to the leading twist approximation (see e.g. 
Ref.~\cite{Barone:2001sp}).
This allows to isolate all the dependence on the vector part of $\vec{b}_T$ in the coefficients of such expansion, leaving a set of scalar functions depending only on the modulus $b_T$.
These scalar functions are the TMDs.
For example, the coefficient of $\gamma^+$ defines the unpolarized TMDs and the Sivers function:
\be \label{eq:unp_Clifford}
\mathcal{V}^+  = {1}/{4}\, 
\mbox{Tr}_{\mbox{\small Dirac}} \, \bigl[ \gamma^+ \coll^{\mbox{\small sub}} \bigr] = 
\begin{cases}
f_1 - \frac{1}{M} |\vec{S}_T \times \vec{k}_T| f_{1 T}^\perp \, &\mbox{ initial state} \,,\\
D_1 - \frac{1}{M} |\vec{S}_T \times \vec{k}_{h,T}| D_{1 T}^\perp \, &\mbox{ final state}\,.
\end{cases}
\ee
Formally, if $C$ is a generic TMD function referring to a collinear factor in the plus direction, then its definition equipped with subtractions is inherithed directly from Eq.~\eqref{eq:sub_coll} and it is given by:
\begin{align}
&\widetilde{C}_{j,\,H}^{\mbox{\small sub}} (\xi,\, b_T; \, \mu, \, y_P - y_1) = 
\left(Z_{\mbox{\tiny TMD}}\right)_j(\mu,\, y_P - y_1) 
\, Z_2 \left(\alpha_S(\mu)\right)
\times \notag \\
&\quad \times
\lim_{y_{u_2} \to -\infty}
\dfrac{\frac{1}{4} \mbox{Tr}_{\mbox{\small Dirac}} \left[
\Gamma \,
\widetilde{\coll}_{j,\,H}^{(0)} (\xi, \,\vec{b}_{T}; \, \mu, \, y_P - y_{u_2})\right]^{\substack{\mbox{\footnotesize leading}\\\mbox{\footnotesize twist coeff.}}}
}
{\ftsoft{2}^{(0)} (b_T; \, \mu, \,y_1 - y_{u_2})}\, = 
\notag \\
&\quad= 
\frac{1}{4} \mbox{Tr}_{\mbox{\small Dirac}} \left[
\Gamma \,
\widetilde{\coll}_{j,\,H}^{\mbox{\small sub}} (\xi, \,\vec{b}_{T}; \, \mu, \, y_P - 
y_{u_2})\right]^{\substack{\mbox{\footnotesize leading}\\\mbox{\footnotesize twist coeff.}}},
\label{eq:fact_defTMDs}
\end{align}
where $\Gamma$ is the proper Dirac matrix combination to extract the desidered TMD and $Z_{\mbox{\tiny TMD}}$ is its own UV counterterm.
The label ``leading twist coeff." means that the TMDs are obtained, after the 
projection onto the Clifford Algebra, as the coefficients of the expansion at 
leading twist.
The operator definition of TMD as given in Eq.~\eqref{eq:fact_defTMDs}, which follows directly from the TMD factorization prescription, will be referred to 
as the {\bf factorization definition}. Notice that within this definition, the 
TMD is a purely collinear object, as all soft sub-divergences have been subtracted out.

\bigskip


\subsection{Evolution Equations for TMDs}\label{subsec:evo_tmds} 


\bigskip

In the factorization definition\footnote{In the following, we will drop the superscript ``sub" since, from now on, we will always refer to subtracted quantities.} of TMDs, Eq.~\eqref{eq:fact_defTMDs}, a $2$-h soft factor appears as a consequence of the subtraction mechanism.
Therefore, we can use the results of Section~\ref{subsec:2h_S} to write the evolution equation (Collins-Soper evolution) for $\widetilde{C}$ with respect to the rapidity cut-off $y_1$.
On the other hand, the evolution with respect to the scale $\mu$ (i.e. the Renormalization Group evolution) is ruled by the 
anomalous dimension $\gamma_C$. 
The equations are given by:
\begin{align}
&\dfrac{\partial \log{\widetilde{C}_{j,\,H}(\xi,\, b_{T}; \, \mu, 
\,\zeta)}}{\partial 
\log{\sqrt{\zeta}}} = 
\frac{1}{2} \widetilde{K}(b_T;\,\mu) \,, \label{eq:CS_evo} \\
&\dfrac{\partial \log{\widetilde{C}_{j,\,H}(\xi,\, b_{T}; \, \mu, \, 
\zeta)}}{\partial 
\log{\mu}} = 
\gamma_C \left(\alpha_S(\mu),\, \frac{\zeta}{\mu^2} \right) \,,\label{eq:RG_evo}
\end{align}
which, for later convenience, have been re-written in terms of a new 
variable, $\zeta$, defined as follows:
\begin{equation} \label{eq:zeta}
\begin{cases}
\zeta = \left(M x\right)^2 \, e^{2 (y_P - y_1)}
&\mbox{ initial state hadron;} \\
\zeta = \left(\frac{M}{z}\right)^2 \, e^{2 (y_P - y_1)}
&\mbox{ final state hadron,}
\end{cases}
\end{equation}
where $M$ is the mass of the reference hadron, 
while $x$ and $z$ are the light-cone fractions of the momentum of the reference parton with respect to the hadron. 
Thanks to the definitions in Eq.~\eqref{eq:xi_def}, in both  initial and final states we can write $\zeta \sim Q^2 e^{-2 y_1}$.
In addition to the previous evolution equations, we also have the RG evolution of $\widetilde{K}$, Eq.~\eqref{eq:gammaK_evo}, and the CS evolution of $\gamma_C$, given by:
\begin{align}
\frac{\partial \gamma_C \left(\alpha_S(\mu),\, {\zeta}/{\mu^2} \right)}
{\partial \log{\sqrt{\zeta}}} = -\frac{1}{2}\gamma_K(\alpha_S(\mu)) ,
\label{eq:gammaG_evo}
\end{align}
which gives:
\begin{equation} \label{eq:gammaG_evo_sol}
\gamma_C \left(\alpha_S(\mu),\, {\zeta}/{\mu^2} \right) = 
\gamma_C \left(\alpha_S(\mu),\,1 \right) - 
\frac{1}{4} \gamma_K (\alpha_S(\mu)) \log{\frac{\zeta}{\mu^2}}.
\end{equation}
With the help of Eqs.~\eqref{eq:gammaK_evo},\eqref{eq:gammaG_evo} and 
\eqref{eq:gammaG_evo_sol}, 
we can rewrite the solution to Eqs.~\eqref{eq:CS_evo} and \eqref{eq:RG_evo}  
as~\cite{Collins:2011zzd}:
\begin{align}
&\widetilde{C}_{j,\,H}(\xi,\, b_{T}; \, \mu, \,\zeta) = \widetilde{C}_{j,\,H}(\xi,\, b_{T}^\star; \, \mu_0, \,\zeta_0) \times 
\notag \\
& \times \,
\mbox{exp} \Big \{
\frac{1}{4} \, \widetilde{K}(b_T^\star;\,\mu_0) \, \log{\frac{\zeta}{\zeta_0}} 
+ 
\int_{\mu_0}^{\mu} \frac{d \mu'}{\mu'} \, \left[ \gamma_C(\alpha_S(\mu'),\,1) - \frac{1}{4} \, 
\gamma_K(\alpha_S(\mu')) \, \log{\frac{\zeta}{\mu'^2}}\right]
\Big \} \times \notag \\
& \times \left(M_C\right)_{j,\,H}(\xi,\,b_T) \mbox{ exp} \Big \{
-\frac{1}{4} \, g_K(b_T) \, \log{\frac{\zeta}{\overline{\zeta}_0}}
\Big \}  \label{eq:tmd_sol}
\end{align} 
where the standard choices for the reference values of the scales are\footnote{Notice 
that the reference value of $\zeta$ is different in the 
perturbative and in the non-perturbative parts. This follows from the 
application of the evolution equation to ${\widetilde{C}(\xi,\, \vec{b}_{T}; \, 
\mu_0, \,\zeta_0)}/{\widetilde{C}(\xi,\, \vec{b}_{T}^\star; \, \mu_0, 
\,\zeta_0)}$, which gives $M_C(b_T) \mbox{ exp} \left( -{1}/{4} g_K(b_T) 
\log{{\zeta_0}/{\overline{\zeta}_0}}\right)$.}:
\begin{align}
&\mu_0 = \mu_b = \frac{2 e^{-\gamma_E}}{b_T^\star} \,; \label{eq:mub} \\
&\zeta_0 = \mu_b^2 \,;  \label{eq:zetab} \\
&\hspace{-.3cm}\begin{cases}
\overline{\zeta}_0 = \left(M x \right)^2
&\mbox{ initial state;} \\
\overline{\zeta}_0 = \left(\frac{M}{z} \right)^2
&\mbox{ final state.} 
\end{cases}\label{eq:zeta_ref_st}
\end{align}
In the solution of the evolution equation the $b_T^\star$ prescription, Eq.~\eqref{eq:bstar}, has been used in 
order to separate the perturbative from the non-perturbative content, in 
complete analogy to what was done for the soft factor in Section~\ref{subsec:2h_S}.
In particular, in Eq.~\eqref{eq:tmd_sol}, the non-perturbative behavior of the 
TMD is described by two functions.
The first is $g_K$, the same function that appears in Eq.~\eqref{eq:soft_asy} in the 
asymptotic behavior of $\ftsoft{2}$. 
The second is the \textbf{TMD model} function  $\left(M_C\right)_{j,\,H}(\xi,\,b_T)$, that embeds the 
genuine 
non-perturbative behavior of the TMD: it depends on the flavor of the reference parton 
and on the reference hadron associated to the collinear part. 
By definition, the model should not influence the TMD at small $b_T$. Furthermore, since the Fourier transform of the TMD has to be well behaved, the model should be sufficiently suppressed at large $b_T$\footnote{Although the function $g_K$ gives a suppression factor in Eq.~\eqref{eq:tmd_sol}, it is modulated by (minus) the logarithm of $\zeta$ and consequently it may create problems when the rapidity cut-off becomes too low.}. These properties restrict the behaviour of the non-perturbative function $M_C$ at small and large $b_T$ as follows
\begin{align}
&\lim_{b_T \to 0} M_C(\xi,\,b_T) = 1 ; 
&\lim_{b_T \to \infty} M_C(\xi,\,b_T) = 0 .
\label{eq:model_prop}
\end{align}

The factorization procedure can be applied either to the full collinear factor or to the TMDs themselves, 
in order to study their behavior at small $b_T$, outside of their natural collinear momentum region.
This is given by a convolution of a finite (calculable in perturbative QCD) hard coefficient $\mathcal{C}$ with the TMD \emph{integrated} over $\vec{k}_T$.
The proof can be found in Chapter 13 of \TheBook.
Hence, TMDs at small $b_T$ can be written as Operator Product Expansions (OPE):
\begin{align}
&\widetilde{C}_{j,\,H}(\xi,\, b_{T}; \, \mu, \,\zeta)
\overset{\mbox{low }b_T}{\sim} \hspace{.1cm}
\widetilde{\mathcal{C}}_j^{\,k}(b_{T}; \, \mu, \,\zeta) \otimes c_{k,\,H} (\mu) ,
= \notag \\
&\quad\hspace{.5cm}=
\begin{cases}
\left(
\widetilde{\mathcal{C}}_j^{\hspace{.2cm}k}(b_{T}; \, \mu, \,\zeta) 
\otimes 
f_{k/H}(\mu ) \right) (x)
&\mbox{ initial state;} \\[15pt]
z^{-2+2\epsilon} \left(
d_{H/k} (\mu)
\otimes 
\widetilde{\mathcal{C}}_{\hspace{.2cm}j}^{k}(b_{T}; \, \mu, \,\zeta) \right) (z)
&\mbox{ final state.} 
\end{cases}
\label{eq:ope_tmds}
\end{align}
where $\widetilde{\mathcal{C}}_j^{\,k}$ are the Wilson Coefficients of the OPE, which are matrices in the flavor space. 
A sum over $k$ is implicit.
In the second line of Eq.~\eqref{eq:ope_tmds} we distinguish the Wilson Coefficients of the initial state from those corresponding to the final state according to the position of their upper and lower flavor indices. 
The convolution $\otimes$ of two generic functions $f$ and $g$ is 
defined as
\begin{align}
\left(f \otimes g\right)(x) = \int_x^1 \, \frac{d \rho}{\rho} f({x}/{\rho}) g(\rho) ,
\label{eq:conv_def}
\end{align}
where we recall that the Wilson Coefficients of the final state have a normalization factor $\rho^{2-2\epsilon}$ when the convolution is made explicit, see Ref.~\cite{Echevarria:2016scs}.
The integrated TMDs are indicated by lowercase letters. In the following, $c_{k,\,H}$ will be a generic integrated TMD, while $f$ will label integrated TMD PDFs and $d$ will refer to integrated TMD FFs.
Thanks to the OPE, the solution of the evolution equations, Eq.~\eqref{eq:tmd_sol}, can be rewritten as
\begin{align}
&\widetilde{C}_{j,\,H}(\xi,\, b_{T}; \, \mu, \,\zeta) =
\left(
\widetilde{\mathcal{C}}_j^{\;k}(b_{T}^\star; \, \mu_0, \,\zeta_0) \otimes c_{k,\,H} (\mu_0) 
\right) (\xi) \,
\times \notag \\
&\quad \times
\mbox{ exp} \Big \{
\frac{1}{4} \, \widetilde{K}(b_T^\star;\,\mu_0) \, \log{\frac{\zeta}{\zeta_0}} 
+
\int_{\mu_0}^{\mu} \frac{d \mu'}{\mu'} \, \left[ \gamma_C(\alpha_S(\mu'),\,1) - \frac{1}{4} \, 
\gamma_K(\alpha_S(\mu')) \, \log{\frac{\zeta}{\mu'^2}}\right]
\Big \} \times \notag \\
&\quad \times
\left(M_C\right)_{j,\,H}(\xi,\,b_T) \mbox{ exp} \Big \{
-\frac{1}{4} \, g_K(b_T) \, \log{\frac{\zeta}{\overline{\zeta}_0}}
\Big \} . \label{eq:tmd_sol_2}    
\end{align}
%
The definition of integrated TMDs coincides with the Fourier transformed TMDs in $b_T = 0$.
Perturbative QCD fails to give the right result in $b_T = 0$ because of the new UV divergences introduced by the integral over the whole range of $k_T$.
In fact, as we explain in Appendix~\ref{subapp:smallbT_tmds}, $\widetilde{C}$ goes to zero as $b_T \rightarrow 0$ (see Eq.~\eqref{eq:tmd_smallbT}) and the usual collinear PDFs and FFs are not recovered.
This problem is completely analogous to that encountered in Section~\ref{subsec:2h_S} and it can be solved in a similar way, by defining a regularization procedure for the definition of the integrated TMDs (see Appendix~\ref{app:smallbT}).

\bigskip


\subsection{Rapidity dilations}\label{subsec:rap_dil} 


\bigskip

 In the operator definition of the TMD, Eq.~\eqref{eq:fact_defTMDs}, we introduced a rapidity cut-off $y_1$, required in the subtraction mechanism of the overlapping between soft and collinear momentum regions; $y_1$ acts as a lower bound for the rapidity of the particles described by the TMD, which are supposed to be collinear, hence very fast moving along the reference direction (plus direction) of the jet.
Therefore, despite it is a full-fledged arbitrary cut-off, its value has to be chosen ``large enough" to preserve the physical meaning of the TMDs.
In fact, in a physical observable $y_1$ should be set in the limit in which the factorization procedure holds, i.e. $y_1 \to \infty$. 
However, TMDs are not physical observables and they depend on $\mu$ as well as $y_1$, hence they can be considered at a fixed, and finite, value of $y_1$.

\bigskip

\begin{figure}[t]
  \centering 
\begin{tabular}{c@{\hspace*{15mm}}c}  
      \includegraphics[width=6.5cm]{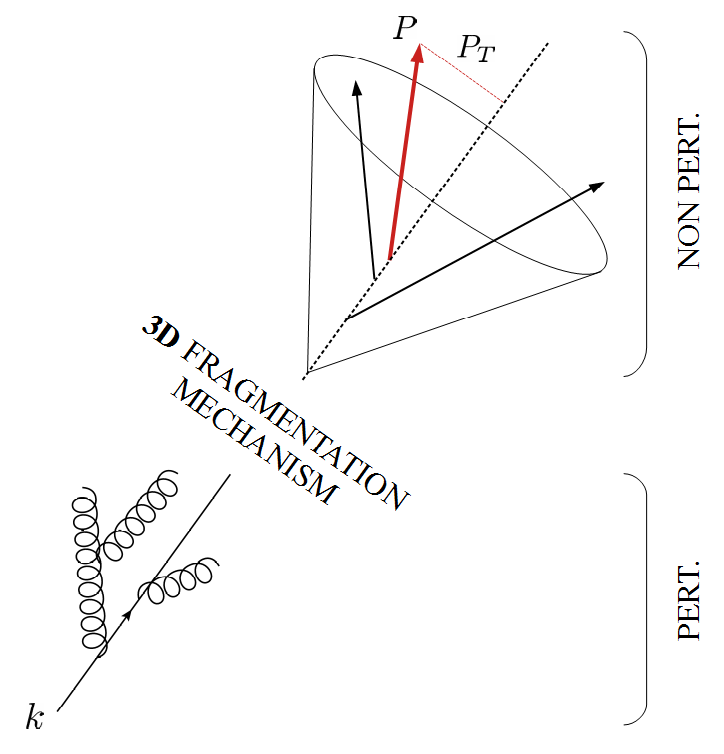}
  &
      \includegraphics[width=6.5cm]{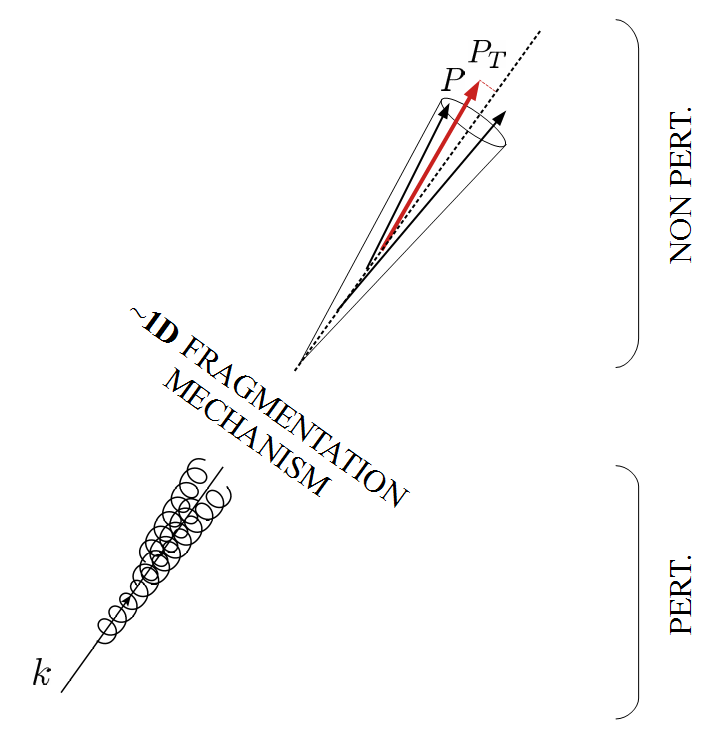}
  \\
  (a) & (b)
\end{tabular}
 \caption{Pictorial representation of a TMD Fragmentation Function, in which the separation between perturbative and non-perturbative regime is explicitly shown, corresponding to two different values of the rapidity cut-off. In panel (a) the rapidity cut-off of the TMD FF is set to a generic value $y_1$. In panel (b) the rapidity cut off has been shifted to $\widehat{y}_1 > y_1$. The two TMDs represent different physical configurations, as the range spanned by the rapidities of the particles belonging to those TMDs are different. This  transformation alters the fragmentation mechanism; in fact at extremely large values of the rapidity cut off, one can reach a quasi 1-dimensional configuration.}
 \label{fig:rap_dil}
\end{figure}

The transformation rule for a shift in the rapidity cut-off can be easily obtained from the solution of the evolution equations in Eq.~\eqref{eq:tmd_sol}.
If we shift $y_1$ to $\widehat{y}_1 = y_1 - \theta$, where $\theta$ is some real number, then (neglecting the dependence on all variables except $\zeta$) the full TMD transforms as:
\begin{align}
\widetilde{C}(\zeta) \mapsto 
&\hspace{.1cm}\widetilde{C}(\widehat{\zeta}) = \widetilde{C}(\zeta) \, \mbox{exp} 
\left[ \frac{1}{2} \theta \, \widetilde{K}\right],
\label{eq:rap_shift_tmd_2}
\end{align}
Therefore, the full effect of this transformation is a \emph{dilation factor} which depends on the soft kernel $\widetilde{K}(b_T,\,\mu)$ and the shift parameter $\theta$.
Notice that the transformed TMD describes a different physical configuration, as the rapidities of the collinear particles have been shrinked to a narrower range.
In particular, as $\widehat{y}_1$ approaches the factorization limit of infinite rapidity, the particles belonging to that TMD become more and more tightly aligned along the reference direction of the collinear group.
As a consequence, in the limit of infinite rapidity cut-off, the collinear particle motion is basically 1-dimensional and the 3D picture of the hadron structure is altered (see Fig.~\ref{fig:rap_dil}).
%
%
%
Since the non-perturbative information about the 3D structure of the hadrons is encoded into the model $M_C$, we can define a transformation that makes the TMD invariant with respect to the shift of the rapidity cut-off by acting simultaneously on the model.
The resulting transformed TMD will describe the same physical configuration of the initial TMD, because the alteration due to the tightened range of rapidity will be totally reabsorbed by the transformed model, that will compensate
for the dilation factor  $\mbox{exp} 
\left[ \frac{1}{2} \theta \, \widetilde{K}\right]$ in Eq.~\eqref{eq:rap_shift_tmd_2}.
Then, for any $\theta < 0$ such transformation is defined as:
\begin{align}
&y_1 \mapsto \mathcal{D}_{\theta} (y_1)  = y_1 - \theta , \label{eq:y_1_rapdil}   \\
&M_C(b_T) \mapsto \mathcal{D}_{\theta} \left(M_C(b_T)\right) =
M_C(b_T)\, \mbox{exp} 
\left[ -\frac{1}{2} \theta \, \widetilde{K} (\mu,\,b_T)\right],
\label{eq:model1_rapdil}  
\end{align}
where only the dependence on $b_T$ has been shown explicitly in $M_C$.
Due to the dilation factor in front of the model, we will refer to the previous transformation $\mathcal{D}_{\theta}$ as a  \textbf{rapidity dilation} (RD), that makes TMDs invariant with respect to the choice of the rapidity cut-off:
\begin{align}
\widetilde{C}(\zeta,\,M_C) \mapsto 
&\hspace{.1cm}\mathcal{D}_{\theta} 
\left( \widetilde{C}(\zeta,\,M_C) \right) = 
\widetilde{C}\left(\zeta e^{2\theta},\,\mathcal{D}_{\theta} M_C\right) =
\widetilde{C}(\zeta,\,M_C) .
\label{eq:rap_dil_tmd}  
\end{align}
The transformed model, Eq.~\eqref{eq:model1_rapdil}, acquires the same  properties of Eq.~\eqref{eq:model_prop}. 
In fact, since $\widetilde{K}$ goes to zero at small $b_T$, then the dilation factor is 1 for $b_T \sim 0$. 
Furthermore, since $\widetilde{K}$ is basically negative, at large $b_T$ the dilation factor give an additional suppression beside those due to the properties of $g_K$ and $M_C$.
Rapidity dilation make TMDs invariant under the choice of the rapidity cut-off $y_1$, which nevertheless has to be considered an arbitrary and \emph{large} parameter.
This is in fact one of the necessary hypothesis at the basis of any factorization formula: all the particles described by a collinear part (and ultimately by a TMD) must have a large and positive rapidity, according to the reference direction of the collinear group.
Hence, a rapidity dilation performed with a very large and positive $\theta$ would contradict the initial hypothesis on the validity of factorization itself.
The correct way to interpret this transformation is to apply it only \emph{after} the factorization formula has been derived, in the limit of $y_1 \rightarrow \infty$.
Roughly speaking, the model associated with a certain choice describes how collinear particles with rapidity in the range\footnote{In the real world, quite different from the massless limit, the upper bound is $y_P$, the large and positive rapidity of the reference hadron.} $y_1 \leq y < \infty$ 
behave in the non-perturbative regime.
Then, rapidity dilations simply balance the perturbative and non-perturbative information encoded in the TMDs according to the choice of rapidity cut-off, in order to keep 
their combination invariant.
Furthermore, rapidity dilations offer a new point of view which helps in the comparison between different physical configurations. Let's consider, for example, those depicted in Fig.~\ref{fig:rap_dil}, where panel (a) and panel and (b) represent TMDs described by $C(\zeta,\,M)$ and $C(\mathcal{D}_\theta \zeta,\,M)$, respectively, with $\mathcal{D}_\theta \zeta < \zeta$. It is interesting to point out that this interpretation is totally equivalent to considering the two TMDs evaluated within the same range of rapidity but associated to two different non-perturbative models, i.e. interpreting the TMD depicted in panel (a) as described by $C(\mathcal{D}_\theta\zeta,\,\mathcal{D}_\theta M)$ and the TMD in panel (b) as described by $C(\mathcal{D}_\theta \zeta,\,M)$.
Notice that rapidity dilations define a group under the multiplication laws:
\begin{align}
\mathcal{D}_{\theta_2} \circ  \mathcal{D}_{\theta_1} =
\mathcal{D}_{\theta_1+ \theta_2} ,
\quad
\mathcal{D}_{\theta} \circ  \mathcal{D}_{-\theta} = 
\mathrm{id} \,.
\label{eq:rapdil_group}
\end{align}
Rapidity dilations are closely reminiscent of a 1-parameter 
gauge transformations for the ``fields" $y_1$ and $M_C$, that make the TMD 
invariant. Here the TMD plays the role of the ``Lagrangian".
In this sense, rapidity dilations might be considered a symmetry for the 
TMDs.
There is an interesting analogy between the action of rapidity dilations 
on the rapidity cut-off, $y_1$, and the action of the Renormalization 
Group (RG) on the energy scale $\mu$. 
Here, an arbitrary $\mu$ allows to regularize the UV divergences, but it 
introduces some arbitrariness in the theory, as $\mu$ can be set to any value. 
Some quantities are independent of the choice of this scale, like cross 
sections, where the RG-variation of the fields is exactly compensated by the 
RG-variation of the external on-shell particles (LSZ mechanism). 
Similarly, rapidity dilations (RD) allow to control the arbitrariness in 
the choice of the rapidity cut-off, $y_1$, that regularizes the rapidity
divergences.
In particular, TMDs defined along the plus direction are RD-invariant, as 
the transformation of the model $M_C$, Eq.~\eqref{eq:model1_rapdil}, exactly 
compensates for the rapidity shift, Eq.~\eqref{eq:y_1_rapdil}. 
However, there are quantities that are not RD-invariant.
As we will show in Sections~\ref{subsubsec:z_refl}, 
the TMD defined along the minus direction $\widetilde{C}_-$ or the 2-h soft factor $\ftsoft{2}$
are examples of such quantities.
On the other hand, combinations as $\widetilde{C}_+ \, \widetilde{C}_- \, \ftsoft{2}$ are RD-invariant,
because the dilation factor in $\widetilde{C}_-$ exactly compensates the variation in $\ftsoft{2}$.  

\bigskip

When a TMD appears in a cross section, its non-perturbative content, i.e. the last line of Eq.~\eqref{eq:tmd_sol_2}, has to be extracted from experimental data and
the result will depend on the choice of the rapidity cut-off, as represented in Fig.~\ref{fig:rap_dil}.
Rapidity dilations ensure that the physical information encoded in the TMDs stays unaltered if the rapidity cut-off is moved toward the limit of infinite rapidity.
If $\widetilde{C}^{\mbox{\tiny NP}}$ denotes the full non-perturbative content of the TMD, then its transformation rule under rapidity dilation is given by:
\begin{align}
&\mathcal{D}_{\theta}\left(
\widetilde{C}^{\mbox{\tiny NP}}_{j,\,H}
\right)
(\zeta,\,M,\,g_K)= 
\widetilde{C}^{\mbox{\tiny NP}}_{j,\,H}
(\mathcal{D}_{\theta}\zeta,\,\mathcal{D}_{\theta}M,\,g_K) =
\notag \\
&\quad= 
\left(\mathcal{D}_{\theta}M\right)_{j,\,H}(b_T) 
\mbox{ exp} \Big \{
-\frac{1}{4} \, g_K(b_T) \, \log{\frac{\mathcal{D}_{\theta}\zeta}{\overline{\zeta}_0}}
\Big \}= 
\notag \\
&\quad= 
\left(M\right)_{j,\,H}(b_T)
\mbox{ exp} \Big \{
-\frac{1}{4} \, g_K(b_T) \, \log{\frac{\zeta}{\overline{\zeta}_0}}
\Big \} \,
\mbox{ exp} \Big \{
-\frac{1}{2} \,\theta \, \widetilde{K}(b^\star_T,\,\mu)
\Big \} = 
\notag \\
&\quad= 
\widetilde{C}^{\mbox{\tiny NP}}_{j,\,H}(\zeta,\,M,\,g_K) \,
\mbox{ exp} \Big \{
-\frac{1}{2} \,\theta \, \widetilde{K}(b^\star_T,\,\mu)
\Big \},
\label{eq:rapdil_NP}
\end{align}
where the second step is given by Eqs.~\eqref{eq:K_star}, \eqref{eq:y_1_rapdil} and \eqref{eq:model1_rapdil}.
In this case, the dilation factor is fully computable in perturbative QCD.
Notice that the function $g_K$ is not affected by the rapidity dilation, as it should. In fact $g_K$ is also involved in the definition of the soft factor $\ftsoft{2}$ (Eq.~\eqref{eq:soft_asy}), which must not depend on the extraction of the TMD.

Due to rapidity dilations, the choice of the model depends on the choice of the rapidity cut-off.
Therefore, in general two independent extractions of TMDs, that use different values of $\zeta$, will feature different models.
However, rapidity dilations allow to relate these independent extractions of TMDs.
The main difficulty here is that theory is devised in the $b_T$-space, while measurements are performed in the transverse momentum space. 

As a practical example, let's suppose we want to compare the TMD extractions of two independent research Groups, A and B, that have analyzed the same sample of data.
They will provide two TMD functions in transverse momentum space $C^{(A)}$ and $C^{(B)}$. 
Since they obtained their result fitting the same experimental data, the two functions have to be compatible within the overlapping of the respective uncertainty bands, built by considering all source of errors (collinear PDFs/FFs uncertainties, experimental errors, fitting uncertainties, etc ...). 
However a meaningful comparison can only be made for small values of transverse momentum, because TMDs turn non-physical at large $k_T$ (see~\ref{subapp:smallbT_tmds}).
Both results can be written as the Fourier transform of their $b_T$ counterparts. Schematically:
\begin{align}
&C^{(A)} (k_T,\,\zeta,\,M^{(A)}) = \int \, \frac{d^2 \vec{b}_T}{(2\pi)^2} \,
e^{i \vec{k}_T \cdot \vec{b}_T} \,
\widetilde{C}^{\mbox{\tiny P}}_{\zeta} (b^\star_T) \, 
\widetilde{C}^{\mbox{\tiny NP}}_{\zeta,\,M^{(A)}} (b_T); \label{eq:compare_extractionsA} \\
&C^{(B)} (k_T,\,\zeta',\,M^{(B)}) = \int \, \frac{d^2 \vec{b}_T}{(2\pi)^2} \,
e^{i \vec{k}_T \cdot \vec{b}_T} \,
\widetilde{C}^{\mbox{\tiny P}}_{\zeta'} (b^\star_T) \, 
\widetilde{C}^{\mbox{\tiny NP}}_{\zeta',\,M^{(B)}} (b_T).
\label{eq:compare_extractionsB}
\end{align}
where only the dependence on the rapidity cut-off and on the model are shown explicitly and $\widetilde{C}^{\mbox{\tiny P}}$ denotes the perturbative content of the TMD.
In principle, also the choice of $g_K$ may be different for the two extractions.
However, Group A and B have to agree also in the estimate of the $\soft{2}$, and this gives further constraints.
Hence, even if $g_K^{(A)}$ and $g_K^{(B)}$ have different functional forms, they should share more or less the same shape. 
For simplicity, in the following we will set $g_K^{(A)} \sim g_K^{(B)}$.

The two rapidity cut-off are different but, supposing $\zeta' < \zeta$,   
a certain real number $\theta < 0$ must exists such that $\zeta' = \zeta e^{2\theta}$.
Hence, Group A can perform a rapidity dilation in order to comply with the choice of Group B.
By using Eq.~\eqref{eq:rapdil_NP}, they can write:
\begin{align}
&C^{(A)} (k_T,\,\zeta,\,M^{(A)}) = \int
\frac{d^2 \vec{b}_T}{(2\pi)^2} \,
e^{i \vec{k}_T \cdot \vec{b}_T} \,
\widetilde{C}^{\mbox{\tiny P}}_{\zeta'} (b^\star_T) \, 
\widetilde{C}^{\mbox{\tiny NP}}_{\zeta',\,\mathcal{D}_\theta \, M^{(A)}} (b_T) = 
\notag \\
&\quad =
\int
\frac{d^2 \vec{b}_T}{(2\pi)^2} \,
e^{i \vec{k}_T \cdot \vec{b}_T} \,
\widetilde{C}^{\mbox{\tiny P}}_{\zeta'} (b^\star_T) \, 
\widetilde{C}^{\mbox{\tiny NP}}_{\zeta,\,M^{(A)}} (b_T)\,
\mbox{ exp} \Big \{
-\frac{1}{2} \,\theta \, \widetilde{K}(b^\star_T)
\Big \}.
\label{eq:rapdil_groupA}
\end{align}
In this way, the two estimates are written with the same perturbative part in $b_T$-space.
If TMDs were valid throughout the whole spectrum of $k_T$s, then the comparison between Eq.~\eqref{eq:compare_extractionsB} and Eq.~\eqref{eq:rapdil_groupA} would lead to $\mathcal{D}_\theta \, M^{(A)} \sim M^{(B)}$.
However, the Fourier transforms in Eqs.~\eqref{eq:compare_extractionsA} and~\eqref{eq:compare_extractionsB} have to be compatible at small $k_T$, but there is no constraint for larger values.
Furthermore, we known that the perturbative content is constant for $b_T$ larger than a certain $b_{\mbox{\tiny SAT}} \geq b_{\mbox{\tiny MAX}}$.
At the same time, the non-perturbative content should be of order $1$ at small/moderate $b_T$, i.e. up to $b_{\mbox{\tiny SAT}}$, in order not to interfere too drastically on the perturbative information.
Therefore, we can roughly split the Fourier transform in two parts as:
\begin{align}
&C (k_T) \sim \int^{b_{\mbox{\tiny SAT}}}
\frac{d^2 \vec{b}_T}{(2\pi)^2} \,
e^{i \vec{k}_T \cdot \vec{b}_T} \,
\left(
\widetilde{C}^{\mbox{\tiny P}}_{\zeta} (b^\star_T) - \widetilde{C}^{\mbox{\tiny P}}_{\zeta} (b_{\mbox{\tiny SAT}})
\right)\, 
\widetilde{C}^{\mbox{\tiny NP}}_{\zeta,\,M} (b_T) 
+ \notag \\
&\quad+
\widetilde{C}^{\mbox{\tiny P}}_{\zeta} (b_{\mbox{\tiny SAT}}) \,
\int \frac{d^2 \vec{b}_T}{(2\pi)^2} \,
e^{i \vec{k}_T \cdot \vec{b}_T} \,
\widetilde{C}^{\mbox{\tiny NP}}_{\zeta,\,M} (b_T).
\label{eq:FT_dec}
\end{align}
The first part is integrated only up to the saturation value $b_{\mbox{\tiny SAT}}$. It is clearly dominated by perturbative information, since $\widetilde{C}^{\mbox{\tiny NP}}$ is not drastically different from $1$ in that range, for any choice of $\zeta$ and $M$.
As a consequence, this part is almost the same for both Eq.~\eqref{eq:rapdil_groupA} and~~\eqref{eq:compare_extractionsB}.
On the other hand, the second part is simply proportional to the Fourier Transform of the non perturbative content of the TMD.
Different choices of $\zeta$ and $M$ can give integrands of the same order up to $b_{\mbox{\tiny SAT}}$  but they can differ on how rapidly they go to zero as $b_T$ goes to infinity: at large $b_T$ they could be very small and at the same time differ for many orders of magnitude. This difference is not evident at small values of $k_T$, since the area under the curve in $b_T$ space, after the saturation value, can be neglected in any case.
However, the differences may be consistent at large $k_T$.
This is not a problem, since TMDs lose their physical meaning in this region.
Hence, Group A can compare its result with that of Group B by Fourier transforming its non-perturbative part {\it after} it has been rapidity-dilated.
Then, the following relation should hold within uncertainties:
\begin{align}
R^{\mbox{\tiny NP}}(k_T) = 
\ddfrac{
\mathcal{FT} \Big[
\widetilde{C}^{\mbox{\tiny NP}}_{\zeta',\,M^{(B)}} (b_T)\Big]}
{
\mathcal{FT} \Big[
\widetilde{C}^{\mbox{\tiny NP}}_{\zeta,\,M^{(A)}} (b_T)\,
\mbox{ exp} \Big \{
-\frac{1}{2} \,\theta \, \widetilde{K}(b^\star_T)
\Big \} \Big]}
\sim 1
\quad \mbox{ at small } k_T.
\label{eq:rapdil_compareFT}
\end{align}

\bigskip


\subsubsection{Rapidity dilation and \texorpdfstring{$z$}{z}-axis reflection \label{subsubsec:z_refl}}


\bigskip

The behaviour under $z$-axis reflection, which simply exchanges the plus and 
minus directions, 
is particularly important for widely studied processes, like SIDIS, Drell-Yan 
and $\epm \to H_A\,H_B \,X$, where two TMDs associated to opposite directions 
are multiplied together. 
If $R_z$ is the Lorentz transformation that reverses the $z$-axis, then the 
rapidity of the reference hadron swaps its sign under the action of $R_z$. 
On the other hand, the rapidity cut-off is not the rapidity of any real 
particle. 
It is just an ad hoc number and hence it is trivially invariant under the action 
of $R_z$. However, the particles belonging to the collinear group associated to 
the TMD in the minus direction should have a very large negative rapidity 
according to the limit $y_1 \rightarrow +\infty$.
Therefore, a proper rapidity cut-off would be $y_2 = - y_1$, as if $y_1$ had changed its sign.
Summarizing:
\begin{align}
\begin{cases}
y_P &\mapsto R_z\left( y_P \right)  = - y_P; \\
y_1 &\mapsto R_z\left( y_1 \right)  = y_1 
\stackrel{def}{=} - y_2 .
\end{cases}
\label{eq:Rz_rapidity}
\end{align}
As a consequence, the variable $\zeta$ for a TMD in the minus direction is obtained by simply replacing $\zeta_{+} \propto \mbox{exp} \left(y_P - y_1\right)$ with 
$\zeta_{-} \propto \mbox{exp} \left(y_2 - y_P\right)$ and the full TMD transforms as:
\begin{equation} \label{eq:Rz_tmd}
\widetilde{C}_{+} (\zeta_{+}) \mapsto 
R_z\left( \widetilde{C}_{+} (\zeta_{+}) \right) = 
\widetilde{C}_{-} (\zeta_{-}) ,
\end{equation}
where only the dependence on the rapidity cut-off has been made explicit.

There is a non trivial interplay between $z$-axis reflection and rapidity dilations, since the two transformations do not commute.
In fact, if the rapidity cut-off $y_1$ of $C_+$ is shifted, then the rapidity cut-off $y_2$ of $C_-$ is shifted as well, but with the sign reversed.
This can easily be seen by a direct computation, with the help of Eqs.~\eqref{eq:y_1_rapdil} and \eqref{eq:Rz_rapidity}:
\begin{align}
\mathcal{D}_{\theta} \left(y_2\right) = 
\mathcal{D}_{\theta} \left(-y_1\right) = -y_1 + \theta = y_2 + \theta.
\label{eq:y2_rapdil}
\end{align}
Therefore, according to Eq.~\eqref{eq:model1_rapdil}, the model of $C_-$ transforms as:
\begin{align}
\mathcal{D}_{\theta} \left(M_{C_-}(b_T)\right) =
M_{C_-}(b_T)\, \mbox{exp} 
\left[ \frac{1}{2} \theta \, \widetilde{K}\right].
\label{eq:model2_rapdil}
\end{align}
However, in the $z$-reversed TMD, $C_-$, the rapidity cut-off appears with the opposite sign with respect to $C_+$.
Hence, there is no more compensation between the rapidity shift and the transformed model, and $C_-$ is not invariant under rapidity dilations.
This can be summarized by saying that the two transformations do not commute:
\begin{align}
\begin{cases}
&R_z\left( \mathcal{D}_{\theta} \left(\widetilde{C}_{+} (\zeta_{+}) \right) \right) =  
R_z\left( \widetilde{C}_{+} (\zeta_{+}) \right) = 
\widetilde{C}_{-} (\zeta_{-}) ;\\
&\mathcal{D}_{\theta}\left( R_z \left(\widetilde{C}_{+} (\zeta_{+}) \right) \right) =
\mathcal{D}_{\theta}\left( \widetilde{C}_{-} (\zeta_{-})  \right) =
\widetilde{C}_{-} (\zeta_{-})\, \mbox{exp} 
\left[\theta \, \widetilde{K}\right] .
\end{cases}
\label{eq:Rz_rapdil_commutation}
\end{align}
Finally, lets consider the behavior of the (asymptotic) $2$-h soft factor, defined in Eq.~\eqref{eq:soft_asy}, under rapidity dilations. 
Since the soft model is not affected by the transformation and $y_1 \rightarrow y_1 - \theta$, while $y_2 \rightarrow y_2 + \theta$, it transforms as:
\begin{equation} \label{eq:2soft_rapdil}
\mathcal{D}_{\theta} \left( \ftsoft{2}(b_T; \, \mu,\,y_1 - y_2) \right) =
\ftsoft{2}(b_T; \, \mu,\,y_1 - y_2) \, \mbox{exp} 
\left[-\theta \, \widetilde{K}\right] .
\end{equation}
Therefore, by exploiting Eqs.~\eqref{eq:rap_dil_tmd}, 
\eqref{eq:Rz_rapdil_commutation}, and \eqref{eq:2soft_rapdil}, 
the combination $\widetilde{C}_+ \, \widetilde{C}_- \, \ftsoft{2}$ 
cross section) 
is invariant under rapidity dilations.

\bigskip


\section{Universality and Process Classification} \label{sec:univ_class}


\bigskip

Process-independent quantities play the most important role in factorized cross sections. 
They are \textbf{universal}, which means that once they have been estimated they can be used in any cross section, regardless of the specific process.
This is particularly useful for those quantities that carry non-perturbative information. 
Since they cannot be computed analytically, they have to be extracted from experimental data.
However, if they are universal, \emph{any} process that allows for their presence in the cross section can be exploited, and we can prefer those with a richer amount of data.
A lack of universality would undermine the predictive power of QCD itself.
In fact, if the non-perturbative quantities had to be extracted again for each individual process, the phenomenological analysis of a hadronic cross sections would be reduced to a mere fit of experimental data.

\bigskip

In general, a factorized cross section is a convolution of three different objects: the hard part, the collinear factors and the soft factor (see Section~\ref{sec:intro}).

The hard part is completely process-dependent.
However, it can be computed in perturbative QCD and its lack of universality does not affect the predictive power of the theory.

Collinear parts and the TMDs, as defined in Section~\ref{sec:coll_parts}  by the factorization definition, Eq.~\eqref{eq:fact_defTMDs}, depend only on their internal variables and hence are completely blind to the kinematics of the process.
Therefore, they can be really considered universal quantities.

On the other hand, the soft factor, defined in Section~\ref{sec:soft_factor}, is not completely process-independent.
In fact, it depends on the number $N$ of the collinear factors involved in the factorized cross section, each related to its reference parton of type  
$j$ and to its reference hadron $H$.
Therefore, they are not insensitive to the kinematics of the process in which they appear, because they depend both on the number of the Wilson lines replacing the collinear parts and also on their color representation, which is fixed by the parton type  
$j$ and differs from quark and gluons. 
However, at fixed $N$ and for reference partons of the same kind, soft factors are actually the same object, modulo crossing symmetry.
As an example, Drell-Yan scattering with two quark-initiated collinear factors in the initial state, 
$\epm \to H_A \, H_B\, X$, with two quark-initiated collinear factors
in the final state, and also SIDIS, with one quark-initiated collinear factor in the initial and one in the final state, share the same soft factor $\soft{2}$ modulo the crossing symmetry that relates the three processes.
Notice that in this case there are only two collinear factors and charge conservation allows only two quarks as reference partons.

Since processes with a different number $N$ of collinear factors have a different soft factor in their factorized cross section, it is possible to classify them according to this number.
This coincides with the number of reference hadrons participating to the hadronic process. The classes derived with this criterion will be called \textbf{hadron classes}.
Formally, a process belongs to the \textbf{$\mathbf{N}$-h class} if it globally involves $N$ collinear parts, which can appear  in the initial and/or in the final state, 
in all possible combinations and for all the allowed kind of reference partons.
Therefore, $\soft{N}$ can be considered universal only \emph{within} the $N$-h 
class, 
modulo crossing symmetry and the possible color representations of its Wilson 
lines.
This is a weaker kind of universality,  that holds only for a limited number of processes.
For instance, processes involving one collinear group belong to the \hclass{1}: 
Deep Inelastic Scattering (DIS), corresponding 
to one reference hadron in the initial state; $\epm \to H\,X$ correspoding to 
one reference hadron in the final state.
Processes involving two collinear groups belong to the \hclass{2}: here we  have 
Drell-Yan like scattering, 
$\epm \to H_A \, H_B\, X$, and SIDIS.

\begin{figure}[t]
  \centering 
\begin{tabular}{c@{\hspace*{15mm}}c}  
      \includegraphics[width=5.5cm]{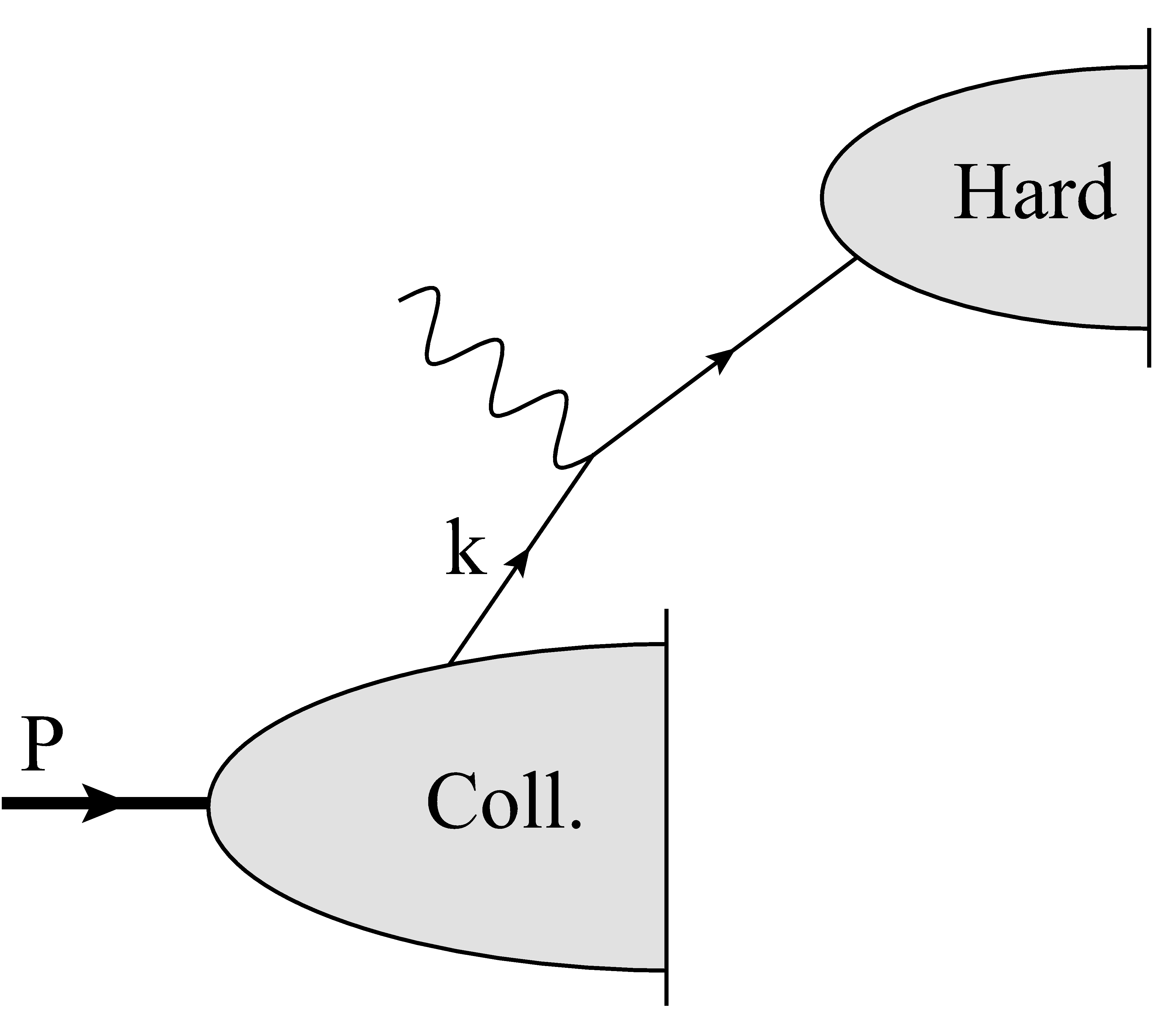}
  &
      \includegraphics[width=4cm]{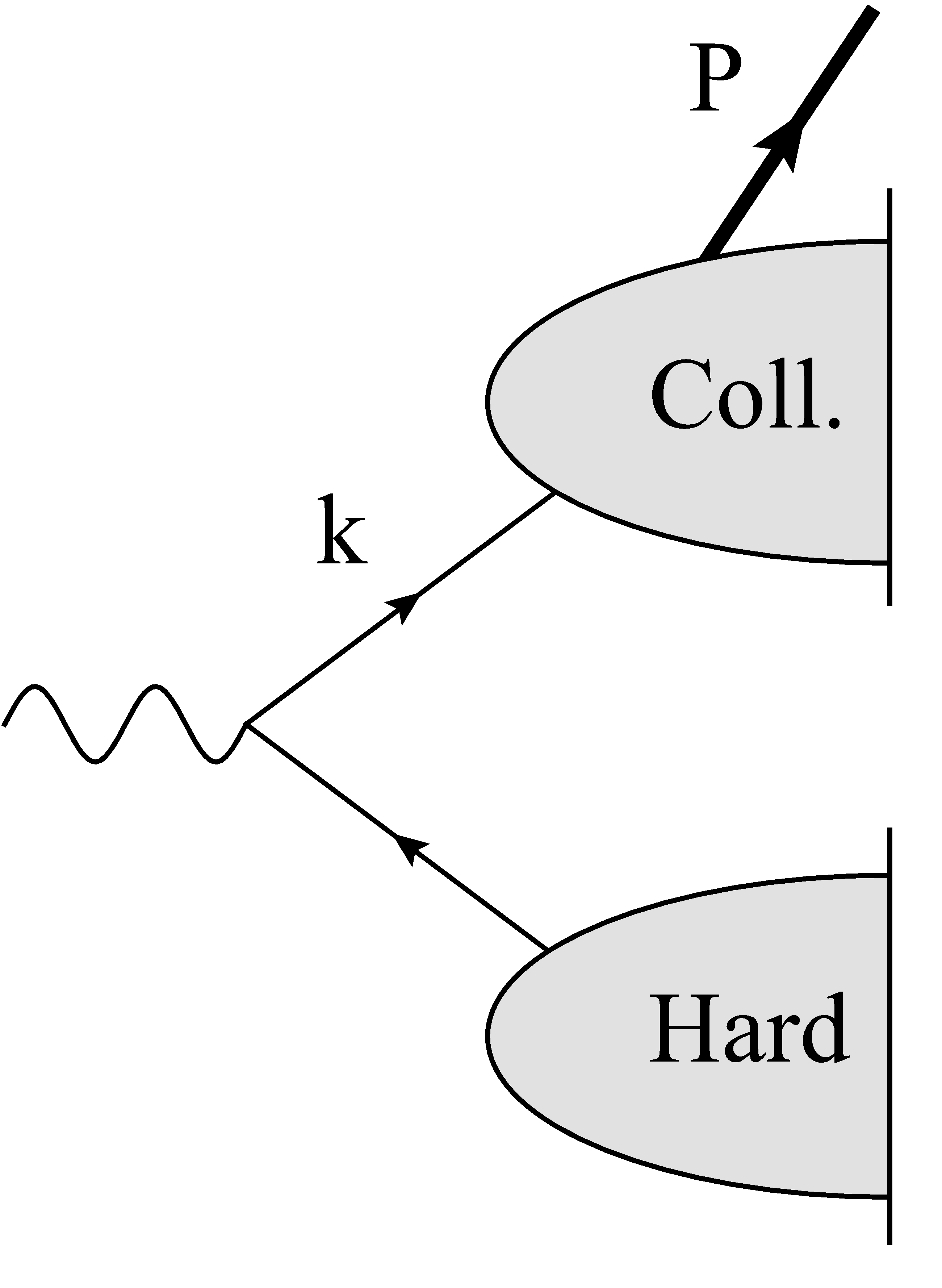}
  \\
  (a) & (b)
\end{tabular}
 \caption{Pictorial representation of DIS (a) and of $\epm \rightarrow H \, X$ (b). In both cases, there is  at least one hard real emission, which produces a collinear factor completely crossed by the final state cut. Consequently  it can be reabsorbed in the hard factor of the cross section.}
 \label{fig:real_emission}
\end{figure}

The classification above has nothing to do with
the nature of the factorization adopted  (collinear or TMD): it depends only 
on the specific kinematics of the processes, that can be different case by case.
However, it is possible to identify common properties within each 
hadron-class that allows to determine, a priori, which factorization scheme 
should be used.
Consider for example the \hclass{1} case. In both DIS and 
$\epm \rightarrow H\,X$ there is \emph{at least} one hard real emission, since 
there is always a fermion leg crossing the final state cut (see 
Fig.~\ref{fig:real_emission}).
%
%
The collinear factor associated to the real emission is totally crossed by the final state cut and hence it does not have any reference hadron.
Therefore, it can be considered far off-shell and part of the hard factor.
All the information about soft transverse momentum is washed away and 
the collinear factorization scheme has 
to be applied. 
The factorized cross section for \hclass{1} processes is then written as a convolution of the collinear part associated to the reference hadron with an hard factor that, once considered together with the hard real emissions, can be interpreted as a partonic cross section, i.e. the partonic counterpart of the process.

In the \hclass{2}, instead, the choice of factorization scheme is 
non-trivial and depends on the specific kinematics of the process.  
It is dictated by the size of one parameter, namely 
the ratio between the modulus of the weak boson transverse momentum $q_T$ and 
the typical energy scale of the process $Q$ (see Ref.~\cite{Collins:2011zzd}).
When ${q_T}/{Q} \ll 1$, TMD factorization has to be applied, while if 
${q_T}/{Q} \gg 1$, collinear factorization will be appropriate \footnote{For 
$\epm \rightarrow H_A\, H_B \, X$ processes, the condition ${q_T}/{Q} \ll 1$ corresponds to having the two hadrons almost back-to-back in the c.m.   
frame. On the contrary, ${q_T}/{Q} \ll 1$ implies that the back-to-back 
configuration cannot be realized, see Section~\ref{subsec:coll_vs_tmd}.}.
The cross sections predicted in these two kinematical ranges, computed within two different approximations, do not automatically match; in fact, the intermediate region, where $q_T\sim Q$, is usually called ``matching region". 
Several studies have been devoted to the implementation of different algorithms to map these kinematics regions and to match the collinear cross sections to the TMD cross section (a problem known as ``matching"), see for example Refs.~\cite{Boglione:2016crp,Echevarria:2018qyi,Collins:2016hqq,Boglione:2016bph,Boglione:2019nwk}.

\bigskip

According to the previous considerations, one can build a hierarchy based on universality. 
The lowest level is occupied by quantities, like the hard part, that are 
completely process dependent but usually fully computable in perturbation
theory.  
At the top of the hierarchy we find quantities, like the collinear factors, 
that are absolutely process independent: they carry non-perturbative 
information but their universality properties guarantee that they can be extracted from one particular process and then used in any other. 
In the middle there are quantities which are only universal within their own $N$-hadron class, like the soft factors. As they carry non-perturbative information, they cannot be computed perturbatively. They too have to be extracted from experimental data, but they can only be used, class by class, for the 
processes involving the same number of collinear groups and for the same kind of reference partons.

\bigskip

In this sense, it is very important to provide a working scheme where objects with different degrees of universality are neatly separated, in such a way to maximize their perturbative content and their universal parts, while reducing the class-dependent factors to the minimum. For the latter, special experimental efforts will be required in order to gather a large number of high quality data corresponding to several different processes, which will then be analyzed simultaneously in a completely consistent framework. The latest analyses of the BELLE Collaboration and the current plans towards the realization of a new Electron Ion Collider (EIC) are indeed moving towards this direction~\cite{Seidl:2019jei,Guan:2018ckx,Accardi:2012qut,Aschenauer:2017jsk}.

\bigskip

The classification introduced above has to be intended as a criterion to classify  processes on the basis of their factorized cross section properties, and of their corresponding soft factor.
Therefore, the number $N$ that labels the classes is not the number of all hadrons involved in the process, in general much greater than the number of collinear factors.
The difference is more evident when we consider the final state of a scattering process.
In general, experimentalists detect a huge number of hadrons, grouped in jets.
The number of jets does not correspond to the number of collinear factors, which instead is the number of reference hadrons, i.e. the number of jets in which the hadron is detected in order to study the jet's fragmentation properties.
The actual topology of the event (e.g. the number of jets) is described by event-shape variables, like thrust.
An example will be presented in Section~\ref{sec:epm_1h}.

\bigskip


\section{The 2-hadron class \label{sec:2h_class}} 


\bigskip

We will now focus on the \hclass{2} of processes.
As mentioned above, in this class the choice of factorization 
scheme depends on a single parameter, the ratio ${q_T}/{Q}$ 
(see Section~\ref{subsec:coll_vs_tmd}). The \hclass{2} plays a crucial role, as its 
soft factor $\soft{2}$ is \emph{exactly} the same object that appears at denominator in the subtracted collinear factor $\coll$, Eq.~\eqref{eq:sub_coll} and, consequently, in the general definition of the TMD, Eqs.~\eqref{eq:FF_Clifford}~and~\eqref{eq:unp_Clifford}.

\bigskip

\bigskip


\subsection{2-h Class Cross Section \label{subsec:2h_xs} }


\bigskip

In Section~\ref{sec:soft_factor} we have provided a useful formalism to decompose the $2$-h class soft factor and the collinear part $\coll$ in a fully perturbative (computable) part and a strictly 
non-perturbative term, which can be modeled through the 
functions $g_K(b_T)$, $M_S(b_T)$ and $M_C(b_T)$, as shown in 
eqs.~\eqref{eq:soft_asy} and \eqref{eq:tmd_sol}. 
At this stage we have achieved all the necessary tools to be able to write an explicit expression for the \hclass{2} cross 
section. 
Its generic structure is analogous to that given in Eq.~\eqref{eq:tmd_fact} for 
$\epm 
\rightarrow H_A\, H_B \, X$, with $H_A$ and $H_B$ in an almost back-to-back 
configuration:
\begin{align}
d \sigma_{\mbox{\small 2-h}} &\sim H \, \times \, 
\mathcal{FT} \Big[ \, \widetilde{C}_+ \, \times \, \widetilde{C}_- \, \times \, 
\ftsoft{2} \, \Big]\sim \notag \\
&\sim H \, \times \, 
\mathcal{FT} \Big[ \, \dfrac{\widetilde{C}_+^{\mbox{\small unsub}}}{\ftsoft{2}} \, 
\times \, 
\dfrac{\widetilde{C}_-^{\mbox{\small unsub}}}{\ftsoft{2}} \, \times \, 
\ftsoft{2} \, \Big]\,, \label{eq:struct_xs2j}
\end{align}
where $C_+$, $C_-$ refers to TMDs defined along the plus and the minus direction, respectively.
The soft factor $\ftsoft{2}$ appearing in 
Eq.~\eqref{eq:struct_xs2j} 
is the same object that appears as subtraction factor in the factorization definition of the TMDs.  
Reorganizing the three  $\ftsoft{2}$ factors and reabsorbing them in the TMD, leads to a different definition of TMDs (see e.g. \TheBook,~\cite{Aybat:2011zv}):
\begin{align}
\widetilde{C}&_+^{\; \mbox{\small sqrt}}(\xi_+,\, \vec{b}_{T}; \, \mu, \, 
y_{P_1} - y_n) = \notag \\
&= \lim \limits_{\substack{y_{u_1} \to +\infty\\y_{u_2} \to -\infty}} \, 
\widetilde{C}_+^{\mbox{\small unsub}} (\xi_+, \,\vec{b}_{T}; \, \mu, \, y_{P_1} 
- y_{u_2}) \, 
\sqrt{\dfrac{\ftsoft{2} (\vec{b}_{T}; \, \mu, \,y_{u_1} - y_n)}
{\ftsoft{2} (\vec{b}_{T}; \, \mu, \,y_{u_1} - y_{u_2}) \, \ftsoft{2} 
(\vec{b}_{T}; \, \mu, \,y_n - y_{u_2})}} \label{eq:sqrt_def_1}
\end{align}
\begin{align}
\widetilde{C}&_-^{\; \mbox{\small sqrt}}(\xi_-,\, \vec{b}_{T}; \, \mu, \, 
y_n-y_{P_2}) = \notag \\
&= \lim \limits_{\substack{y_{u_1} \to +\infty\\y_{u_2} \to -\infty}} \, 
\widetilde{C}_-^{\mbox{\small unsub}} (\xi_-, \,\vec{b}_{T}; \, \mu, \, y_{u_1} 
- y_{P_2}) \, 
\sqrt{\dfrac{\ftsoft{2} (\vec{b}_{T}; \, \mu, \,y_n - y_{u_2})}
{\ftsoft{2} (\vec{b}_{T}; \, \mu, \,y_{u_1} - y_{u_2}) \, \ftsoft{2} 
(\vec{b}_{T}; \, \mu, \,y_{u_1} - y_n)}} . \label{eq:sqrt_def_2}
\end{align}
This definition of TMDs is often referred to as the \textbf{square root 
definition}.

There are many advantages to it. 
First of all, a single rapidity cut-off $y_n$ is sufficient to regularize all  
rapidity divergences, the perturbative computations are much easier and the 
evolution equation 
are unified and symmetrized, see Ref.~\cite{Collins:2011zzd}.
Moreover, as mentioned above, the square root definition allows to solve 
the soft factor problem in the \hclass{2}. In fact, according to this 
definition, the cross section assumes a ``Parton-Model"-like structure, where 
all soft gluons are reabsorbed in the TMD definition, very  convenient for phenomenological applications: 
\begin{align}
d \sigma_{\mbox{\small 2-h}} \sim H \, \times \, 
\mathcal{FT} \Big[ \, \widetilde{C}_+ \, \times \, \widetilde{C}_- \, \times \, 
\ftsoft{2} \, \Big]\sim
 H \, \times \, 
\mathcal{FT}  \Big[ \, \widetilde{C}_+^{\; \mbox{\small sqrt}} \times 
\widetilde{C}_-^{\; \mbox{\small sqrt}} \, \Big]\,, \label{eq:struct_xs2j_sqrt}
\end{align}
As an example, the unpolarized cross section for $\epm \rightarrow H_A\, H_B 
\, X$ for almost back-to-back spinless hadrons,  
Eq.~\eqref{eq:tmd_fact}, becomes:
\begin{align}
W^{\mu\,\nu}(Q,\,p_A,\,p_B) &= \frac{8 \pi^3 z_A z_B}{Q^2} \! \sum_f 
H^{\mu\,\nu}_{f,\overline{f}}(Q) \!
\int d^2 \vec{b}_T  \ftsoft{2}(\vec{b}_T) 
\widetilde{D}_{1,H_A / f} (z_A, \vec{b}_T)  
\widetilde{D}_{1 H_B / \overline{f}}(z_B, \vec{b}_T)  \notag \\
&= \frac{8 \pi^3 z_A z_B}{Q^2} \! \sum_f H^{\mu\,\nu}_{f,\,\overline{f}}(Q) \!
\int d^2 \vec{b}_T 
\widetilde{D}_{1,H_A / f}^{\; \mbox{\small sqrt}} (z_A, \vec{b}_T)  
\widetilde{D}_{1, H_B / \overline{f}}^{\; \mbox{\small sqrt}} (z_B, 
\vec{b}_T) \,. \label{eq:tmd_fact_sqrt}
\end{align}

Despite its numerous advantages, the square root definition lowers the 
degree of universality of the TMD, as it relates it to the $2$-h soft 
factor which, by definition, is only universal within its corresponding $2$-h class.
In other words, the square root definition is \emph{optimal} for 
the \hclass{2}, as it beautifully simplifies the $2$-h cross section making 
it suitable for phenomenological applications; its drawback, however, is that it 
ceases to be valid outside the \hclass{2}.
On the other hand, abandoning the square root definition of the TMDs in favor 
of the factorization definition, Eq.~\eqref{eq:FF_Clifford}, will force us to 
face the soft factor problem and take a new (and potentially very hard) 
challenge: reformulating the way we do phenomenology, in terms of newly 
defined fundamental objects, where the soft factors are modeled explicitly 
rather than absorbed in the definition of the TMDs. 

We will attempt such a strategy, adopting the factorization 
definition of the TMD, Eq.~\eqref{eq:fact_defTMDs}, and relying on the results 
of Sections~\ref{sec:soft_factor} and \ref{sec:coll_parts} for the decomposition of the 
collinear and soft factors in terms of their perturbative and non-perturbative 
parts.

Using the solution of the evolution equations for the TMDs,  
Eq.~\eqref{eq:tmd_sol_2}, and the soft factor, Eq.~\eqref{eq:soft_asy},
it is possible to write the \hclass{2} cross section in terms of perturbative 
and non-perturbative functions. Apart from the hard factor and a Fourier 
transform, the relevant structure is given by:
\begin{align}
&\widetilde{C}_+(\xi_+,\,  \vec{b}_{T}; \, \mu, \, \zeta_1) \; 
\widetilde{C}_-(\xi_-,\, \vec{b}_{T}; \, \mu, \,\zeta_2) \;
\ftsoft{2}(b_T; \, \mu,\,y_1 - y_2) = \notag \\
&\quad= \widetilde{C}_+(\xi_+,\, \vec{b}_{T}^\star; \, \mu_0, \,\mu_0^2) \;
\widetilde{C}_-(\xi_-,\, \vec{b}_{T}^\star; \, \mu_0, \,\mu_0^2) \times \notag\\
&\quad\times\mbox{ exp} \Big \{ 
\frac{1}{4} \, \widetilde{K}(b_T^\star;\,\mu_0) \, \log{\frac{\zeta_1 
\zeta_2}{\mu_0^4}} + 
\int_{\mu_0}^{\mu} \frac{d \mu'}{\mu'} \, \left[ 
2 \gamma_C(1) - \frac{1}{4} \, \gamma_K(\mu') \, \log{\frac{\zeta_1 
\zeta_2}{\mu'^4}}
\right] \Big \} \times \notag \\
&\quad\times M_{C_+}(\xi_+,\,b_T) \,  M_{C_-}(\xi_-,\,b_T) 
\mbox{ exp} \Big \{ 
-\frac{1}{4} \, g_K(b_T)  \, \log{\frac{\zeta_1 \zeta_2}{\overline{\zeta_1}_0 
\overline{\zeta_2}_0}}
\Big \} \times \notag \\
&\quad\times \mbox{ exp} \Big \{ 
\frac{y_1 - y_2}{2} \left[\widetilde{K}(b_T^\star;\,\mu_0)  - 
\int_{\mu_0}^{\mu}\, \frac{d \mu'}{\mu'}\, \gamma_K(\mu) \right]
\Big \} \times \notag \\
&\quad\times M_S(b_T) \mbox{ exp} \Big \{ 
-\frac{y_1 - y_2}{2} g_K(b_T)
\Big \} \,,\label{eq:2jxs_evo_1}
\end{align}
where the reference values of the scales can be set to standard choices, 
Eqs.~\eqref{eq:mub},~\eqref{eq:zetab}, \eqref{eq:zeta_ref_st} 
and the errors due to evolution equations are neglected,   
since they are suppressed by $\mathcal{O}\Big( e^{-(y_1 - y_2)} \Big)$.
From Section~\ref{subsubsec:z_refl}, the product of the two rapidity cut-off gives $\zeta_1 \, \zeta_2 \sim Q^4 e^{-2(y_1-y_2)}$, hence
the 
second and the third lines in 
Eq.~\eqref{eq:2jxs_evo_1} generate contributions that  exactly cancel the 
fourth line and the exponential of the fifth line, respectively.
Therefore, we simply have:
\begin{align}
&\widetilde{C}_+(\xi_+,\, \vec{b}_{T}; \, \mu, \, \zeta_1) \; 
\widetilde{C}_-(\xi_-,\, \vec{b}_{T}; \, \mu, \,\zeta_2) \; 
\ftsoft{2}(b_T; \, \mu,\,y_1 - y_2) = \notag \\
&\quad= \widetilde{C}_+(\xi_+,\, \vec{b}_{T}^\star; \, \mu_0, \,\mu_0^2) \;
\widetilde{C}_-(\xi_-,\, \vec{b}_{T}^\star; \, \mu_0, \,\mu_0^2) \times \notag\\
&\quad\times \mbox{ exp} \Big \{ 
\widetilde{K}(b_T^\star;\,\mu_0) \, \log{\frac{Q}{\mu_0}} + 
\int_{\mu_0}^{\mu} \frac{d \mu'}{\mu'} \, \left[ 
2 \gamma_C(1) - \gamma_K(\mu') \, \log{\frac{Q}{\mu'}}
\right]\Big \}  \notag \times \\
&\quad\times M_{C_+}(\xi_+,\,b_T) \,  M_{C_-}(\xi_-,\,b_T) \, M_S(b_T) 
\mbox{ exp}\Big \{ 
-g_K(b_T) \, \log{\frac{Q}{\sqrt{\overline{\zeta_1}_0 \overline{\zeta_2}_0}}}
\Big \}
\,.\label{eq:2jxs_evo_2}
\end{align}
As expected, in the previous equation there is no residual dependence on the 
rapidity cut-offs $y_1$ and $y_2$, hence we can simply set $\zeta_{1,\,2} = Q^2$.
Needless to say, this result is compatible with rapidity dilations since, as shown 
in Section~\ref{subsubsec:z_refl}, the combination $\widetilde{C}_+ \widetilde{C}_- \ftsoft{2}$ 
is invariant on the choice of the rapidity cut-off.

\bigskip


\subsection{Factorization Definition vs. Square Root Definition \label{subsec:sqrt_def} }


\bigskip

We can now compare the factorization definition with the square root definition 
of the TMDs.
Ref.~\cite{Collins:2011zzd} shows that the unsubtracted TMDs 
$\widetilde{C}_i^{\mbox{\small unsub}}$ ($i = 1,\,2$), are the same in the 
two definitions. Hence we can compute their ratio 
(here we pick the plus direction):
\begin{align}
& \dfrac{\widetilde{C}_+^{\; \mbox{\small sqrt}}(\xi_+,\, \vec{b}_{T}; \, \mu, 
\, y_{P_1} - y_n)} {\widetilde{C}_+ (\xi_+,\, \vec{b}_{T}; \, \mu, \, y_P -y_1)} 
= \notag \\
&= \lim 
\limits_{\substack{y_{u_1} \to +\infty\\y_{u_2} \to -\infty}} 
\sqrt{\dfrac{\ftsoft{2} (b_{T}; \, \mu, \,y_{u_1} - y_n)}
{\ftsoft{2} (b_{T}; \, \mu, \,y_{u_1} - y_{u_2}) \, \ftsoft{2} (b_{T}; \, \mu, 
\,y_n - y_{u_2})}} \, 
\ftsoft{2} (b_{T}; \, \mu, \,y_1 - y_{u_2}) = \notag \\
&= \sqrt{\ftsoft{2}(b_T; \, \mu_0, \, 0) \, \mbox{exp} \Big( (y_1 - y_n) \, 
\widetilde{K}(b_T;\,\mu) \Big)} = \notag \\
&= \sqrt{M_S(b_T)} \times e^{\frac{(y_1 - y_n)}{2} \, 
\widetilde{K}(b_T^\star;\,\mu)} \, e^{-\frac{(y_1 - y_n)}{2} \, g_K(b_T)} \,, 
\label{eq:def_comparison_1}
\end{align}
where in the second line we used the asymptotic part of the solution to the 
evolution equations for the $2$-h soft factor, Eq.~\eqref{eq:soft_evo_sol}, 
which is the only part that survives in the large rapidity cut-off limits, while in the last 
step we used Eq.~\eqref{eq:soft_asy} in order to separate the perturbative from 
the 
non-perturbative content.
Obviously, a perfectly analogous result holds for the TMD relative to 
the opposite direction, $\widetilde{C}_-$ .

As we are interested in the comparison of the two TMD definitions at the same value 
of the rapidity cut-off, we can take $y_1=y_n$ so that in 
Eq.~\eqref{eq:def_comparison_1} the dependence on the soft kernel 
$\widetilde{K}$ disappears, leaving only a square root of the soft model 
$M_S(b_T)$. Therefore we have:
\begin{equation} \label{eq:def_comparison_2}
\widetilde{C}^{\; \mbox{\small sqrt}}(\xi,\, \vec{b}_{T}; \, \mu, \, y_P - y_n) 
=  \sqrt{M_S(b_T)} \times 
\widetilde{C}(\xi,\, \vec{b}_{T}; \, \mu, \, y_P - y_n) \,,
\end{equation}
which clearly holds for both $\widetilde{C}_+$ and $\widetilde{C}_-$.
This is a very important result, as it shows that the choice of TMD 
definition (square root or factorization definition)  
only affects the non-perturbative content of the TMDs, while having no impact 
on the perturbative part. 
Consequently, $C^{\; \mbox{\small sqrt}}$ will differ from $C$ mainly in the 
small $k_T$ region.

According to Eq.~\eqref{eq:def_comparison_2}, the square root definition is 
obtained from Eq.~\eqref{eq:tmd_sol} by multiplying the TMD defined through 
the factorization definition by a square root of the soft model.
In other words, the contribution of the soft physics just acts on $M_C(\xi,\,b_T)$:
\begin{equation} \label{eq:models_comparison}
M_C^{\; \mbox{\small sqrt}} (\xi,\,b_T)= M_C(\xi,\,b_T) \times \sqrt{M_S(b_T)}\,.
\end{equation}
To conclude, we can compare the effect of using either one of two different 
TMD definitions in the cross section. Had we used the square root 
definition, its net effect in Eq.~\eqref{eq:2jxs_evo_2} would have 
been the replacement:
\begin{equation}
 M_{C_+}(\xi_+,\,b_T) \,  M_{C_-}(\xi_-,\,b_T) \, M_S(b_T) \rightarrow 
M_{C_+}^{\; \mbox{\small sqrt}} (\xi_+,\,b_T) \, 
M_{C_-}^{\; \mbox{\small sqrt}} (\xi_-,\,b_T) 
\,. \label{eq:sqrt_2jxs}
\end{equation}
Clearly the square root definition offers an ideal framework to perform 
the phenomenological study of the $2$-h class of processes: it solves the 
soft problem by reabsorbing the soft factor in the TMD definition and allows 
to extract the model functions $M_{C_{1,\,2}}^{\;\mbox{\small sqrt}}$ from 
experimental data. However, this operation makes it impossible to disentangle 
the non-perturbative soft effects due to $M_S$ which, instead, remains explicit 
when using the factorization definition for the TMD.

Eq. \eqref{eq:sqrt_2jxs} is particularly important from the phenomenological point 
of view, as it relates the TMDs obtained from data analyses based on the square 
root definition (which has been very widely used in the last ten years) to the TMDs 
extracted using the factorization definition. In this regard, the methodology 
proposed in this paper allows to profit of the past experience and to benefit of all the 
results obtained in previous analyses, while extending the scheme to all those 
processes which could not be considered before, because they belong to a different 
hadron class.
A rather straightforward example of this strategy will be the combined analysis of 
the BELLE measurements of the polarization of $\Lambda$ hyperons~\cite{Guan:2018ckx} 
in $\epm \to \Lambda \pi (K) X$ processes (2-h class), already studied in 
Ref.~\cite{DAlesio:2020wjq} within a generalized-parton model approach, and in 
$\epm \to \Lambda X$, i.e. in a 1h-class process.
This will be presented in a forthcoming paper. 

\bigskip


\section{Factorization of \texorpdfstring{$\epm \to H\,X$}{e+e- --> HX} \label{sec:epm_1h}}


\bigskip

In this section we will focus on the $\epm \to H\,X$ process, which belongs to the \hclass{1} according to the classification of  Section~\ref{sec:univ_class}. Here we have only one true collinear part, associated to the reference hadron $H$, which can be quark- or gluon-initiated; beside, in any case, there is always at least one hard real emission that gives a collinear contribution crossing the final state cut, hence included in the hard factor, which can then be interpreted as a partonic cross section.
The soft factor of the process is unity according to the collinear factorization scheme, see~\TheBook.

The thrust $T$ will be included in the derivation of the final result.
It is an event-shape variable that describes the topology of the final state, i.e. the number of  observed jets.
It can take values from $0.5$ to $1.0$: the lower limit corresponds to a spherical distribution of particles in the final state, while the upper limit indicates an exact two-jet configuration (pencil-like events).
Among all jets, only one is related to the collinear part, while the others have to be included in the partonic cross section.
Therefore, the value of thrust will determine  which Feynman graphs have to be considered in the calculation of the hard part, that will acquire a non-trivial dependence on $T$.

\bigskip

Similar cross sections have been studied in the framework of Soft Collinear Effective Theories (SCET), within a TMD-like factorization scheme. The relations connecting the SCET to the CSS  definition of TMDs have been investigated in Refs.~\cite{Collins:2012uy,Echevarria:2012js}, where perfect equivalence has been found with the square root definition, see Eq.~\ref{eq:sqrt_def_2}. 
In particular, the cross section for $\epm \to H \mbox{(jet)} \, X$  has been considered in Refs.~\cite{Kang:2017mda},~\cite{Neill:2016vbi} and~\cite{Kang:2017glf}. In these papers  the dependence on thrust $T$ is not considered; instead the radius $R$ of the jet is introduced
as a reference for the transverse momentum of the detected hadron. 
For the case of $\epm$ event shape angularities, but with no dependence on transverse momentum, one could refer for example to Ref.~\cite{Bell:2018gce}.

\bigskip

In the (thrust dependent) cross section of $\epm \to H\,X$, the leptonic tensor $L_{\mu\,\nu}$,  corresponding to the initial state contribution, is Lorentz contracted with the hadronic tensor $W^{\mu\,\nu}_H$, associated to the final state  (see for example Ref.~\cite{Collins:2011zzd}).
The cross section is then written as:
\begin{equation} \label{eq:epm_crosssec}
\frac{d \sigma}{\lips{P} \, dT} = \frac{2 \alpha^2}{Q^6} L_{\mu \, \nu} 
W^{\mu \, \nu}_H (T).
\end{equation}
Since the coupling of QED is much smaller than $\alpha_S$, the leptonic tensor can be well approximated by its lowest order:
\begin{equation} \label{eq:lept_tens}
 L^{\mu \, \nu} = l_1^\mu l_2^\nu +  l_2^\mu l_1^\nu - g^{\mu \, \nu} l_1 \cdot 
l_2,
\end{equation}
where $l_1$ and $l_2$ are the momenta of the incoming electron and positron, and the electron mass is neglected.

The hadronic tensor $W^{\mu \, \nu}_H$ depends on the momentum $P$ of the outgoing hadron and on the momentum $q$ of the boson connecting the initial with the final state.
Furthermore, it depends on thrust, $T$.
Its definition is:
\begin{align}
&W^{\mu \, \nu}_H(P,\,q,\,T) = 
4 \pi^3 \, \sum_X \delta^{(4)} 
\left( p_X +P - q \right) \,
\langle 0 | \,
j^\mu(0) | \,
P,\,X,\,\mbox{out} \,
\rangle_T \,
{}_T \langle P,\,X,\,\mbox{out} |
j^\nu(0) | \,
0 \rangle = 
\notag \\
&\quad= 
\frac{1}{4\pi} \, \sum_X \, \int d^4 z\,
e^{i q \cdot z}
\langle 0 | \,
j^\mu\left(z/2\right) | \,
P,\,X,\,\mbox{out} \,
\rangle_T \,
{}_T \langle P,\,X,\,\mbox{out} |
j^\nu\left(-z/2\right) | \,
0 \rangle,
\label{eq:current_def_W}
\end{align}
where $j^\mu$ are the electromagnetic currents for the hadronic fields and the final states have been labeled by ``T" in order to recall that their topology is fixed by the value of thrust.
The factor $1/(4\pi)$ in the last line coincides with the normalization choice of~\TheBook.
The hadronic tensor can be decomposed in terms of   
structure functions:
\begin{align} 
W^{\mu \, \nu}_H = 
\left(-g^{\mu \, \nu} + 
\frac{q^\mu q^\nu}{q^2} \right) 
F_{1,\,H} \, +
\frac{\left( P^\mu - q^\mu \frac{P \cdot q}{q^2} \right) 
\left( P^\nu - q^\nu \frac{P \cdot q}{q^2} \right)}
{P \cdot q} \,
F_{2,\,H}.
\label{eq:had_tens}
\end{align}
Thanks to this decomposition and by using the definition of the fractional energy $z = 2 \, {P  \cdot q}/{Q^2}$, see Eq.~\eqref{eq:z_def}, we can easily compute the projections: 
\begin{align}
&-g_{\mu\,\nu}W^{\mu \, \nu}_H = 3
F_{1,\,H} - 
\left( \frac{2}{z} \, \frac{M^2}{Q^2} - \frac{z}{2}
\right) \, F_{2,\,H} = 
3 F_{1,\,H} + \frac{z}{2} \, F_{2,\,H} + 
\mathcal{O} \left( \frac{M^2}{Q^2} \right); 
\label{eq:W_proj1} \\
&\frac{P_\mu P_\nu}{Q^2} W^{\mu \, \nu}_H = 
\left(
-\frac{M^2}{Q^2} + \left( \frac{z}{2} \right)^2
\right) \, F_{1,\,H} + 
\left(
\frac{2}{z} \frac{P^4}{Q^4} - z \frac{P^2}{Q^2} + 
\left( \frac{z}{2} \right)^3
\right) \, F_{2,\,H} = \notag \\
&\quad= \left( \frac{z}{2} \right)^2 \, \left[ 
F_{1,\,H} + 
\frac{z}{2} \, F_{2,\,H} \right] + 
\mathcal{O} \left( \frac{M^2}{Q^2} \right).
\label{eq:W_proj2}
\end{align}

\bigskip


\subsection{Factorization of the hadronic tensor \label{subsec:W_fact}}


\bigskip

\begin{figure}[t]
 \centering 
\begin{tabular}{c@{\hspace*{15mm}}c}  
     \includegraphics[width=6.3cm]{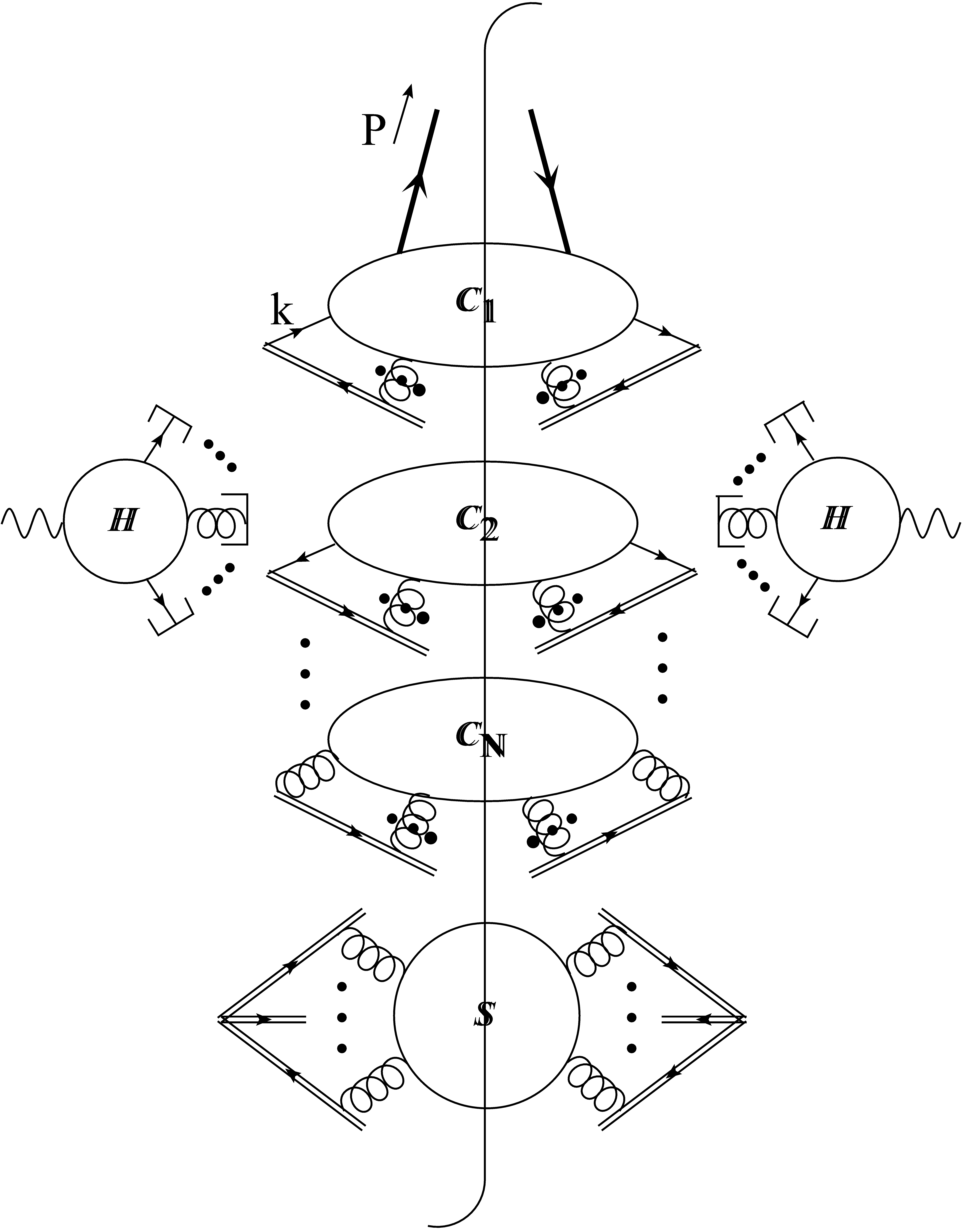}
&
     \includegraphics[width=6.3cm]{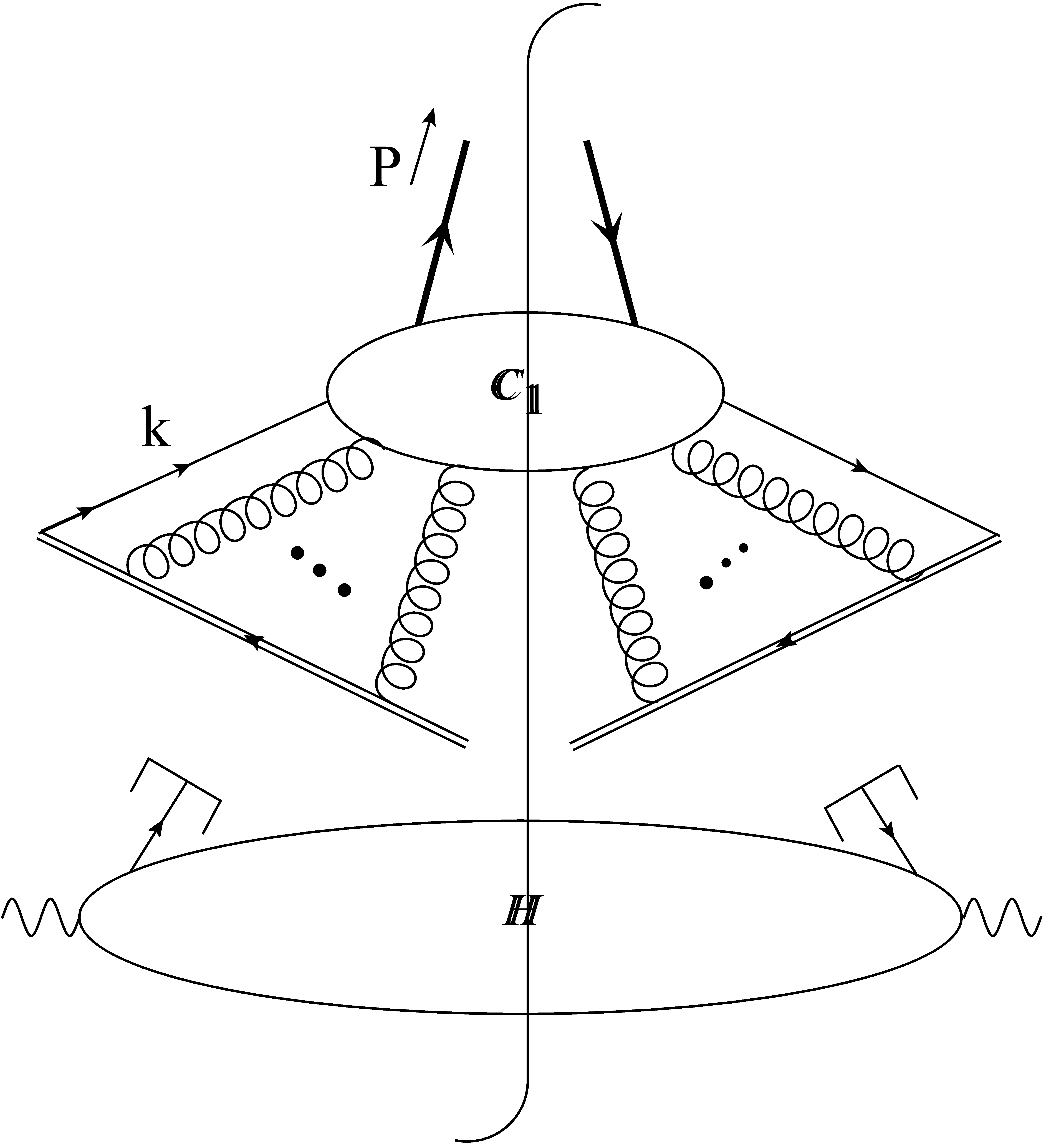}
 \\
 (a) & (b)
\end{tabular}
\caption{(a): Leading momentum regions for the hadronic tensor $W^{\mu \, \nu}_H$ as they appear in the first step of factorization.
(b): Actual representation of the leading momentum regions of the hadronic 
tensor $W^{\mu \, \nu}_H$. All the collinear factors corresponding to real 
emissions have been included into the hard part. The soft factor of the 
process is equal to one, as expected for a \hclass{1} process (see 
Section~\ref{subsec:coll_vs_tmd}).}
\label{fig:epm_regions}
\end{figure}

The factorization procedure allows to factorize the hadronic tensor $W^{\mu \, \nu}_H$ into hard, collinear and soft parts, 
as shown in Fig.~\ref{fig:epm_regions}~(a).
According to Ref.~\cite{Collins:2011zzd} and by using dimensional 
regularization, 
we have:
\begin{align}
&W^{\mu \, \nu}_H(P,\,q,\,T) = \sum_{N \geq 2} \, 
\sum_{j_1} \int \kmeas{k}{1} \, \sum_{j_2} \int \kmeas{k}{2} \prod_{\alpha = 
3}^N \,  
\int \kmeas{k}{\alpha} \coll_{\alpha}(k_\alpha)_{j_\alpha} \times \notag \\
&\mbox{Tr}_D \Big \{ 
\mathcal{P}_1 \coll_{1}(k_1,\,P)_{j_1,\,H} \overline{\mathcal{P}}_1 
\mathbb{H}_{j_1,\dots j_N}^\mu(\widehat{k}_1,\dots\widehat{k}_N,\,T) \,
\mathcal{P}_2 \coll_{2}(k_2)_{j_2} \overline{\mathcal{P}}_2 \,
(\mathbb{H}^\dagger)_{j_1\, \dots j_N}^\nu
(\widehat{k}_1,\dots \widehat{k}_N,\,T)
\Big \} \notag \\
&\times \int \kmeas{k}{S} \soft{N}{}_{j_1\, \dots j_N}(k_S) \, \delta^{(n)}(q - 
\widehat{k}_1 - \widehat{k}_2 - \sum_{\alpha} \widehat{k}_\alpha). 
\label{eq:W_prelim}
\end{align}
In Eq.~\eqref{eq:W_prelim} the collinear parts are represented by 
$\coll_j$: they depend only on the entering total momentum $k_j$ and on the 
type $j$ of the corresponding parton (either a gluon or a 
quark/antiquark of flavor $j$/$\Bar{j}$), and they are averaged over the color of 
the initiating parton. Among them, $\coll_1$ and $\coll_2$ are associated 
to the fermionic legs of the quark and the antiquark, hence they appear 
associated to the fermionic projectors, $\mathcal{P}_j$ and 
$\overline{\mathcal{P}}_j$, which connect them to the 
hard parts and make the jet partons on-shell.
Since the hard part and the collinear parts are computed in the same frame (the 
$h$-frame, as defined in Appendix \ref{app:kin}) the expressions for these projectors are simply
\begin{equation} \label{eq:projectors}
\mathcal{P} = \frac{\gamma^- \, \gamma^+}{2}
\quad\text{and}\quad 
\overline{\mathcal{P}} = \frac{\gamma^+ \, \gamma^-}{2}.
\end{equation}
Furthermore, by charge conjugation, $j_2 = \overline{j}_1$.
The projectors defined above will be fundamental in extracting the leading twist 
FFs of the quark and the anti-quark in the cross section.
All the other collinear parts, $\coll_\alpha$, are generated by gluons.
In this case, the role of the fermionic projectors of Eq.~\eqref{eq:projectors} 
is played by a gluon density matrix $\rho_{j'\,j}$ that encodes the information 
about the gluon polarization. In the following, we will consider the case of a 
fragmenting quark, corresponding to the collinear part $\coll_1$ as depicted 
in Fig.~\ref{fig:epm_regions}. 

In Eq.~\eqref{eq:W_prelim} the hard parts are represented by $\mathbb{H}$ and 
its hermitian conjugate, $\mathbb{H}^\dagger$: they encode the kinematics of the process. 
Momentum conservation is ensured by the appropriate delta function. 
However, in the hard contributions, the parton momenta are \emph{approximated}, 
in that only their leading components are considered, as stressed 
by the ``$^\wedge$" hats on them.
In practice, the momentum $\widehat{k}_\alpha$ is $k_\alpha$ projected onto the (unknown) direction of its corresponding collinear part:
\begin{equation} \label{eq:approx_momenta}
\widehat{k}_\alpha = w_{\alpha} \, \frac{k_\alpha \cdot 
\widetilde{w}_\alpha}{w_{\alpha} \cdot \widetilde{w}_\alpha},
\end{equation}
where $w_{\alpha}$ and $\widetilde{w}_\alpha$ are the light-like vectors 
corresponding to the plus and the minus directions, respectively, in the 
reference frame of $\coll_\alpha$.
The approximated momentum of the fragmenting quark is simply:
\begin{equation} \label{eq:appr_k1}
\widehat{k}_1 = \left( \widehat{k}_{1,\,h}^+,\,0,\,\vec{0}_T \right)_h,
\end{equation}
where $\widehat{k}_{1,\,h}^+ = k_{1,\,h}^+$, since the reference frame of the 
fragmenting parton corresponds (by definition) with the hadron frame.
Furthermore, 
kinematics impose constraints on the possible values that $k_{1,\,h}^+$ can assume, since $P_h^+ < k_{1,\,h}^+ < {P_h^+}/{z}$ (see Eqs.~\eqref{eq:z_coll},~\eqref{eq:upper_limit} and~\eqref{eq:lower_limit}).

\begin{figure}[t]
\centering
\includegraphics[width=8.5cm]{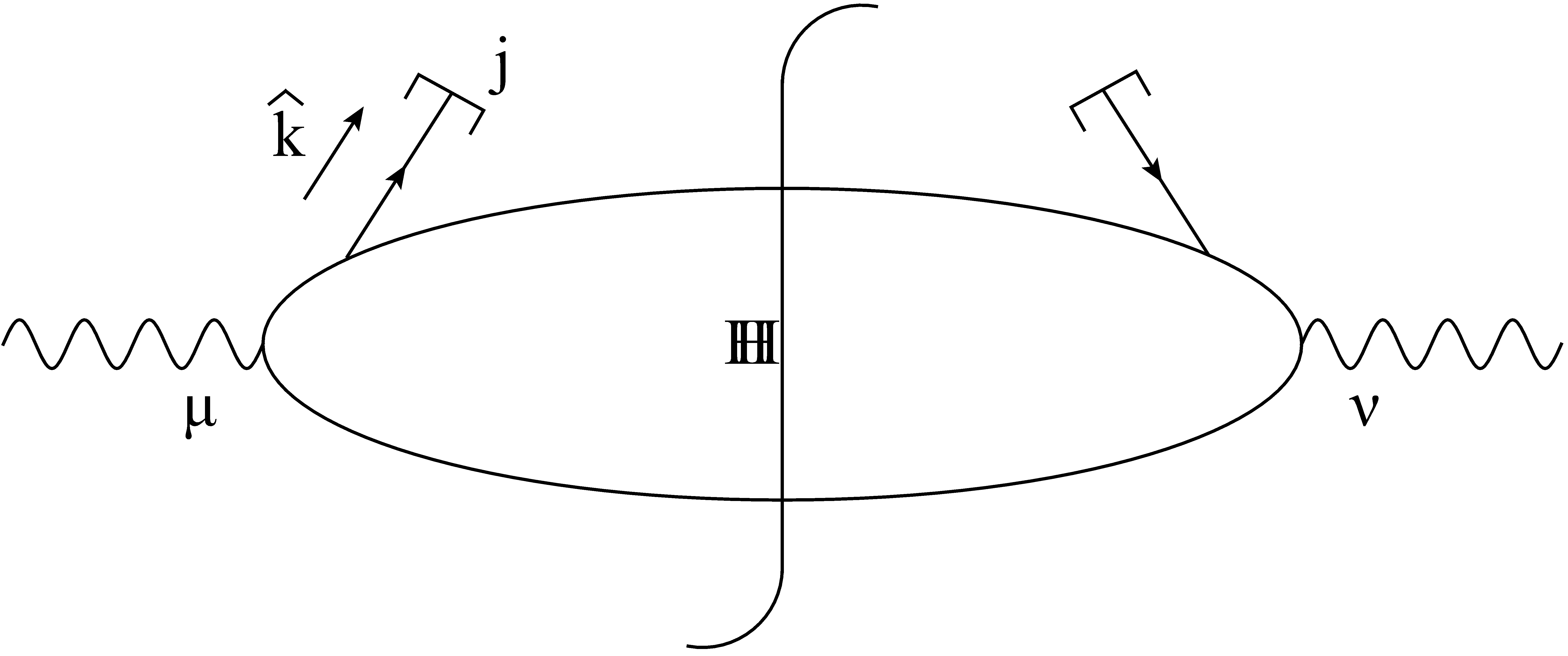}
\caption{Pictorial representation of $\widehat{W}_j^{\mu \, \nu}$.}
\label{fig:W_partonic}
\end{figure}

Finally, the soft contribution is a $N$-h soft factor, where $N$ is the total 
number of partons exiting the hard scattering.
It depends on the collinear parton type only 
through their color representation.
Notice that the total soft momentum $k_S$ cannot be involved in the kinematics 
of the process, since it is washed out by the real hard emission (at least one, $\coll_2$). 
In fact, none of the $k_S$ components appear in the conservation delta.
As a consequence, $\soft{N}$ is integrated over \emph{all} the components of 
$k_S$ and its contribution becomes trivial. 
As expected for a process belonging to the \hclass{1}, the soft factor is unity 
and can be omitted in the leading region representation~\cite{Collins:2011zzd}. 
All the collinear parts except $\coll_1$, whose reference hadron is the detected hadron $H$, are actually hard contributions.
Therefore, they can be included in one single (larger) hard factor that will 
involve not only $\mathbb{H}$ and  $\mathbb{H}^\dagger$, but also all the $\coll_\alpha$.
The final result, depicted in Fig.~\ref{fig:epm_regions}~(b), is given by:
\begin{align}
W^{\mu \, \nu}_H(P,q,T) = \sum_{j_1} \int d \widehat{k}_{1,\,h}^+ 
\int \frac{d k_{1,\,h}^- \, d^{D-2}\vec{k}_{1,\,T,\,h}}{(2 \pi)^D}
\, \mbox{Tr}_D \Big \{ 
\mathcal{P}_1 \coll_{1}(k_1,\,P)_{j_1,\,H} \overline{\mathcal{P}}_1 
\mathcal{H}_{j_1}^{\mu \, \nu}(Q,\,\widehat{k}_{1,\,h}^+,\,T)
\Big \}. \label{eq:W_prelim_2}
\end{align}
In the above equation, all the hard contributions have been collected in 
the hard coefficient $\mathcal{H}^{\mu \, \nu}$.
Notice that, while the collinear part $\coll_1$ depends on all the components 
of $k_1$, the hard contribution depends only on its leading component, 
$k_{1,\,h}^+$.
Then,  $\coll_1$ and $\mathcal{H}^{\mu \, \nu}$ are not completely 
disentangled, because a convolution over $k_{1,\,h}^+$ will survive.
In the following we will drop the index ``1" related to the fragmenting parton, 
which has become redundant.
Applying the fermionic projectors and parity conservation, 
the only surviving contribution in the case of 
$\epm \to H\,X$ is given by the coefficient of $\gamma^-$ in the expansion of Eq.~\eqref{eq:FF_Clifford}:
\begin{align}
\mathcal{P} \coll(k,\,P)_{j,\,H} \overline{\mathcal{P}} = \gamma^- \; 
\frac{\mbox{Tr}_D}{4}  \Big \{ 
\gamma^+ \coll(k,\,P)_{j,\,H}
\Big \}. \label{eq:proj_coll}
\end{align}
The Dirac trace of $\gamma^+ \coll(k,\,P)_{j,\,H}$ defines two TMD FFs (as in Eq.~\eqref{eq:unp_Clifford}):
\begin{align}
&\frac{1}{\widehat{z}} \, \int \frac{d k_{h}^-}{(2 \pi)^D} \frac{\mbox{Tr}_D}{4} 
\Big \{ \gamma^+ \coll(k,\,P)_{j,\,H}\Big \} = \notag \\
&\quad= D_{1,\,H/j}(\widehat{z},\,\vert -\widehat{z} \, \vec{k}_{h,\,T} \vert) - 
\frac{1}{M} |\vec{S}_T \times \vec{k}_{h,\,T}| 
D_{1T,\,H/j}^{\perp}(\widehat{z},\,\vert -\widehat{z} \, \vec{k}_{h,\,T} \vert), 
\label{eq:TMDs_projectors}
\end{align}
where $M$ and $\vec{S}_T$ are the mass and the transverse spin of the detected 
hadron, while $\widehat{z} = {P_h^+} / {k_h^+}$.
The function $D_{1,\,H/j ^\perp}$ is the unpolarized TMD FF, while $D_{1T,\,H/j}$ is the Sivers-like TMD FF.
For simplicity, in the following we will collectively indicate with $D_{j,\,H}(\widehat{z},\,-\widehat{z} \, 
\vec{k}_{h,\,T})$ the sum 
of the two contributions in the r.h.s. of Eq.~\eqref{eq:TMDs_projectors}.
Therefore:
\begin{align}
&W^{\mu \, \nu}_H(P,\,q,\,T) = \sum_{j} \int d \widehat{k}_h^+ \; \widehat{z} \,
\int d^{D-2}\vec{k}_{h,\,T} \; D_{j,\,H}(\widehat{z},\,-\widehat{z} \, 
\vec{k}_{h,\,T})\,
\mbox{Tr}_D \Big \{ \gamma^- \mathcal{H}^{\mu \, 
\nu}(Q,\,\widehat{k}_h^+,\,T)_j\Big \} = 
\notag \\
&= \sum_{j} \int_z^1 \, d \widehat{z} 
\; \widehat{W}_j^{\mu \, \nu}({z}/\widehat{z},\,Q,\,T) \,
\int d^{D-2}\vec{k}_{h,\,T} \; 
D_{j,\,H}(\widehat{z},\,-\widehat{z} \, 
\vec{k}_{h,\,T}) ,
\label{eq:W_final_kT}
\end{align}
where the kinematics constraints over $\widehat{z}$ have been taken into account.
The role of the hard factor in the previous equation is played by the function $\widehat{W}_j^{\mu \, \nu}$, which is the partonic analogue of the full hadronic tensor $W^{\mu \, \nu}_H$.
It is defined as:
\begin{equation} \label{eq:partonic_W}
\widehat{W}_j^{\mu \, \nu}(\widehat{k},\,q,\,T) = \mbox{Tr}_D \Big \{ 
\widehat{k}_h^+ \gamma^- \mathcal{H}_j^{\mu \, \nu}(Q,\,\widehat{k}_h^+,\,T)\Big \},
\end{equation}
Notice that since the approximated parton momentum has only a plus component, 
we can write $\widehat{k}_h^+ \gamma^- = 
\slashed{\widehat{k}} = \sum_{\mbox{ \small 
spin}} u( \, \widehat{k} \, ) \overline{u}( \, \widehat{k} \, ) $.
Therefore, $\widehat{W}_j$ is the algebraic expression 
corresponding to the pictorial representation given in 
Fig.~\ref{fig:W_partonic}.
Its actual definition has to be equipped with the subtraction of the double counting due to the overlapping with the collinear momentum region (see Section~\ref{subsec:xs_div}).
%
%
In Eq.~\eqref{eq:W_final_kT}, the dependence on the parton transverse momentum is only in the collinear part and, in principle, the integrand of $W^{\mu \, \nu}_H$ could be defined as the hadronic tensor differential in $\vec{k}_{h,\,T}$.
However, although the parton transverse momentum is not a physical observable, kinematics relates $\vec{k}_{h,\,T}$ with the transverse momentum $\vec{P}_{p,\,T}$ of the outgoing hadron in the parton frame, i.e. measured with respect to its final state jet axis, that we identify with the thrust axis  (see Appendix~\ref{app:kin}).
This can be measured (as it has been done by the BELLE Collaboration,  Ref.~\cite{Seidl:2019jei}) and the definition of the hadronic tensor differential in $P_{p,\,T}$ is obtained by the change of variables $\vec{k}_{h,\,T} = -\frac{1}{\widehat{z}} \; \vec{P}_{p,\,T}\left[1 + 
\mathcal{O}( \frac{P_{p,\,T}^2}{Q^2} )\right]$ (see Eq.~\eqref{eq:P_p}).
Therefore:
\begin{align}
\frac{d W_H^{\mu \, \nu}(z,\,Q,\,T)}{d^2 \vec{P}_{p,\,T}} = 
\sum_{j} \int \frac{d \widehat{z}}{\widehat{z}^{2}} \,
\widehat{W}_j^{\mu \, \nu}({z}/\widehat{z},\,Q,\,T) \,
D_{j,\,H}(\widehat{z},\, \vec{P}_{p,\,T}) \, 
 \left[1 + \mathcal{O}(\frac{P_{p,\,T}^2}{Q^2} )\right].
\label{eq:W_final_pT_diff_1}
\end{align}
The differential of $\vec{P}_{p,\,T}$ carries information about two variables: the modulus $P_{p,\,T}$ and the azimuthal angle $\beta$ in the $x\,y$-plane of the parton frame.
While the first can be measured, the angle $\beta$ cannot be determined experimentally.
In fact, an angular dependence in the TMD contribution $D_{j,\,H}$ can originate from the Sivers-like 
contribution $|\vec{S}_T \times \vec{P}_{p,\,T}|$ (see Eq.~\eqref{eq:TMDs_projectors}).
However, as explained in Ref.~\cite{Guan:2018ckx}, the transverse 
spin of the hadron is orthogonal to its transverse momentum with respect to the axis of the jet, identified with the
thrust axis.
Hence $|\vec{S}_T \times \vec{P}_{p,\,T}| = \pm S_T \, P_{p,\,T}$ for any choice of the $x$-axis in the parton frame.
Therefore, the integration over $\beta$ is trivial and results just in a $2\pi$ factor on the r.h.s of Eq.~\eqref{eq:W_final_pT_diff_1}:
\begin{align}
\frac{d W_H^{\mu \, \nu}(z,\,Q,\,T)}{d P^2_{p,\,T}} = 
\pi \, 
\sum_{j} \int \frac{d \widehat{z}}{\widehat{z}^{2}} \,
\widehat{W}_j^{\mu \, \nu}({z}/\widehat{z},\,Q,\,T) \,
D_{j,\,H}(\widehat{z},\, P_{p,\,T}) \, 
 \left[1 + \mathcal{O}(\frac{P_{p,\,T}^2}{Q^2} )\right],
\label{eq:W_final_pT_diff_2}
\end{align}
where:
\begin{align}
&D_{j,\,H}(\widehat{z},\, P_{p,\,T}) = D_{1,\,H/j}(\widehat{z},\,P_{p,\,T}) 
\mp \frac{\widehat{z}}{M} S_T \, P_{p,\,T} \,
D_{1T,\,H/j}^{\perp}(\widehat{z},\,P_{p,\,T}).
\label{eq:TMDs_pT}
\end{align}

\bigskip


\subsection{Factorized Cross Section \label{subsec:xs_fact}}


\bigskip

The full cross section is obtained by contracting the hadronic tensor of Eq.~\eqref{eq:W_final_pT_diff_2}, and its partonic counterpart, Eq.~\eqref{eq:partonic_W}, with the leptonic tensor, as in 
Eq.~\eqref{eq:epm_crosssec}:
\begin{align}
\frac{d \sigma}{(\lips{P}) \, dP^2_{p,\,T} \,dT} = 
\pi
\sum_{j} \int_z^1 \frac{d \widehat{z}}{\widehat{z}^{2}} \, 
\frac{d \widehat{\sigma}_j}{\lips{\widehat{k}}\,dT} \,
D_{j,\,H}(\widehat{z},\, P_{p,\,T}) \, 
 \left[1 + \mathcal{O}(\frac{P_{p,\,T}^2}{Q^2} )\right],
\label{eq:xsec_fully_differential_1}
\end{align}
where the dependence on thrust has been made explicit.
Let's focus on the r.h.s. of the previous equation.
The only non-zero component of the approximated parton momentum $\widehat{k}$ is in the plus direction, as defined in Eq.~\eqref{eq:appr_k1}.
Therefore, its Lorentz invariant phase space measure can only be written as:
\begin{align}
&\lips{\widehat{k}} =
\frac{1}{2} d |\vec{\widehat{k}}| 
\, |\vec{\widehat{k}}|\,  d\!\cos{\theta} \, d \phi =
\frac{Q^2 }{8}  \,\frac{z}{\widehat{z}} \, d \left(\frac{z}{\widehat{z}} \right) 
\, d\!\cos{\theta} \, d \phi,
\label{eq:lips_k_appr}
\end{align}
and carries information about the polar angle $\theta$ and the azimuthal angle $\phi$ with respect to the beam axis (LAB frame, see Appendix~\ref{app:kin}).
On the l.h.s the same variables have to appear explicitly.
Hence, the Lorentz invariant phase space of the detected hadron is written in the LAB frame as well:
\begin{align}
&\lips{P} = 
\frac{Q^2 }{8}  \,z \, d z \, d\!\cos{\theta} \, d \phi \,
\left[1 + \mathcal{O} \left(\frac{M^2}{Q^2} \right) \right] .
\label{eq:lips_P_LAB}
\end{align}
Finally, the cross section is given by:
\begin{align}
&\frac{d \sigma}{dz \,d\!\cos{\theta} \,d \phi\, dP^2_{p,\,T} \,dT} = \notag \\
&\quad= 
\pi
\sum_{j} \int_z^1 \frac{d \widehat{z}}{\widehat{z}} \, 
\frac{d \widehat{\sigma}_j}
{d({z}/{\widehat{z}})\,d\!\cos{\theta} \,d \phi\,dT} \,
D_{j,\,H}(\widehat{z},\, P_{p,\,T}) 
\, \left[1 + \mathcal{O}(\frac{P_{p,\,T}^2}{Q^2} ,\; \frac{M^2}{Q^2})\right].
\label{eq:xsec_fully_differential_2}
\end{align}
There are five independent observables:
\begin{enumerate}
\item The fractional energy $z = {2 |\vec{P}|}/{Q}$.
\item The polar angle $\theta$ of the outgoing hadron with respect to the electron.
\item The azimuthal angle $\phi$ of the outgoing hadron with respect to the $x$-axis in the LAB frame.
This is significant only if such axis can be defined unambiguously, as in the case of polarized leptons.
Otherwise, we can simply drop $d\phi$ on both sides of Eq.~\eqref{eq:xsec_fully_differential_2} as a result of integration, which is our case.
\item The thrust $T$, defined in Eq.~\eqref{eq:hadron_thrust}.
\item The (modulus of the) transverse momentum of the outgoing hadron $P_{p,\,T}$ with respect to its final state jet axis, that we identify with the thrust axis.
\end{enumerate}
Common scenarios are those in which experiments  provide two or three of the variables listed above:
\begin{itemize}
\item $z$ and $\theta$ are measured, but the thrust axis is not reconstructed,  
hence $P_{p,\,T}$ is unknown.
In this case, in addition to the integration over $T$, the previous cross section has to be integrated over all possible values of the transverse momentum  $P_{p,\,T}$, 
restoring the integrated TMDs as in Eq.~\eqref{eq:W_final_kT}:
\begin{align}
&\frac{d \sigma}{dz \,d\!\cos{\theta}} = 
\sum_{j} \int_z^1 \frac{d \widehat{z}}{\widehat{z}} \,
\frac{d \widehat{\sigma}_j}{d {z}/{\widehat{z}} \, d\cos{\theta}} \,
d_{j/H}(\widehat{z})\,
\left[1 + \mathcal{O}(\frac{M^2}{Q^2})\right],
\label{eq:xsec_theta_diff}
\end{align}
where we used the results of Appendix~\ref{subapp:smallbT_tmds}.
This result coincides with the cross section presented in Chapter 12 of Ref.~\cite{Collins:2011zzd}.
The convolution over $\widehat{z}$ is between renormalized quantities, as we did for the OPE of TMDs at small $b_T$ in Eq.~\eqref{eq:ope_tmds_renorm}. 

The dependence on $\theta$, both in the partonic and in the full cross section, can expressed in terms of longitudinal (L) and transverse (T) contributions:
\begin{align} 
\frac{d \sigma}{d x \, d \cos{\theta}} = 
\frac{3}{8} (1 + \cos^2{\theta})
\frac{d \sigma_T}{d x} + 
\frac{3}{4} \sin^2{\theta}  \frac{d \sigma_L}{d x}\,,
\label{eq:xs_theta}
\end{align}
where $x$ can be $z$ in the full cross section, or ${z}/{\widehat{z}}$ in its partonic counterpart.
The structure functions 
are related to the transverse and the 
longitudinal component of the cross section as 
follows:
\begin{align}
&\frac{d \sigma_T}{d x} = \frac{4 \pi \alpha^2}{3 Q^2} x F_1(x, Q^2), 
\label{eq:xsec_T}\\
&\frac{d \sigma_L}{d x} = \frac{\pi \alpha^2}{3 Q^2} \left[ 2 x F_1(x, Q^2) + 
x^2  F_2(x, Q^2)\right]. \label{eq:xsec_L}
\end{align}
\item $z$ and $P_{p,\,T}$ are measured, but the polar angle $\theta$ of the outgoing hadron with respect to the beam axis is integrated over.
Indeed, the measurement of the transverse momentum $P_{p,\,T}$ has to be done with respect to the jet axis, which for our purposes coincides with the thrust axis.
Therefore, if the cross section is differential in $P_{p,\,T}$, it also has to be differential in $T$ (or in an analogous variable that allows to determine the axis of the jet)\footnote{On the contrary, clearly, it is possible to measure $T$ regardless of $P_{p,\,T}$.}.

The integration of the partonic cross section with respect to $\theta$ is straightforward and follows from Eqs.~\eqref{eq:xs_theta},~\eqref{eq:xsec_T} and~\eqref{eq:xsec_L}:
\begin{align}
&\int_{-1}^1 \, d \cos{\theta} \, 
\frac{d \widehat{\sigma}_j}{d x \, d \cos{\theta} \, dT} = \frac{d \sigma_T}{d z} + 
\frac{d \sigma_L}{d z} = 
\notag \\
&\quad=
\frac{4 \pi \alpha^2}{3 Q^2} \, x \, 
\left( \frac{3}{2} \, F_{1,\,j}(x,\,Q^2,\,T) + 
\frac{x}{4} \, F_{2,\,j}(x,\,Q^2,\,T)\right).
\label{eq:xs_int_theta}
\end{align}
For simplicity, in the following we will collectively indicate with ${d\sigma}/{d x}$ the sum 
of the two contributions on the r.h.s. of Eqs.~\eqref{eq:xs_int_theta}.
Therefore:
\begin{align}
&\frac{d \sigma}{dz\, dP^2_{p,\,T} \,dT} = 
\pi
\sum_{j} \int_z^1 \frac{d \widehat{z}}{\widehat{z}} \,
\frac{d \widehat{\sigma}_j}
{d({z}/{\widehat{z}})\,dT} \,
D_{j,\,H}(\widehat{z},\, P_{p,\,T}) 
\, \left[1 + \mathcal{O}(\frac{P_{p,\,T}^2}{Q^2} ,\; \frac{M^2}{Q^2})\right].
\label{eq:xsec_PT_differential_1}
\end{align}
Since TMDs are defined in the Fourier conjugate space, see
Eq.~\eqref{eq:fact_defTMDs}, it is more convenient to write the cross section 
using their $b_T$-space counterparts: 
\begin{align}
&\int d^{D-2} \vec{P}_{p,\,T} \; 
e^{i \, \vec{P}_{p,\,T} \cdot \vec{b}_T} 
D_{j,\,H}(z, \vec{P}_{p,\,T}) \,
\left[1 + \mathcal{O}(\frac{P_{p,\,T}^2}{Q^2})\right]=
\notag \\
&\quad=
z^{D-2}  \int d^{D-2} \vec{k}_{T,\,h}  \; e^{i \, \vec{k}_{T,\,h} \cdot (- z \vec{b}_T)} 
D_{j,\,H}(z, -z \vec{k}_{T,\,h}) =
z^{D-2} \widetilde{D}_{j,\,H}(z, \, - z \, \vec{b}_T),
\label{eq:FT_PT}
\end{align}
where $D_{j,\,H}$ is actually only a function of the modulus of $\vec{P}_{p,\,T}$.
Hence:
\begin{equation} \label{eq:antiFT_PT}
D_{j,\,H}(z, P_{p,\,T}) \,
\left[1 + \mathcal{O}(\frac{P_{p,\,T}^2}{Q^2})\right]= 
\int \frac{d^{2} \vec{b}_T}{(2 \pi)^{2}} \, 
e^{i \, \frac{\vec{P}_{p,\,T}}{z} \cdot 
\vec{b}_T} \, \widetilde{D}_{j,\,H}(z, \,b_T).
\end{equation}
Notice that all definitions in  Eqs.~\eqref{eq:fact_defTMDs}  and~\eqref{eq:tmd_sol_2} hold for the Fourier transformed TMD FFs $\widetilde{D}_{j,\,H}$.
Finally, the cross section in its final form is given by:
\begin{align}
\frac{d \sigma}{dz\, dP^2_{p,\,T} \,dT} &= 
\pi
\sum_{j} \int_z^1 \frac{d \widehat{z}}{\widehat{z}} \,
\frac{d \widehat{\sigma}_j}{d({z}/{\widehat{z}})\,dT} \,
\times \notag \\
&\quad\times \int \frac{d^{2} \vec{b}_T}{(2 \pi)^{2}} \, 
e^{i \, \frac{\vec{P}_{p,\,T}}{\widehat{z}} \cdot  \vec{b}_T} \, 
\widetilde{D}_{j,\,H}(\widehat{z}, \,b_T)
\, \left[1 + \mathcal{O}(\frac{M^2}{Q^2})\right].
\label{eq:xsec_PT_differential_2}
\end{align}
As for the cross section in Eq.~\eqref{eq:xsec_theta_diff} the convolution is between renormalized quantities, as we will discuss in the next Section.
Furthermore, in constrast to Eq.~\eqref{eq:xsec_PT_differential_1}, the cross section written in terms of the Fourier transform is a function defined for \emph{any} value of $P_{p,\,T}$ and in fact the errors are only sized as ${M^2}/{Q^2}$.
However, the physical meaning is lost for large values of the transverse momentum of the outgoing hadron, as the TMDs themselves become non physical in the large $P_{p,\,T}$ region (see the discussion at  the end of Appendix~\ref{subapp:smallbT_tmds}).
Hence, the cross section of Eq.~\eqref{eq:xsec_PT_differential_2} can only be trusted where $P_{p,\,T} \ll Q$ or, more precisely, where $P_{p,\,T} \ll P^+ = z \, {Q}/{\sqrt{2}}$, which is the actual condition that allows to consider the outgoing hadron as a collinear particle, according to the power counting rules.
\end{itemize}

\bigskip


\subsection{Subtraction Mechanism \label{subsec:xs_div}}


\bigskip

As it is clear from Eq.~\eqref{eq:xsec_PT_differential_2}, the $\epm \to HX$ cross section, differential in $z$, in thrust $T$ and in the transverse momentum of the detected hadron with respect to the thrust axis, $P_{p,\,T}$,  offers a direct probe of the transverse motion of partons.
Recently the BELLE Collaboration has provided high statistics experimental data corresponding to such cross section \cite{Seidl:2019jei}.
Although the final result 
of Eq.~\eqref{eq:xsec_PT_differential_2} is simply the Fourier Transform of the convolution of a TMD FF and a thrust-dependent hard factor, i.e. the partonic cross section integrated over $\theta$, 
the phenomenological application of Eq.~\eqref{eq:xsec_PT_differential_2} requires special care.

\bigskip

The final cross section is RG invariant if the anomalous dimension of the hard factor is exactly equal and opposite to that of the TMD FF, order by order in perturbation theory.
This argument applies to renormalized quantities, i.e. functions provided of the proper UV counterterm.
Furthermore, the hard factor in 
Eq.~\eqref{eq:xsec_PT_differential_2} has to be 
properly subtracted to avoid double counting due to the overlapping with the collinear momentum region.
Therefore, the hard factor of the final cross section is defined in two steps: first it is equipped with subtractions, then it is renormalized.

The unsubtracted analogue 
of the hard factor in Eq.\eqref{eq:xsec_PT_differential_2} is the partonic version of the full cross section. 
Being a partonic quantity, it is completely unaware of the outgoing hadron.
It describes the process at partonic level, which means $\epm \rightarrow  f \, X$, where $f$ is a parton of type $f$ that replaces the detected hadron.
The most convenient frame where to compute $\widehat{\sigma}^{\mbox{\small unsub}}$ is the analogue of the hadron frame, where the momentum of $f$ lies along the plus direction. 
The expression of its final state tensor $\widehat{W}_f^{\mu\,\nu}{}^{\mbox{\small , unsub}}$
is obtained from the integrand of Eq.~\eqref{eq:W_final_kT}.
For a given value of $T$, the phase space available for real emissions is restricted, because only the final state topology associated with that particular value of thrust can be reached.
A simple way to force the phase space to describe only the region of interest is introducing sharp cut-offs that shrink the available range of values of T.
For instance, if we are interested in the quasi $2$-jet limit, we can force $T$ to remain in the neighborhood of $1$ by defining a minimal value of $T$,  $T_{\mbox{\tiny MIN}} \leq 1$.
In practice, the unsubtracted final state tensor is obtained by computing the contribution of all the Feynman graphs needed when $T$ lies in the range defined by the topology cut-offs, in the massless limit and by setting all the soft/collinear divergent quantities to their lowest order (in the language of \TheBook, this corresponds to the application of the hard approximator $T_H$, modified to include the introduction of the cut-offs for thrust).
Since the lowest order for the collinear part is just a product of delta function that sets the momentum of the fragmenting parton to be equal to that of the outgoing parton, and the lowest order for the soft factor is 
unity, in $k_T$-space we simply have:
\begin{align}
\frac{d \widehat{W}_f^{\mu\,\nu}{}^{\mbox{\small , unsub}}(\epsilon;\,z,\,T;\,T_c)}
{d^{2-2\epsilon} \vec{k}_T} = 
\sum_j \, \int_z^1 \frac{d \widehat{z}}{\widehat{z}} \,
\widehat{W}_j^{\mu\,\nu}{}^{\mbox{\small , unsub}}(\epsilon;\,{z}/{\widehat{z}},\,T;\,T_c) \,
\delta_{j\,f} \,
\delta(1-\widehat{z}) \,
\delta^{2-2\epsilon}(\vec{k}_T),
\label{eq:unsub_W_kTspace}
\end{align}
where $T_c$ stands for a generic topology cut-off for thrust.
As a consequence, the Fourier transform of the previous expression does not depend on $b_T$.
In Eq.~\eqref{eq:unsub_W_kTspace} we explicitly showed the dependence on $\epsilon$, which is the regulator used in dimensional regularization (where the spacetime dimension is set to $D = 4 - 2\epsilon$). 
In fact, the unsubtracted final state tensor is collinear divergent and presents poles in $\epsilon$.
The standard subtraction procedure removes the overlapping with the momentum region described by the collinear part, that represents the boundary of the phase space corresponding to the emissions along the direction of the outgoing parton $f$, and hence it also removes the collinear divergences of $\widehat{W}_f^{\mu\,\nu}{}^{\mbox{\small , unsub}}$.
The subtraction term is obtained by considering the collinear approximation of the unsubtracted final state tensor and it coincides with the partonic version of the (bare) TMD FFs, that will be denoted by $D^{(0)}_{f/j}$ (in the language of \TheBook, the subtraction term is obtained by applying $T_A T_H$, i.e. applying both the approximator collinear to the outgoing particle and the hard approximator).
However, due to the presence of topology cut-offs, represented by $T_c$ in Eq.~\eqref{eq:unsub_W_kTspace}, we have to slightly modify the subtraction mechanism. 
In particular, it is not the whole partonic TMD has to be subtracted out, but only the part that actually overlaps.
This coincides with the contribution given by the transverse momenta that lie in the power counting momentum region, i.e. where $k_T$ is at most of order $\lambda$, with $\lambda$ being some IR energy scale. 
Order by order we can relate $\lambda$ and $T_c$ by a precise kinematic relation.
As a consequence, the Fourier transform of the partonic TMD cannot cover the whole range of $k_T$, but it stops when $k_T$ reaches $\lambda$. The resulting quantity does not depend on $b_T$, nevertheless it shows an explicit dependence on the transverse momentum cut-off and, ultimately, on $T_c$.
Summarizing, the (bare) Fourier transformed subtraction term is defined by:
\begin{align}
&\widetilde{D}^{(0),\,(\lambda)}_{f/j}
(\epsilon;\,z,\,\lambda(T_c),\,\zeta) = 
\int d^{2-2\epsilon} \vec{k}_T \, 
e^{-i \, \vec{k}_T \cdot \vec{b}_T} \, 
D^{(0)}_{f/j} (\epsilon;\,z,\,k_T,\,\zeta)\,
\theta\left(\lambda(T_c) - k_T\right)
\label{eq:cutoff_TMDs}
\end{align}
This quantity is both collinear and UV divergent.
Since the poles of $\widetilde{D}^{(0),\,(\lambda)}_{f/j}$ are the same that would be obtained by a complete Fourier transform, the UV divergence is renormalized by using the same UV counterterm that heals the UV divergence in the usual TMDs (i.e. those defined without a cut-off on transverse momentum).
In this regard, as a consequence of the mechanism of subtractions, the \emph{subtracted} final state tensor acquires the same UV divergences of $\widetilde{D}^{(0),\,(\lambda)}_{f/j}$, but with opposite signs.
Then we can easily renormalize it by using the inverse of the TMD UV counterterm.
In $b_T$-space we have the following factorization formula:
\begin{align}
&\underbrace{\widehat{W}_f^{\mu\,\nu}{}^{\mbox{\small , unsub}}
(\epsilon; \,z, \,T, \,T_c)}_\text{coll. divergent}  =
\notag \\
&\quad=
\sum_{j} \int_z^1 \, \frac{d \widehat{z}}{\widehat{z}} \; 
\underbrace{\widehat{W}_j^{\mu \, \nu}{}^{\mbox{\small , sub,} (0)}
(\epsilon;\,{z}/\widehat{z},\,T,\,\lambda(T_c),\, \zeta )}_\text{UV divergent} 
\,
\underbrace{\widehat{z} \; 
\widetilde{D}^{(0),\,(\lambda)}_{j,\,f}
(\epsilon; \, \widehat{z},\,\lambda(T_c),\,\zeta)}_\text{UV divergent and
coll. divergent}
=
\notag \\
&\quad=
\sum_{j} \int_z^1 \, \frac{d \widehat{z}}{\widehat{z}} \; 
\left\{
\widehat{W}_k^{\mu \, \nu}{}^{\mbox{\small , sub,} (0)}
(\epsilon;\,{z}/\widehat{z},\,T,\,\lambda(T_c),\, \zeta)
Z_{\mbox{\tiny TMD}}^{-1}{}^k_{\hspace{.2cm}j}
(\epsilon;\,\mu,\,\zeta)
\right\} \times
\notag \\
&\quad \times
\left\{
\widehat{z} \,
Z_{\mbox{\tiny TMD}}{}_{j}^{\hspace{.2cm}l}
(\epsilon;\,\mu,\,\zeta)\,
\widetilde{D}^{(0),\,(\lambda)}_{l,\,f}
(\epsilon; \, \widehat{z},\,\lambda(T_c),\,\zeta)
\right\} =
\notag \\
&\quad=
\sum_{j} \int_z^1 \, \frac{d \widehat{z}}{\widehat{z}} \; 
\widehat{W}_j^{\mu \, \nu}{}^{\mbox{\small , sub}}
({z}/\widehat{z},\,T, \,\mu,\,\lambda(T_c), \, \zeta )
\,
\underbrace{\widehat{z} \; 
\widetilde{D}^{(\lambda)}_{j,\,f}
(\epsilon; \, \widehat{z},\,\mu,\,\lambda(T_c),\,\zeta)}_\text{
coll. divergent}
,
\label{eq:partonicW_unsub}
\end{align}
where we simply used the associative property of convolutions and a sum over repeated upper-lower flavor indices is implicit.
Notice that at this stage the renormalized, subtracted final state tensor has acquired a  dependence on both the topology cut-off $T_c$ and on the rapidity cut-off $\zeta$.
Order by order in perturbation theory, the functions $\widehat{W}_j^{\mu \, \nu}{}^{\mbox{\small , sub}}$ are determined recursively by using:
\begin{align}
&\widehat{W}_j^{\mu \nu,\,{[n]}}{}^{\mbox{\small , sub}}
(z,\,T,\,\mu,\,\lambda(T_c),\, \zeta ) = 
\widehat{W}_f^{\mu \nu,\,{[n]}}
{}^{\mbox{\small , unsub}} (\epsilon; \,z, \,T, \,T_c) + 
\notag \\
&-
\sum_j \, 
\sum_{m = 1}^n \;
\int_z^1 \frac{d\widehat{z}}{\widehat{z}}\;
\widehat{W}_{j,\;R}^{\mu \nu,\,{[n-m]}}
{}^{\mbox{\small , sub}} 
({z}/{\widehat{z}},\,T,\,\mu,\,\lambda(T_c),\, \zeta)
\,
\left[ 
\widehat{z} \, 
\widetilde{D}_{j,\,f}^{[m],\,(\lambda)} 
(\epsilon; \, \widehat{z},\,\mu,\,\lambda(T_c),\,\zeta)
\right],
\label{eq:hard_coeff_pert}
\end{align}
In the previous result, we used the fact that the lowest order of the partonic TMDs equipped with $\lambda$ is just a delta function, 
$\widetilde{D}_{f/j}^{[0],\,(\lambda)}\left(
\widehat{z} \,
\right) = \delta_{f \; j} \, \delta(1-\widehat{z})$, as in the case in which there is no cut-off on transverse momenta.
From now on, when the labels ``sub" and ``R" are not explicitly indicated,  $\widehat{W}_j$ will be implicitly considered both subtracted and renormalized.
Once the expression of $\widehat{W}_j$ is known, the full subtracted, renormalized cross section is computed straightforwardly through the partonic structure functions $\widehat{F}_{1,\,j}$ and $\widehat{F}_{2,\,j}$, obtained as in Eqs.~\eqref{eq:W_proj1} and~\eqref{eq:W_proj2}.

\bigskip

The partonic cross section resulting from the previous subtraction procedure obeys the following RG evolution equation:
\begin{align}
\frac{\partial}{\partial \log{\mu}} 
\log{\left(\frac{d \widehat{\sigma}_j 
\left(
\mu,\,\lambda(T_c),\,\zeta
\right)}
{dz \, dT}\right)} = 
-\gamma_{D,\,j} \left( \alpha_S(\mu),\,{\zeta}/{\mu^2} \right),
\label{eq:partonic_xs_RGevo}
\end{align}
where $\gamma_{D,\,j}$ is the anomalous dimension of the TMD FF of flavor $j$.
The RG invariance of the full cross section follows straightforwardly, as the anomalous dimensions of the partonic cross section and of the full TMD FF appearing in  Eq.~\eqref{eq:xsec_PT_differential_2} are equal and opposite.
The derivative of the partonic cross section with respect to the rapidity cut-off $\zeta$ plays the same role of the CS evolution for the TMD (see Eq.~\eqref{eq:CS_evo}). 
Then, in analogy with the soft kernel $\widetilde{K}$, we define the rapidity-independent kernel that determines the CS-evolution for the partonic cross section as:
\begin{align}
\frac{\partial}{\partial \log{\sqrt{\zeta}}} 
\log{\left(\frac{d \widehat{\sigma}_j 
\left(
\mu,\,\lambda(T_c),\,\zeta
\right)}
{dz \, dT}\right)} = 
\frac{1}{2} \widehat{K}
\left(\alpha_S(\mu),\, {\mu^2}/{\lambda(T_c)^2}\right).
\label{eq:partonic_xs_CSevo}
\end{align}
The kernel $\widehat{K}$ has an additive anomalous dimension:
\begin{align}
\frac{\partial}{\partial \log{\mu}} \,
\widehat{K}\left( \alpha_S(\mu),\,{\mu^2}/{\lambda(T_c)^2}\right) = \gamma_K\left(\alpha_S(\mu)\right).
\label{eq:RG_evo_partonicK}
\end{align}
This anomalous dimension is equal and opposite to $\gamma_K$, which is associated to the soft kernel $\widetilde{K}$ (see Eq.~\eqref{eq:gammaK_evo}). 
Finally, the partonic cross section shows an explicit dependence on the scale $\lambda(T_c)$ used to constrain the transverse momentum of the fragmenting parton. The corresponding evolution has no analogue in the TMDs. It is given by:
\begin{align}
\frac{\partial}{\partial \log{\lambda}}\,
\log{\left(\frac{d \widehat{\sigma}_j 
\left(
\mu,\,\lambda(T_c),\,\zeta
\right)}
{dz \, dT}\right)} = 
G\left(\alpha_S(\mu),\,
{\mu^2}/{Q^2},\,{\zeta}/{\mu^2},\,
{\mu^2}/{\lambda(T_c)^2}\right).
\label{eq:partonic_xs_lambdaevo}
\end{align}
The $\lambda$-evolution kernel $G$ is RG invariant. Furthermore, it obeys the following CS-evolution:
\begin{align}
\frac{\partial}{\partial \log{\sqrt{\zeta}}}
G\left(\alpha_S(\mu),\,
{\mu^2}/{Q^2},\,{\zeta}/{\mu^2},\,
{\mu^2}/{\lambda(T_c)^2}\right) = 
\frac{1}{2} \,
\frac{\partial}{\partial \log{\lambda}}
\widehat{K}\left( \alpha_S(\mu),\,{\mu^2}/{\lambda(T_c)^2}\right).
\label{eq:CS_evo_G}
\end{align}
Finally, the solution to the evolution equations Eqs.~\eqref{eq:partonic_xs_RGevo},~\eqref{eq:partonic_xs_CSevo} and~\eqref{eq:partonic_xs_lambdaevo} gives:
\begin{align}
&\frac{d \widehat{\sigma}_j
(\mu,\,\lambda(T_c),\,\zeta)}
{dz \, dT} = 
\left.\frac{d \widehat{\sigma}_j}
{dz \, dT}\right\rvert_{
\substack{\text{ref.}}}\,
\mbox{exp}
\Bigg\{
\int_\mu^Q \,\frac{d\mu'}{\mu'}\,
\gamma_D\left( \alpha_S(\mu'),\,{\zeta}/{(\mu')^2}\right)
\Bigg\} \times
\notag \\
&\quad\times
\mbox{exp}
\Bigg\{
\frac{1}{4}\,
\widehat{K}\left( \alpha_S(Q),\,1\right)\,
\log{\frac{\zeta}{Q^2}} -
\int_{\lambda(T_c)}^Q \,\frac{d\lambda'}{\lambda'}\,
G\left(\alpha_S(Q),\,
1,\,{\zeta}/{Q^2},\,
{Q^2}/{(\lambda')^2}\right)
\Bigg\},
\label{eq:xs_evo_sol}
\end{align}
where we have also used the RG-invariance of the kernel $G$. The label ``ref." stands for the energy scales at the reference values $\mu = Q$, $\zeta = Q^2$ and $\lambda(T_c) = Q$.

So far, we have considered $\lambda$ and $\zeta$ as independent. 
As a consequence, the evolution of the partonic cross section can be written in perfect analogy to the TMD evolution. Furthermore, this approach makes the RG-invariance of the final cross section explicit, see Eq.~\eqref{eq:partonic_xs_RGevo}.
However, the correct separation between hard and collinear momentum regions, represented by the partonic cross section and the TMD FFs respectively, can only be obtained by setting $\zeta = \lambda(T_c)^2$.
This is due to the presence of an upper boundary for the transverse momentum of the fragmenting parton, which automatically reflects onto a lower limit for the rapidity of the produced particles.
Therefore, the only cut-off left in the final cross section is the topology cut-off $T_c$.
It enters in the final formula of Eq.~\eqref{eq:xsec_PT_differential_2}
 through the partonic cross section 
and through the rapidity cut-off of the TMD FFs.
Its value has to be chosen according to the topology of the final state and, ultimately, it depends on the kinematics of the process.

\bigskip


\subsection{A simple example of the Rapidity Dilation mechanism \label{subsec:LO_example}}


\bigskip

\begin{figure}[t]
\centering
\includegraphics[width=5.5cm]{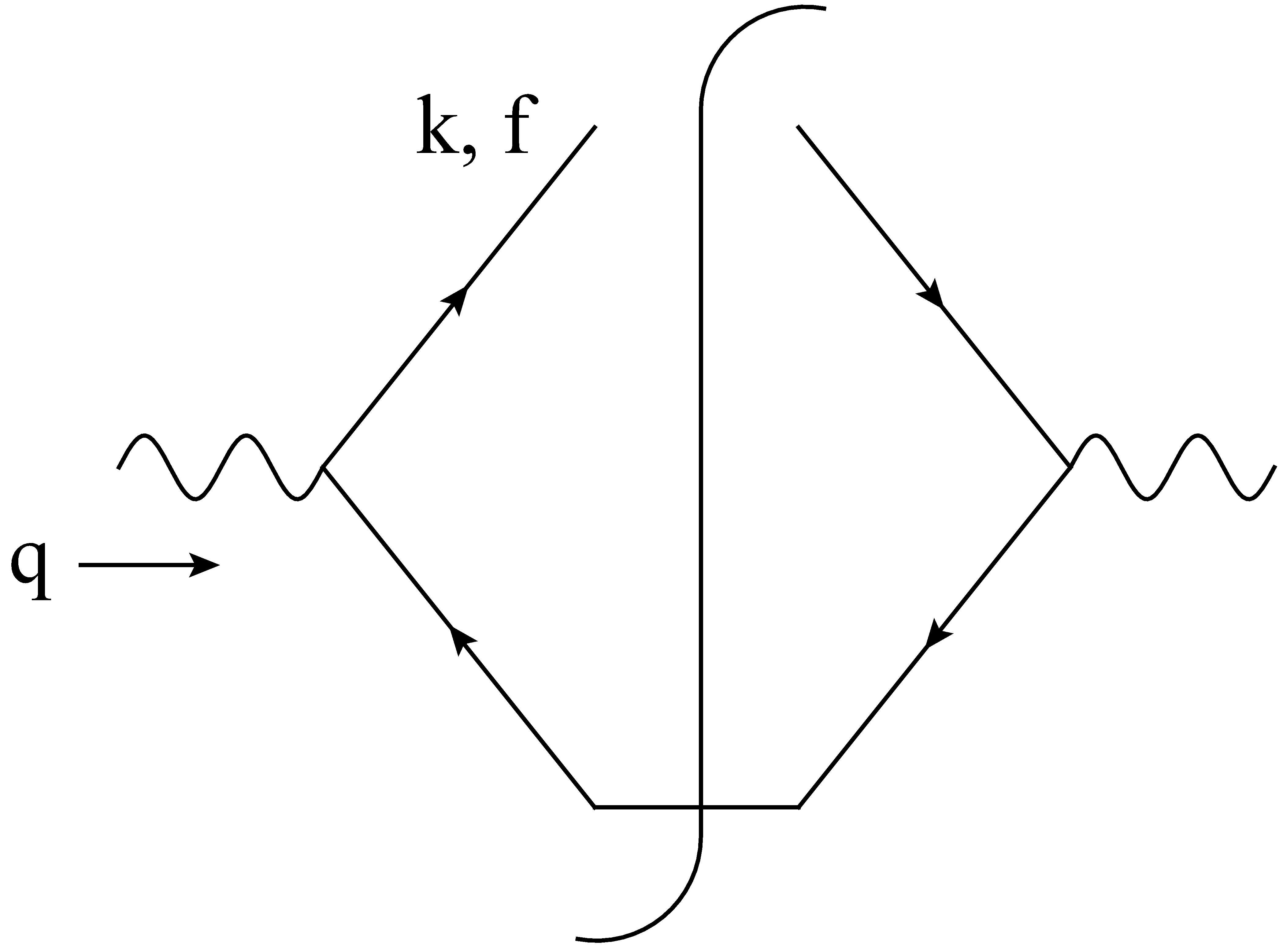}
\caption{Amplitude squared for the LO partonic tensor, in the limit $T \rightarrow 1$.}
\label{fig:LO_hard}
\end{figure}

A crude example of the thrust-dependent cross section of $\epm \rightarrow H \, X$ process can be obtained from a basic QCD approximation in the $2$-jet limit, i.e. $T \rightarrow 1$.
At lowest order, the subtraction mechanism is trivial and the subtracted, renormalized hard coefficient are easily computed from Eq.~\eqref{eq:hard_coeff_pert}:
\begin{align}
&\left(\widehat{W}_f^{\mu \, \nu}\right)^{[0]} (z,\,T) = \left(\widehat{W}_f^{\mu\,\nu}{}^{\mbox{\small , unsub}}
\right)^{[0]} (z,\,T).
\label{eq:LO_hard_coeff}
\end{align}
In the $2$-jet limit, the 
only Feynman diagram contributing to the l.h.s of the previous equation is given by Fig.~\ref{fig:LO_hard}.
It is an exact $2$-jet configuration, hence $T = 1$.
As a consequence, the phase space integration is trivial and does not require the introduction of the topology cut-off $T_c$.
%
%
The actual computation is easier for the projections (see
Eqs.~\eqref{eq:W_proj1} and~\eqref{eq:W_proj2}):
\begin{align}
&-g_{\mu\,\nu}
\left(\widehat{W}_f^{\mu\,\nu}\right)^{[0]} (z,\,T) = 
(1-\epsilon)
\, \delta_f^{\hspace{.2cm}q} e_q^2 \; 2 N_C \, \delta(1-z) \, \delta(1-T) ; \\
&\frac{k_\mu k_\nu}{Q^2}
\left(\widehat{W}_f^{\mu\,\nu}\right)^{[0]} (z,\,T)= 0 .
\label{eq:W_proj_LO}
\end{align}
Notice that the gluon contribution is always suppressed in a $2$-jet configuration.
Then we can compute the lowest order subtracted, renormalized structure functions:
\begin{align}
&\widehat{F}^{[0]}_{1,\,f}(z,\,T) =  
\delta_f^{\hspace{.2cm}q} e_q^2 \; 
N_C \, \delta(1-z) \, \delta(1-T) ; \\
&\widehat{F}^{[0]}_{2,\,f}(z,\,T) = -\frac{2}{z} \; \widehat{F}^{[0]}_{1,\,f}(z,\,T) .
\label{eq:struct_fun_LO}
\end{align}
Finally, by using Eqs.~\eqref{eq:xsec_T},~\eqref{eq:xsec_L} and~\eqref{eq:xs_int_theta}, the LO  subtracted and renormalized partonic cross section appearing in the final result of Eq.~\eqref{eq:xsec_PT_differential_2} is given by:
\begin{align}
&\frac{d \widehat{\sigma}_f^{[0]}}{dz\,dT}= 
a_T \, \frac{4 \pi \alpha^2}{3 Q^2} \, z \, \widehat{F}^{[0]}_{1,\,f}(z,\,T) = 
\notag \\
&\quad=
a_T \, \frac{4 \pi \alpha^2}{3 Q^2} \, N_C \,
\delta_f^{\hspace{.2cm}q} e_q^2 \; 
\delta(1-z) \, \delta(1-T),
\label{eq:LO_hard_coeff_explicit}
\end{align}
where the factor $a_T$ accounts for the limited acceptance in the polar angle $\theta$.
In the following, the detected hadron will be considered spinless for simplicity.
Hence, the Sivers-like contribution disappears and in the cross section will remain only the unpolarized TMD FF $D_1$.
Its crudest estimate is the Leading Log (LL) approximation, given by:
\begin{align}
&\widetilde{D}_{1\;j/H}^{LL}(z,\, b_{T}; \, Q, \,\zeta) =
\frac{1}{z^2} \, d_j (z,\,\mu_b) \times 
\notag \\
&\quad \times 
\mbox{ exp} \left \{
L_b \, g^{\mbox{\tiny LL}}_1 \left( a_S (Q) L_b \right) +
g^{\mbox{\tiny LL}}_2 \left( a_S (Q) L_b , \, \log{\left({\zeta}/{Q^2}\right)} \right)
\right \} \times \notag \\
&\quad \times \left(M_{D_1}\right)_{j,\,H}(z,\,b_T) \mbox{ exp} 
\left \{
-\frac{1}{4} \, g_K(b_T) \, 
\log{\left( z^2 \, \frac{\zeta}{M_H^2} \right)}
\right \},
\label{eq:D1_LL}
\end{align}
where $L_b = \log{({Q}/{\mu_b})}$ and the functions $g^{\mbox{\tiny LL}}_1$, $g^{\mbox{\tiny LL}}_2$ are given in Eqs.~\eqref{eq:LL_functions}. 
Notice that, since there is no need for a topological cut-off, the rapidity cut-off $\zeta$ is  unconstrained in the LO, LL approximation.
This is a consequence of the low degree of information encoded in the perturbative part of the TMD. Basically, all the constraints on the rapidity of the collinear particles are contained in the non-perturbative part of the TMD FFs.
Therefore, any modification of $\zeta$
has to be traced back to a modification of the non-perturbative model $M_{D_1}$ that describes the fragmentation mechanism.
The rapidity dilation transformation discussed in Section~\ref{subsec:rap_dil} allows to choose the rapidity cut-off consistently with the choice of the model. 
Therefore,
the LO, LL cross section is written in terms of a generic $\zeta$. It is given by:
\begin{align}
&\frac{
d \sigma_{2\mbox{\small -jet}}
^{[0],\;LL}}
{dz\, dP^2_T \,dT} = 
\notag \\
&\quad= 
\pi
\sum_{j} \int_z^1 \frac{d \widehat{z}}{\widehat{z}} \,
\frac{d \widehat{\sigma}_j^{[0]}}{d({z}/{\widehat{z}})\,dT} \,
\int \frac{d^2 \vec{b}_T}{(2 \pi)^2} \, 
e^{i \, \frac{\vec{P}_{p,\,T}}{\widehat{z}} \cdot  \vec{b}_T} \, 
\widetilde{D}^{LL}_{j,\,H}(\widehat{z}, \,b_T,\,Q,\,\zeta)
\, \left[1 + \mathcal{O}(\frac{M^2}{Q^2})\right] = 
\notag \\
&\quad=
a_T \, \frac{4 \pi^2 \alpha^2}{3 Q^2} \, N_C \,
\delta(1-T) \, 
\sum_q \, e_q^2 \,
\int \frac{d^2 \vec{b}_T}{(2 \pi)^2} \, 
e^{i \, \frac{\vec{P}_{p,\,T}}{z} \cdot  \vec{b}_T} \, 
\frac{1}{z^2} \, d_q (z,\,\mu_b) \times 
\notag \\
&\quad \hspace{.5cm} \times
\mbox{ exp} \left \{
L_b \, g^{\mbox{\tiny LL}}_1 \left( a_S (Q) L_b \right) +
g^{\mbox{\tiny LL}}_2 \left( a_S (Q) L_b , \, \log{\left({\zeta}/{Q^2}\right)} \right)
\right \} \times \notag \\
&\quad \hspace{.5cm} \times 
\left(M_{D_1}\right)_{q,\,H}(z,\,b_T) \mbox{ exp} 
\left \{
-\frac{1}{4} \, g_K(b_T) \, 
\log{\left( z^2 \, \frac{\zeta}{M_H^2} \right)}
\right \}
\, \left[1 + \mathcal{O}(\frac{M^2}{Q^2})\right]
\label{eq:LO_LL_xs}
\end{align}
Notice that this formula represents the simplest, non trivial approximation beyond the parton model picture. It holds valid to LO in the perturbative expansion and at $T=1$, hence it only has illustrative purposes. A  reliable phenomenological analysis should not rely on Eq.~\eqref{eq:LO_LL_xs}, but rather on the full NLO expression, with the appropriate accuracy in the order of logarithms, which will soon be presented in Ref.~\cite{Boglione-Simonelli:2020}. In the following, we will give a prototypical application of this LO cross section formula to a small sub-sample of the BELLE data~\cite{Seidl:2019jei}, which should only serve as an example of the rapidity dilation mechanism discussed in Section~\ref{sec:coll_parts}. The simplicity of its usage and the small number of free parameters involved in the fitting procedure are indeed the points of strength of Eq.~\eqref{eq:LO_LL_xs}. 
%
\begin{figure}[t]
\centering
\hspace{-1.3cm}
\includegraphics[width=13.7cm]{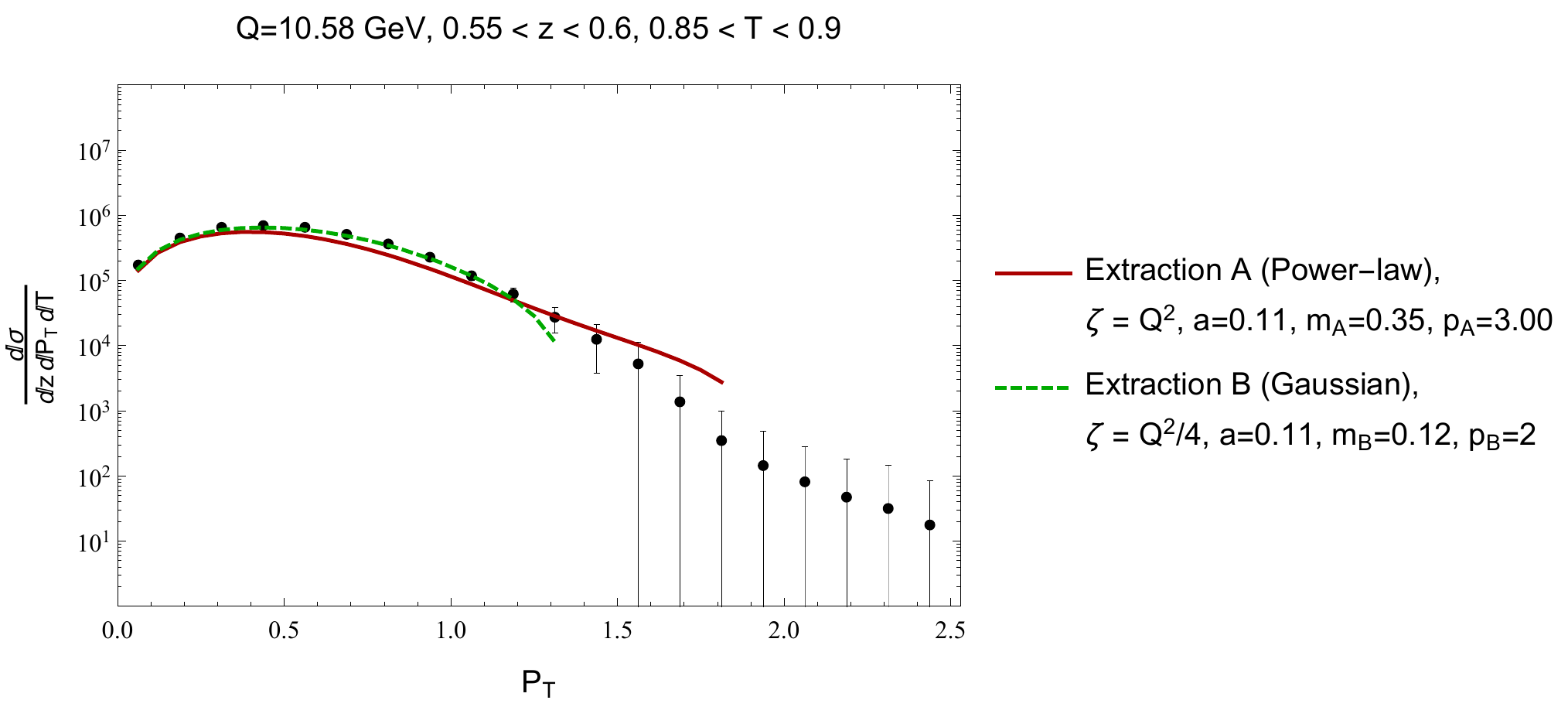}
\caption{$e^+e^- \to H X$ LO cross section computed according to Eq.~\eqref{eq:LO_LL_xs} using two different parameterizations for its non-perturbative part, and at two different values of the rapidity cut-off ($\zeta=Q_0^2$ solid black line, $\zeta=Q_0^2/4$ dashed green line). See text for more details.}
\label{fig:belle_xsec}
\end{figure}
%

For our example, we will consider only the subset of the BELLE $e^+e^- \to HX$ cross sections, corresponding to $0.55<z<0.6,0.85<T<0.90$, in 20 $P_T$ bins ranging from $0.06$ to $2.5$ GeV.  For the BELLE experiment $Q=10.58$ GeV. This data sub-sample is shown in Fig.~\ref{fig:belle_xsec}. Statistical and systematical errors are added in quadrature. The analysis will be performed using the NNFF10 fragmentation function set at LO~\cite{Bertone:2017tyb}, and fixing the values of $b_{\mbox{\tiny MIN}}$ and $b_{\mbox{\tiny MAX}}$ as follows:  $b_{\mbox{\tiny MIN}}=C1/Q \sim 0.1$ GeV$^{-1}$ and $b_{\mbox{\tiny MAX}}=1$ GeV$^{-1}$.

Let's now suppose that, somewhere around the globe, Group A performs a phenomenological analysis of the above BELLE data subset using a power-law parameterization of the model in $P_T$-space 
which, in the $b_T$ space, corresponds to a Bessel-K function, normalized in such a way that it is 1 at $b_T=0$:
\be
M_A (b_T,\,m,\,p) = 
\frac{2^{2-p}}{\Gamma(-1+p)} \, (b_T \, m)^{-1+p} \, K_{-1+p}(b_T \,m)
\ee
where $K_{-1+p}$ is the modified Bessel function of the second kind.
This model was successfully used in Ref.~\cite{Boglione:2017jlh} to fit the $\epm \to HX$ cross sections measured by the TASSO and MARKII Collaborations~\cite{Braunschweig:1990yd,Petersen:1987bq}.
Group A knows that the TMD cross section will become unphysical as $P_T$ grows larger, as it is only valid in the TMD region where $P_T << P^+$ (here $P^+ = z\,Q/\sqrt{2}\sim 4.3$ GeV). Therefore they fix $P_{T,\mbox{\tiny MAX}} = 1.8$ GeV. After this point the cross section will rapidly fall to zero and become negative.
Having set their rapidity cut-off at $\zeta=Q^2$, Group A best fit returns $m_A=0.35$ and $p_A=3.00$ for their two free parameters.
The resulting cross section is shown in Fig.~\ref{fig:belle_xsec} (red, solid line).
%
\begin{figure}[t]
\centering
\includegraphics[width=15cm]{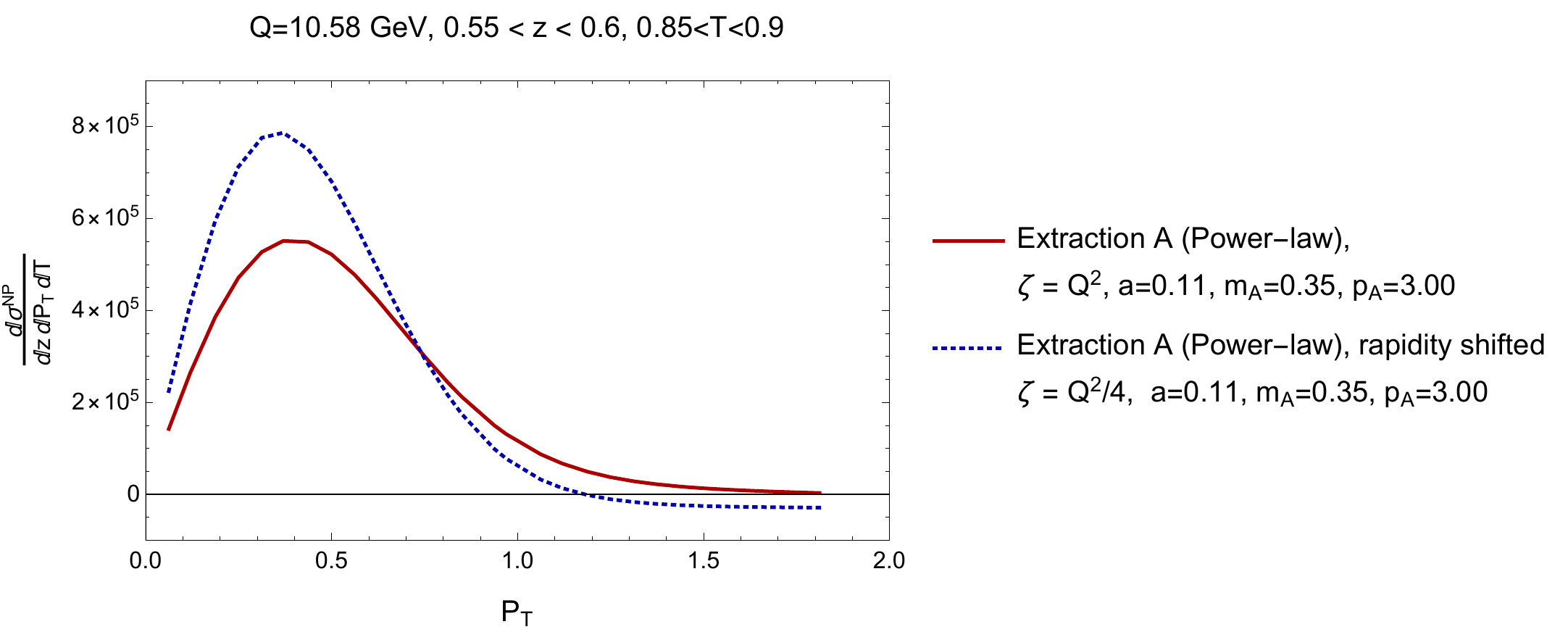}
\caption{The effect of a rapidity shift from $\zeta=Q^2$ to $\zeta=Q^2/4$ on the cross section  extracted by Group A. As expected the cross section is not invariant under this transformation.}
\label{fig:rap-shift}
\end{figure}
%

On the other side of the planet Group B, totally unaware of the work of Group A, performs a fit on the same data sample, but they choose a Gaussian parameterization for the model 
of their cross section (clearly the perturbative part has the same functional form 
in both cases)
\be
M_B(b_T,\,m, \,p) = e^ {-m \, {b_T}^{2} }\,.
\label{eq:gauss}
\ee
Here there is only one free parameter, $m$, as the power $p$ has been fixed to 2 to obtain a Gaussian form. They set their rapidity cut-off to $\zeta=Q^2/4$ and decide to be conservative on their TMD-regime requirement, so they fix $P_{T,\mbox{\tiny MAX}} = 1.3$ GeV. Their fit has only one free parameter, $m_B$, which the $\chi ^2$ minimization procedure sets to $0.12$. The corresponding cross section is shown in Fig.~\ref{fig:belle_xsec}, by the green dashed line.  

Notice that, in principle, there is at least one more free parameter in both analyses, which is used to model the $g_K$ function, see Eq.~\eqref{eq:LO_LL_xs}. As explained in Section~\ref{sec:coll_parts}, however, $g_K$ does not depend on the rapidity cut-off, nor on the flavour $j$ of the fragmenting quark. Therefore, it does not play any active role in a rapidity dilation and is not relevant in this example. We will therefore suppose it to be the same for Group A and B and parameterize it as $g_K = a \, b_T^2$ with $a=0.11$, fixed ``a priori". 

%
\begin{figure}[t]
\centering
\includegraphics[width=15cm]{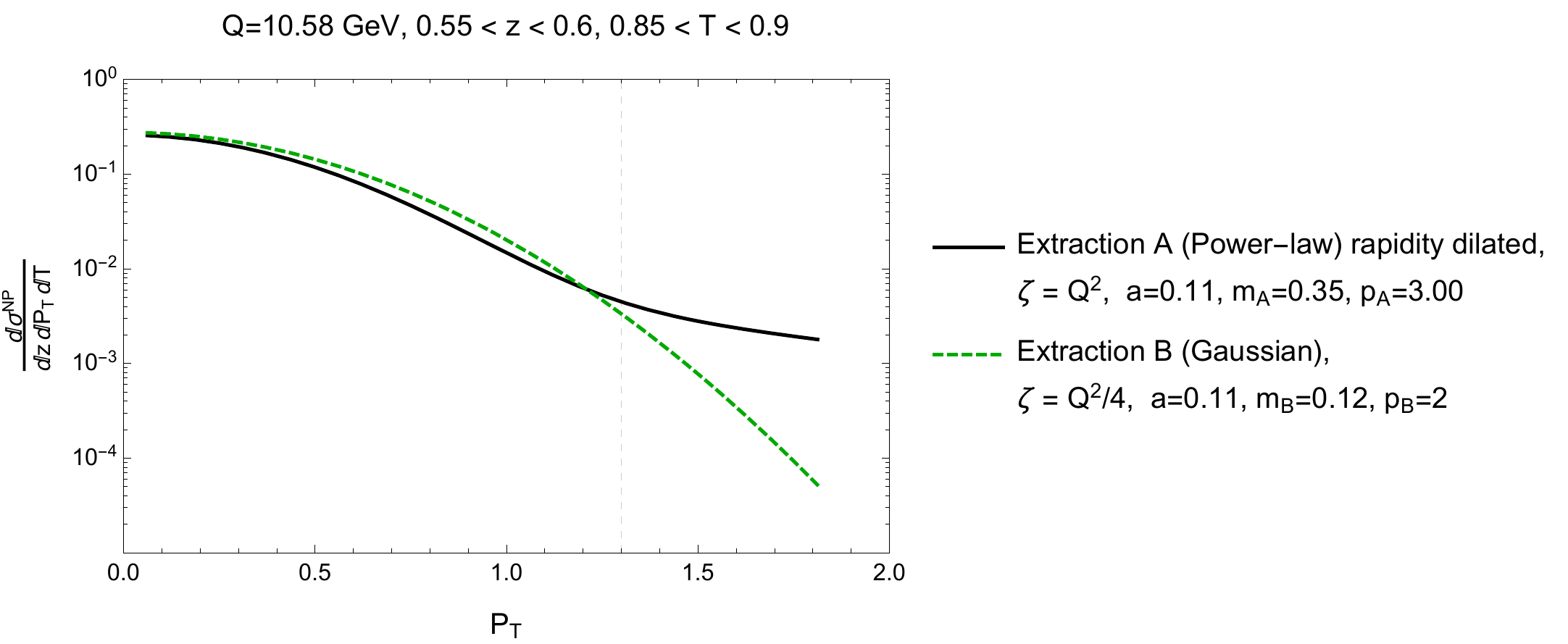}
\caption{Left panel: Non perturbative contribution to the LO cross section, corresponding to the same choice of rapidity cut-off. The solid black line represents the extraction of Group B, while the dashed green line is obtained from the extraction of group A by applying a rapidity dilation, i.e. through a transformation that brings $\zeta=Q_0^2$ to $\zeta=Q_0^2/4$ and compensates this variation by changing the value of the free parameters of the model $M_A$. See text for more details.}
\label{fig:sigmaNP}
\end{figure}
%
As it is clearly shown in Fig.~\ref{fig:belle_xsec}, the results obtained by Group A and B are consistent, within errors, as they fit the same data sample. 
Similarly, also the TMD fragmentation functions extracted by the two groups will be consistent at small $P_T$, where 
they carry a truly physical information about the transverse motion of the  hadronizing parton.
In $b_T$-space, the two TMDs are very similar at small $b_T$ but they may differ in their large $b_T$ behaviour, because of the different choices of models, $M_A(b_T)$ and $M_B(b_T)$.

It is at this point that Eq.~\eqref{eq:rapdil_compareFT} becomes crucial: in fact, it allows the two Groups to relate their independent extractions through a rapidity dilation.
The two extractions will not correspond to a one-to-one relation in $b_T$-space, nevertheless rapidity dilations preserve the physical meaning of TMDs. 
First of all, Group A performs a rapidity shift on their extraction: as expected the cross section is not invariant for a variation of the rapidity cut-off. This is illustrated in fig.~\ref{fig:rap-shift}. However, by applying a full rapidity dilation, i.e. transforming their TMD according to Eq.~\eqref{eq:rap_dil_tmd}, Group A can match their results to those obtained by Group B, in the range of small $P_T$ where the TMD approximation holds valid and where information from the experimental data is able to constrain the model. In fact, according to Eq.~\eqref{eq:rapdil_compareFT}, here we have: 
\be
\mathcal{FT} \Big[
\sigma^{\mbox{\tiny NP}}_{\zeta',\,M_B} (b_T) \Big]
\sim
\mathcal{FT} \Big[
\sigma^{\mbox{\tiny NP}}_{\zeta,\,M_A} (b_T)\,
\exp \Big\{
-\frac{1}{2} \,\theta \, \widetilde{K}(b^\star_T)
\Big\} 
\Big]
\quad \mbox{ at small } P_T.
\label{eq:rapdil_compareFT_pheno}
\ee
Here $\theta=\log 2$.

This is shown in Fig.~\ref{fig:sigmaNP}, where the black solid line represents the results of Group B for the non-perturbative contribution to the full cross section (left hand side of Eq.~\eqref{eq:rapdil_compareFT_pheno}), while the green line corresponds to the results of Group A for the analogous quantity {\it after} the application of a rapidity dilation (right hand side of Eq.~\eqref{eq:rapdil_compareFT_pheno}). 
Notice that $\sigma^{NP}_A$ is related to  $\sigma^{NP}_B$ by a factor 
which is purely perturbative and therefore calculable and totally model independent, see Eq.~\eqref{eq:rapdil_compareFT_pheno}.

\begin{figure}[t]
\centering
\includegraphics[width=8cm]{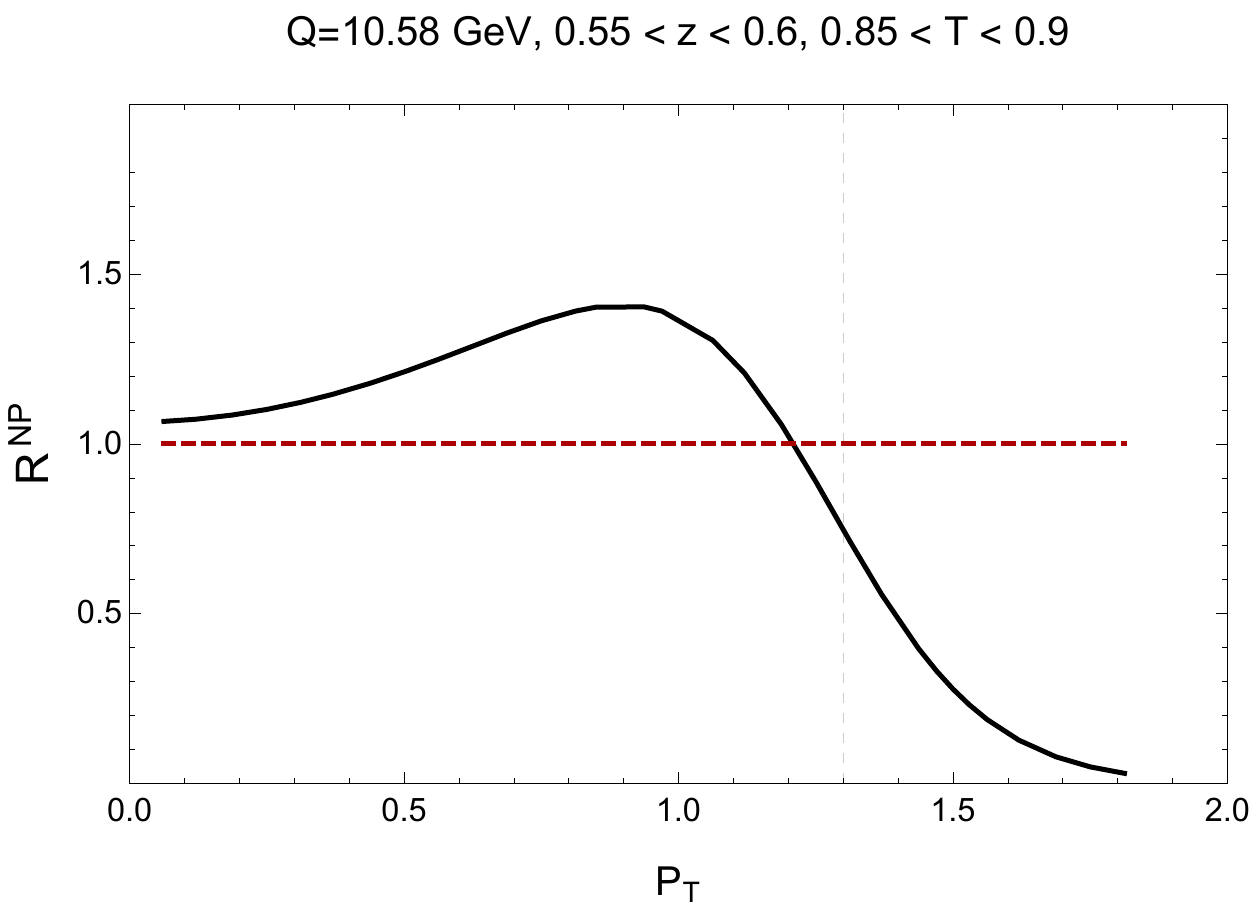}
\caption{The ratio between the non pertubative contributions to the cross sections calculated according to the extraction of Group B and the rapidity dilated extraction of Group A.}
\label{fig:sigmaNPratio}
\end{figure}

Fig.~\ref{fig:sigmaNPratio} shows the ratio of these two curves as a function of $P_T$, $R^{\mbox{\tiny NP}}$.
In an ideal world, where all extraction  converged to the same model, $R^{\mbox{\tiny NP}}$ would be 1 at all values of $P_T$ (dashed red line). However, in a realistic case $R^{\mbox{\tiny NP}}$ is very close to 1 only at small $P_T$, as it should, and it starts deteriorating as $P_T$ grows larger. 
It is not by chance that it stays close to 1 up to  $P_T \sim 1.3$, which corresponds to the value of $P_{T,\mbox{\tiny MAX}}$ set by group B. After that point, the cross section starts to become unphysical and the ratio itself becomes meaningless. In Fig.~\ref{fig:sigmaNPratio} a thin gray vertical line marks $P_{T,\mbox{\tiny MAX}} = 1.3$ GeV. 
Notice that the invariance under rapidity dilation is considerably powerful: it allows to preserve the physical part of the cross section, embodied by the TMD function at small $P_T$, even in a realistic situation in which a very limited range of $P_T$ is constrained by  experimental information,  while compensating for the variation of the rapidity cut-off in the perturbative part by a transformation of the non-perturbative model. 
%
%

\section{Conclusions}

In this paper we have extended the TMD factorization mechanism to processes belonging to different hadron classes. This is potentially a very powerful tool, as it allows us to exploit the same definition of TMD parton densities 
in different processes, which up to now could not be used in a simultaneous data 
analysis. With this extended definition of TMD, in particular, we have been able 
to apply the TMD formalism to the process of one hadron production from $e^+e^-$ 
scattering, belonging to the $1$-h hadron class. Within this scheme, the TMD FFs 
extracted from a phenomenological analysis of the $P_T$ dependent $\epm \to H X$ 
cross sections, can be related to the analogous TMD FFs as extracted in a $2$-h 
class process, like SIDIS 
or $\epm \to H_A \, H_BX$.

Clearly the extension of the factorization scheme comes to a price, a price that in this case turns out to be rather large and two-folded. 
First of all the soft factor, which is responsible for a (partial) breaking of 
universality, cannot be included in the definition of the TMD, 
as it is elegantly done in the standard TMD factorization through the 
``square-root" TMD definition. Freed by its soft contribution, TMD becomes truly 
universal and can be used in any class of processes. The soft factor, however, 
assumes a fundamental role as it becomes a pivotal ingredient of the factorized 
cross section, where the non-perturbative effects of soft physics are encoded 
in the soft model $M_S$.
It will have to be extracted within its corresponding hadron class and should 
only be used within that class. The process $\epm \to H X$ is a slightly 
exceptional case, as the soft factor here becomes unity, as shown in 
Section~\ref{sec:epm_1h}.

Having recovered a solid and truly universal definition of TMD, 
we can factorize cross sections as that of  $\epm \to H X$, where 
there is only one single TMD embodying the long-distance 
contributions. The all-order expression of this cross section has been obtained in Section~\ref{sec:epm_1h} 
following a factorization scheme derived from the CSS factorization procedure.
As a rude, first estimate of the final cross section we presented the result obtained to leading order (LO) and leading log (LL) accuracy. Here we had 
to face an additional problem:
the arbitrariness in the choice of the rapidity cut-off reflects in the 
LO result, undermining its predictive power.
To make the TMD independent of the choice of the rapidity cut-off, they have to 
be made invariant under a specific transformation, which we call ``rapidity 
dilation". 

Such transformations 
regulate how the perturbative and 
non-perturbative contributions are balanced within the TMD itself.
In fact, 
in physical observables the cut-off has to be taken very large ($y_1 
\to \infty$) but 
in the TMDs alone (which are not physical observables) there is total arbitrariness in 
choosing its particular value.
Rapidity dilations control this arbitrariness by acting both on
the rapidity cut-off and on the model $M_C$. The larger $y_1$ the more $M_C$ is 
suppressed, and the TMD is, basically, only perturbative. Less extreme 
values of $y_1$, instead, will correspond to a more dominant non-perturbative 
contribution.  

Separating perturbative and non-perturbative contributions is a highly 
non-trivial problem, which affects any phenomenological analysis. 
For example, ambiguities originate when we have to fix the value of $b_{MAX}$, 
which marks the critical value of the impact parameter at which 
non-perturbative contributions start becoming non negligible. TMDs are well 
defined within the approximation in which the partonic $k^+$ is very large while 
$k_T$ is small (i.e. collinear according to power counting), they should 
therefore correspond to partons with a very large rapidity and very small 
transverse momentum with respect to  the jet axis. The rapidity cut-off $y_1$, 
formally, will have to be taken to infinity but, in practice, the specific size 
of $y_1$ will determine how far we stretch the perturbative content of the TMD 
and where the non-perturbative contribution will become  dominant.  

To clarify the practical relevance of rapidity dilation invariance, in the last 
Section of this paper we have 
presented a simple 
example to show how rapidity dilations can 
offer a tangible help in relating phenomenological analyses performed using 
different non-perturbative model assumptions and different values of the 
rapidity cut-off, and a solid basis for the interpretation of the results of 
independent TMD extractions.

Finally, we want to stress that the scheme we are proposing does not require 
a new start in the phenomenological analysis of all classes of hadronic processes. 
In fact, we can relate the TMDs obtained from data analyses based on the square 
root definition to the TMDs extracted using the factorization definition. 
This allows us to benefit all previous phenomenological analyses and extend them 
to $1$-h class processes. This is indeed the strategy we are planning to pursue 
in the near future.

\bigskip

\section*{Acknowledgements}
We are very grateful to Osvaldo Gonzalez for his questions and suggestions, 
which prompted us to explore in more dept some details of the rapidity 
dilation mechanism. He has been a constant source of inspiration for the 
completion of this project.
We thank Ted Rogers, Leonard Gamberg and Paolo Torrielli for very 
useful discussions.\\
This project has received funding from the European Union’s Horizon 2020 
research and innovation programme under grant agreement No 824093.

\bigskip

\appendix


\section{Wilson Lines \label{app:WL}}


\bigskip

A Wilson line (or a gauge link) is a path-ordered exponential operator defined by:
\begin{align}
W_\gamma = P \left \{ \exp \left[
-i g_0 \, \int_0^1 ds \,
\dot{\gamma}^\mu(s) A^a_{(0)\,\mu}(\gamma(s)) t_a 
\right]
\right \},
\label{eq:wilsdef_1}
\end{align}
where $\gamma$ is a generic path and $P$ denotes the path ordering (i.e. when the exponential is expanded the fields corresponding to higher values of $s$ are to be placed to the left).
The coupling constant and the gluon field are bare quantities, as indicated by the label ``0".
In the previous formula, $t_a$ are the generating matrices of the gauge group, in the appropriate representation.
The Wilson lines guarantee that PDFs and FFs (in both collinear and TMD cases) are gauge invariant, by linking the quark to the anti-quark fields 
in the definition of the collinear factor (see Eq.~\eqref{eq:sub_coll} ).
The Wilson line represents the (all order) propagation of a particle strongly boosted in some direction $n$. If this direction is a straight line the Wilson line depends only on the endpoints of the path and can be written in a compact way as:
\begin{align}
W_n\left(x_2,\,x_1,\,n\right) = P \left \{ 
\exp \left[
-i g_0 \, \int_{x_1}^{x_2} d\lambda \,
n^\mu A^a_{(0)\,\mu}(\lambda n) t_a 
\right]
\right \},
\label{eq:wilsdef_2}
\end{align}
If the strongly boosted particle is a quark, the associated Feynman rules are:
\begin{align}
\begin{gathered}
\includegraphics[width=1.5cm]{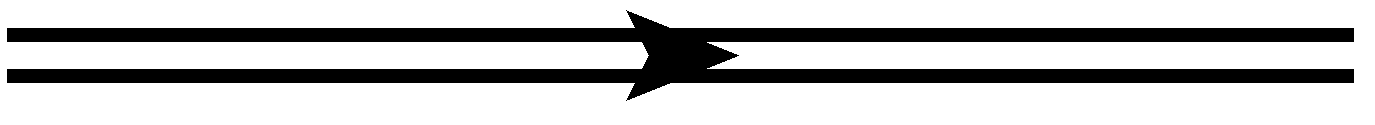}
\end{gathered} 
&= \frac{i}{k \cdot n + i 0} ; \\
\begin{gathered}
\includegraphics[width=1.5cm]{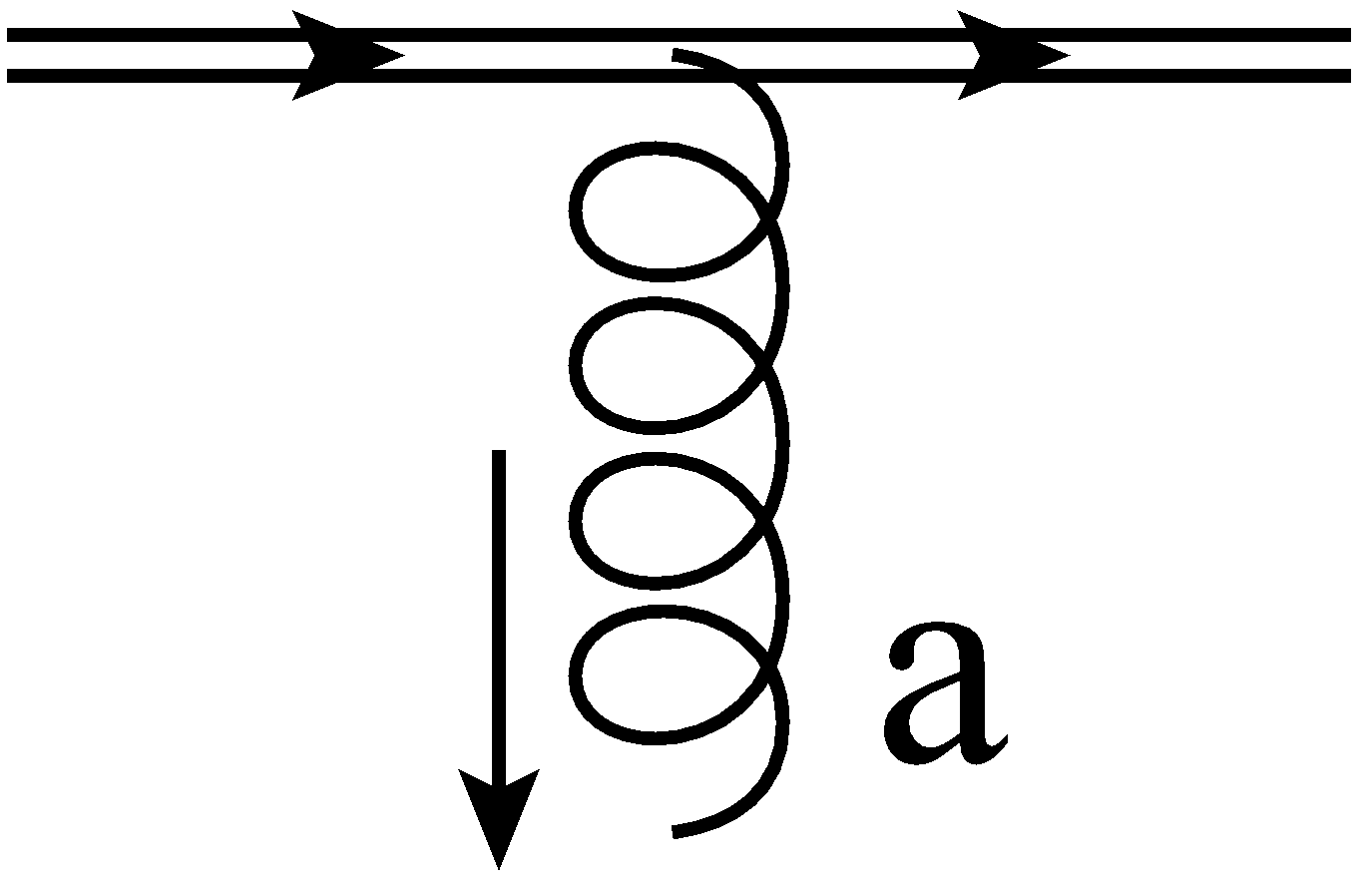}
\end{gathered} 
&= -i g_0 \, n^\mu \, t_a .
\label{eq:wils_feynrules}
\end{align}
More details can be found for instance in Chapter 7 of~\TheBook.

\bigskip


\section{Perturbative QCD and small \texorpdfstring{$b_T$}{bT} region
\label{app:smallbT}}


\bigskip

Soft factors and collinear parts are well defined functions only over a 
 rather small region in the transverse momentum space, according to
 power counting rules.
The Fourier transform to the impact parameter space can be regarded as a kind of analytic continuation, because at fixed $b_T$ we can roughly access all transverse momenta with $k_T \leq \frac{1}{b_T}$, even trespassing the original momentum region.
In particular, the small $b_T$ region is associated with large transverse momenta, where perturbative QCD can be applied and a power expansion in $\alpha_s$ allows us to perform explicit calculations.
This can be proved by a direct application of the factorization procedure to the small $b_T$ approximation of the Fourier transformed function.
For the soft factor this can be found in Section~\ref{sec:soft_factor}, while for collinear parts we refer to Chapter 13 of \TheBook ~and to Ref.~\cite{Collins:2017oxh}.

Despite the undeniable advantage provided by the possibility to perform explicit calculations in the small $b_T$ region, perturbative QCD is not enough to reproduce integrated quantities, which correspond to the Fourier transformed functions evaluated in $b_T = 0$.
These can be recovered from the operator definitions, that obviously give a non-perturbative, all-order point of view. Therefore in $b_T = 0$, Eqs.~\eqref{eq:ft_soft} and~\eqref{eq:ft_soft2} simply confirm that the integrated soft factor is the identity matrix, while Eq.~\eqref{eq:ft_coll} reproduces the integrated PDFs and FFs.
The failure of perturbative QCD in $b_T = 0$ is due to the fact that the integral over $\vec{k}_T$ is intrinsically ill defined, since it extends well beyond the physical momentum region where the TMDs and the soft factor are defined.
As a consequence, new UV divergences arise and the counterterms in Eqs.~\eqref{eq:ft_soft} and~\eqref{eq:ft_coll} are not sufficient to cancel them.
Therefore, the perturbative approach lead to definition of integrated functions as \emph{bare} quantities and they need a renormalization in order to acquire physical meaning and reproduce the correct results.
In the following, such renormalization procedure will be investigated for both the $2$-h soft factor and the TMDs.

\bigskip


\subsection{Small \texorpdfstring{$b_T$}{bT} behaviour of \texorpdfstring{$2$}{2}-h Soft Factor\label{subapp:smallbT_2h_S}}


\bigskip

%
\begin{figure}[t]
  \centering 
\begin{tabular}{c@{\hspace*{15mm}}c}  
      \includegraphics[width=4.2cm]{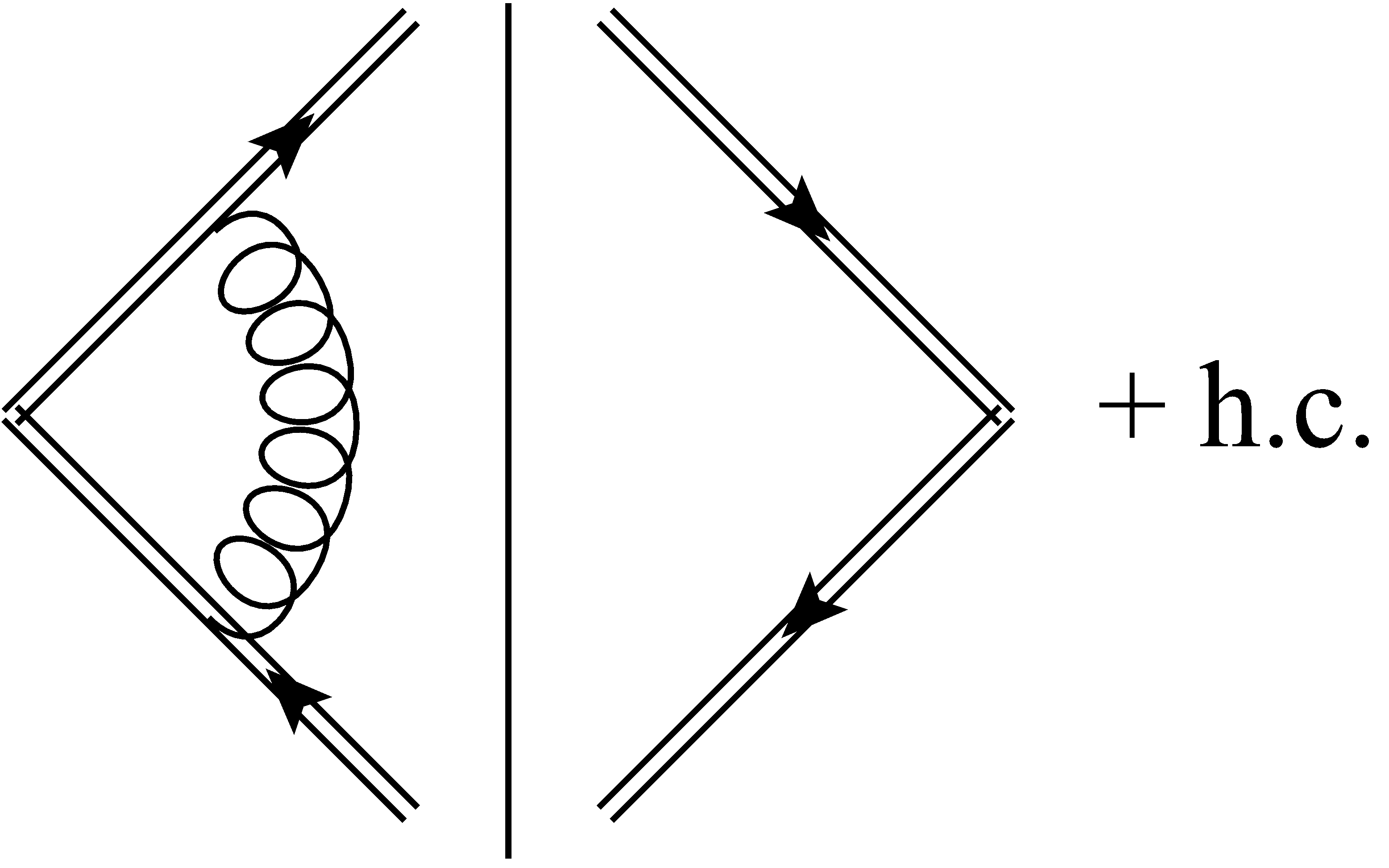}
  &
      \includegraphics[width=5cm]{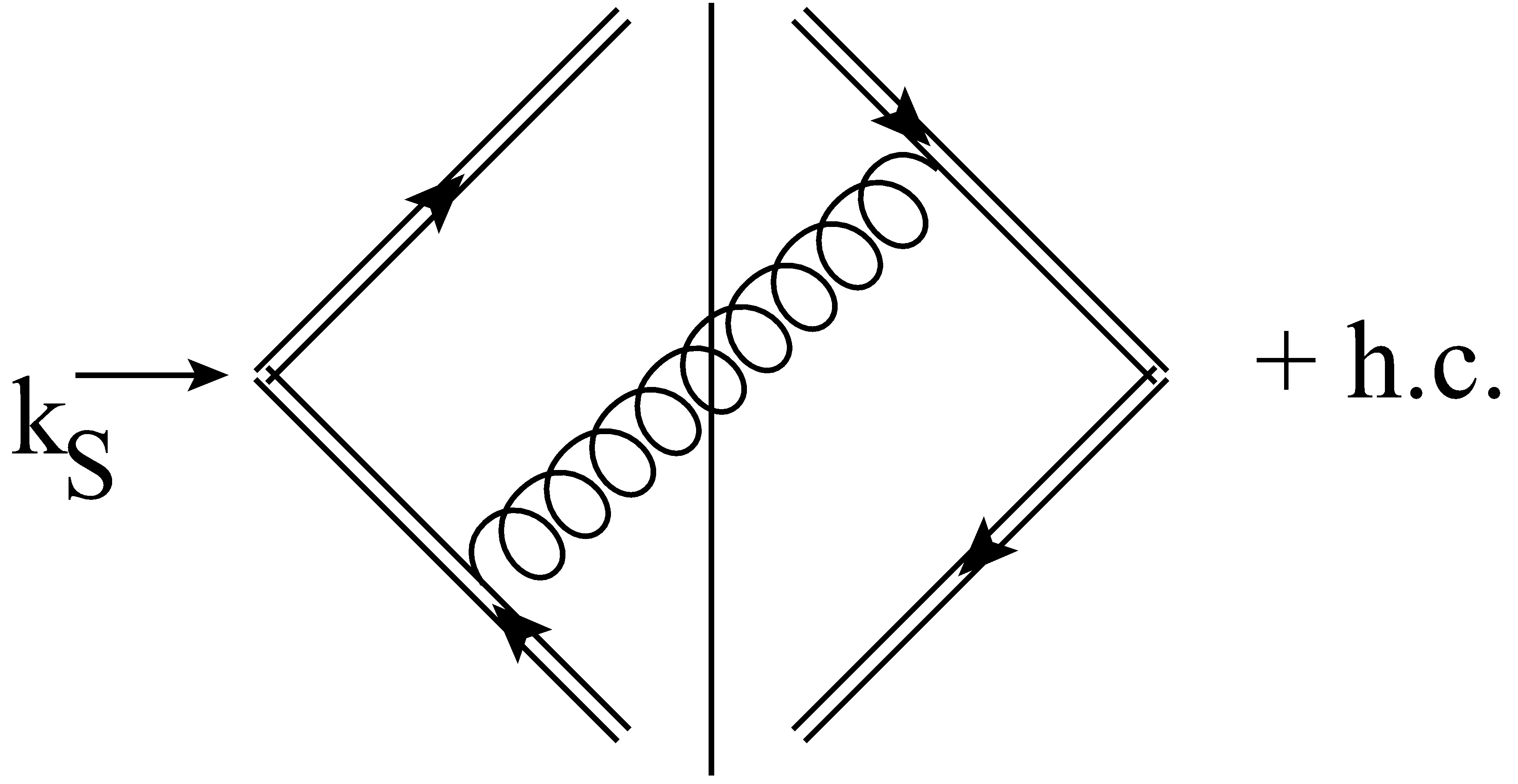}
  \\
  (a) & (b)
\end{tabular}
 \caption{Feynman graphs contributing to the small $b_T$ behavior of the 1 loop soft factor $\ftsoft{2}$. (a): Virtual diagrams (zero in dimensional regularization). (b) Real diagrams.}
\label{fig:soft_1loop}
\end{figure}
%
%
%
The Feynman graphs in Fig.~\ref{fig:soft_1loop} show  that in the small $b_T$ region the (renormalized) $2$-h soft factor is given by:
\begin{align}
\ftsoft{2}(b_T,\,\mu,\,y_1 - y_2) = 
1 - \frac{\alpha_S(\mu)}{4\pi} \, 8 C_F \,(y_1 - y_2) \, 
\log{\frac{\mu \, b_T}{C_1}} + 
\mathcal{O} \left( \alpha_S^2, \; e^{-(y_1-y_2)} \right),
\label{eq:soft_1loop}
\end{align}
where $C_1 = 2 e^{-\gamma_E}$.
The perturbative expansion of the previous equation should be valid at small $b_T$; however in this region $\log (\mu b_T/C_1)$ becomes large and sufficiently near to $b_T = 0$ it completely oversizes $\alpha_S$ so that the expansion becomes meaningless.
Resummation in principle solves this problem.
The soft kernel can be directly obtained from Eq.~\eqref{eq:soft_1loop} by using the definition of Eq.~\eqref{eq:soft2evo_1} or of Eq.~\eqref{eq:soft2evo_2}:
\begin{align}
\widetilde{K}(b_T,\,\mu) = 
- \frac{\alpha_S(\mu)}{4\pi} \, 16 C_F \, 
\log{\frac{\mu \, b_T}{C_1}} + 
\mathcal{O} \left( \alpha_S^2 \right).
\label{eq:kernel_1loop}
\end{align}
This expressions implies that $\widetilde{K}(b_T,\,\mu)$ is large and positive as $b_T$ decreases.
Therefore, the resummed soft factor of Eq.~\eqref{eq:soft_asy} vanishes in $b_T = 0$.
An improvement can be reached by using a leading log estimate of $\widetilde{K}$ by using its evolution equation solution, Eq.~\eqref{eq:K_evo_sol}. 
Actually, it is inappropriate to count the logs of a quantity, like the soft kernel, which is already the result of a resummation procedure.
Despite this, we can apply the same recipe and set all terms to order $\alpha_S^0$ except $\gamma_K$, which has to be taken to 1 loop.
This gives:
\begin{align}
\widetilde{K}^{LL}(b_T,\,\mu) = 
\frac{\gamma_K^{[1]}}{2 \beta_0} \, 
\log{\left(
1 - \frac{\alpha_S(\mu)}{4\pi} \log{\frac{\mu \, b_T}{C_1}}
\right)},
\label{eq:kernel_LL}
\end{align}
that coincides with Eq.~\eqref{eq:kernel_1loop} in the limit $\alpha_S \rightarrow 0$. With this estimate, the divergence of $\widetilde{K}$ is much less severe but it is still there.
An easy way to solve the problem and ensure that the perturbative QCD computation agrees with the operator definition prediction is to introduce a cut-off that prevents the soft transverse momentum to reach the UV region when it is integrated out.
This can be implemented in $b_T$-space by introducing a new parameter $b_{\mbox{\tiny MIN}} \neq 0$ that provides a minimum value for $b_T$.
A modification of the $b^\star$ prescription, Eq.~\eqref{eq:bstar}, is a simple way to insert this cut-off directly in the definition of the soft factor.
For example, we can use the modified $b^\star$ prescription of Ref.~\cite{Collins:2016hqq}:
\begin{equation} \label{eq:mod_bstar}
\vec{b}_T^\star\left(b_c (b_T)\right) = 
\vec{b}_T^\star\left(\sqrt{b_T^2+b^2_{\mbox{\tiny MIN}}}\right).
\end{equation}
%
%
\begin{figure}[t]
  \centering 
\includegraphics[width=12cm]{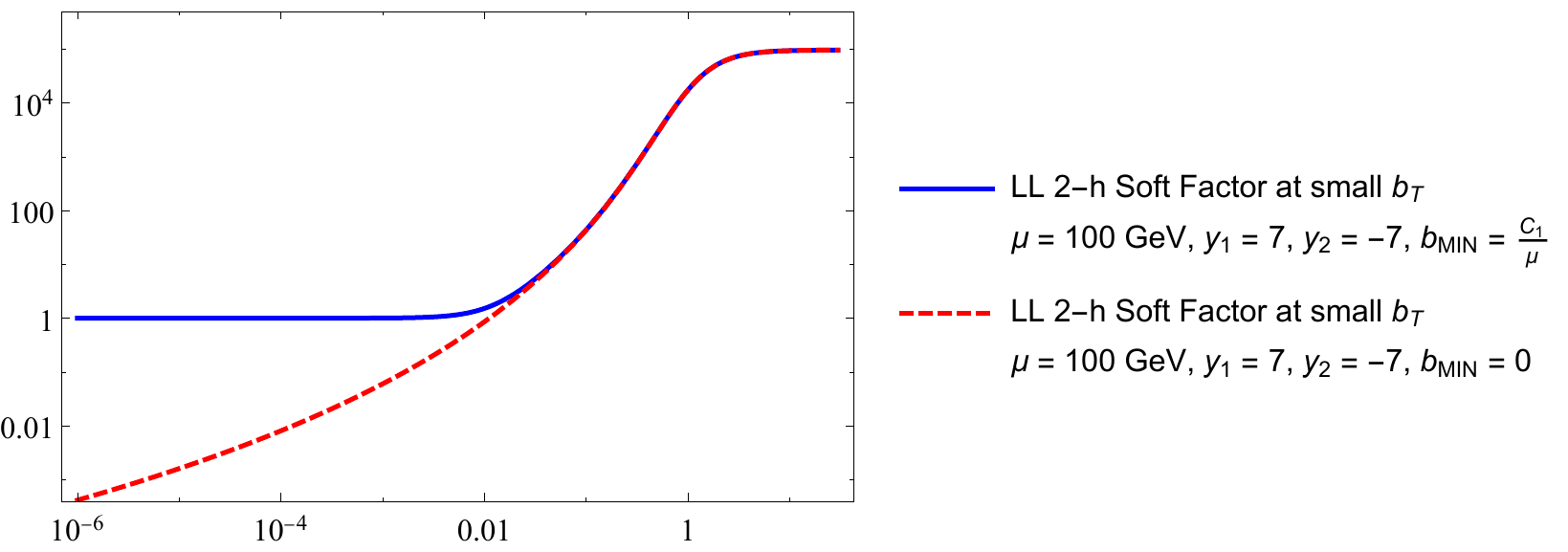}
 \caption{Leading Log (LL) Fourier Transformed $2$-h Soft Factor in the small $b_T$ region. The introduction of $b_{\mbox{\tiny MIN}}$ (blue solid line) allows to recover a full agreement with the operator definition of Eq.~\eqref{eq:ft_soft2}. If the regularization is not introduced (red dashed line), $\ftsoft{2}$ vanishes in $b_T = 0$. Here $b_{\mbox{\tiny MAX}} = 1$ GeV$^{-1}$.}
\label{fig:2hSoft}
\end{figure}
%
Then, the integrated soft factor is given by the unintegrated $\ftsoft{2}$ evaluated in $b_T^\star\left(b_c (0)\right) = b_{\mbox{\tiny MIN}}$.
If $\mu$ can be considered a large energy scale (e.g. if it can be set equal to the hard energy scale $Q$ of the process) 
then we can set $b_{\mbox{\tiny MIN}} \propto {1}/{\mu}$.
Consequently, all logs in Eqs.~\eqref{eq:soft_1loop} and~\eqref{eq:kernel_1loop} are heavily suppressed and the soft kernel is zero at small $b_T$, while the soft factor is unity, see Fig.~\ref{fig:2hSoft}.
Despite this kind of regularization has been devised for the $2$-h soft factor, it applies equally well to the general soft factor $\soft{N}$, where $N$ can be any integer.

\bigskip


\subsection{Small \texorpdfstring{$b_T$}{bT} behaviour of TMDs \label{subapp:smallbT_tmds}}


\bigskip

Formally, the integrated TMD is the Fourier transformed TMD computed at $b_T = 0$. 
In order to recover this result from Eq.~\eqref{eq:tmd_sol_2} by applying perturbative QCD, the Fourier transformed TMD has to be renormalized, otherwise it would vanish in $b_T = 0$. This result can be proved by following the procedure described in Ref.~\cite{Collins:2016hqq}.
First of all, thanks to the properties of the model $M_C$, Eq.~\eqref{eq:model_prop}, and of $g_K$, we can neglect all the non-perturbative content in Eq.~\eqref{eq:tmd_sol_2} at small $b_T$.
Furthermore, in this region we can approximate $b_T^\star$ with $b_T$.
Then, it is a standard result that the $\alpha_S$ expansion of Wilson coefficients can be written as:
\begin{align}
\widetilde{\mathcal{C}}_f^{\,j}(\rho,\, b_T; \, \mu, \,\zeta) = 
\sum_{n=0}^{\infty} \left(\frac{\alpha_S(\mu)}{4\pi}\right)^n \,
\sum_{k=0}^{2 n} \, \sum_{l=0}^{[{k}/{2}]}
\widetilde{\mathcal{C}}_f^{\,j\hspace{.1cm}[n,\,k-l,\,l]}(\rho)
\left(\log{\frac{\mu}{\mu_b}} \right)^{k-l} \, \left(\log{\frac{\zeta}{\mu_b^2}} \right)^l,
\label{eq:wils_coeff_aS}
\end{align}
where $[{k}/{2}]$ denotes the integer part of $k$.
If the scales are fixed according to the standard choices of Eqs.~\eqref{eq:mub} and \eqref{eq:zetab}, all the logs disappear and the only $b_T$ dependence in the Wilson coefficients is given by $\alpha_S (\mu_b)$.
Since $\mu_b \propto {1}/{b_T}$, when $b_T \to 0$ the energy scale becomes very large 
and $\alpha_S$ can be considered a small parameter.
For example, at 1 loop:
\begin{align}
\frac{\alpha_S(\mu_b)}{4\pi} 
\overset{\mbox{low }b_T}{\sim}
\frac{1}{2 \beta_0 \log{{\mu_b}/{\Lambda_{\mbox{\tiny QCD}}}}}.
\label{eq:aS_1loop}
\end{align}
Then, the Wilson coefficients evaluated at the scales $\mu_b$, $\zeta_b$ are well approximated at small $b_T$ by their lowest order term, which is simply a delta function:
\begin{align}
\widetilde{\mathcal{C}}_j^{\;k}(\rho,\, b_T; \, \mu_b, \,\zeta_b) =
\delta_j^{\;k} \delta(1-\rho)+ \mathcal{O} \left( 
\frac{1}{\log{\frac{\mu_b}{\Lambda_{\tiny QCD}}}}
\right)
\label{eq:wils_coeff_smallbT}
\end{align}
 On the other hand, $\widetilde{K}$, which is at exponent, allows for a different number of logarithms in front of each power of $\alpha_S$:
\begin{align}
&\widetilde{K}(b_T; \, \mu) = 
\sum_{n=1} 
\left(\frac{\alpha_S(\mu)}{4\pi}\right)^n \,
\sum_{l=0}^{n}
\widetilde{K}^{[n, l]} \, 
\left(\log{\frac{\mu}{\mu_b}} \right)^{l}.
\label{eq:Ktilde_aS} 
\end{align}
As in the previous case, all the explicit dependence on $b_T$ vanishes if $\mu = \mu_b$ and $\widetilde{K}$ at small $b_T$ is well approximated by its lowest order term. However, since in this case the series starts from $\mathcal{O} \left( \alpha_S(\mu_b) \right)$, we can simply neglect this contribution.
Finally, the anomalous dimensions have a simple expansion in $\alpha_S$:  
\begin{align}
&\gamma_{C}(\alpha_S(\mu), 1) = \sum_{n=1} \left(\frac{\alpha_S(\mu)}{4\pi}\right)^n \,
\gamma_{C}^{[n]} ,
\label{eq:gammaG_aS}\\
&\gamma_{K}(\alpha_S(\mu)) = \sum_{n=1} \left(\frac{\alpha_S(\mu)}{4\pi}\right)^n \,
\gamma_{K}^{[n]}
\label{eq:gammaK_aS} .
\end{align}
The easiest way to study the behavior of their contribution in Eq.~\eqref{eq:tmd_sol_2} at small $b_T$ is to consider its derivative with respect to $\log{b_T}$.
Since ${\partial}/{\partial \log{b_T}} = - {\partial}/{\partial \log{\mu_b}}$, we can compute with the help of Eq.~\eqref{eq:aS_1loop}:
\begin{align}
&\frac{\partial}{\partial \log{b_T}} 
\int_{\mu_b}^{\mu} \frac{d \mu'}{\mu'} \,
\sum_{n=1} 
\left(\frac{\alpha_S(\mu')}{4\pi}\right)^n \,
\left[ \gamma_C^{[n]} - \frac{1}{4} \, 
\gamma_K^{[n])} \, \log{\frac{\zeta}{\mu'^2}}\right] = 
\notag \\
&\quad= \sum_{n=1} 
\left(\frac{\alpha_S(\mu_b)}{4\pi}\right)^n \,
\left[ \gamma_C^{[n])} - \frac{1}{4} \, 
\gamma_K^{[n]} \, \log{\frac{\zeta}{\mu_b^2}}\right] 
\notag \\
&\quad= \frac{1}{4} \frac{\gamma_K^{[1]}}{\beta_0} + 
\mathcal{O} \left(
\frac{1}{\log{b_T \, \Lambda_{\mbox{\tiny QCD}}}}
\right) .
\end{align}
This behavior affects the whole TMD at small $b_T$, giving:
\begin{align}
&\widetilde{C}(\xi,\, b_{T}; \, \mu, \,\zeta) 
\overset{\mbox{low }b_T}{\sim}
\left(b_T\right)^{\frac{1}{4} \frac{\gamma_K^{[1]}}{\beta_0}}
\times \mbox{log corrections } .
\label{eq:tmd_smallbT}
\end{align}
From Eq.~\eqref{eq:kernel_1loop} $\gamma_K^{[1]} = 16 \, C_F$, then the TMDs goes to zero when $b_T \rightarrow 0$ with a power-law behavior.

This is also confirmed by a direct computation of the leading log (LL) estimate of Eq.~\eqref{eq:tmd_sol_2}.
In this approximations, all the quantities are taken at order $\alpha_S^0$, except $\gamma_K$ which instead is computed at 1 loop.
The result is:
\begin{align}
&\widetilde{C}_{f,\,H}(\xi,\, b_{T}; \, \mu, \,\zeta) 
\overset{\mbox{low }b_T}{\sim}
\left( \delta_f^{\hspace{.2cm}j} \otimes c_j (\mu_b) \right) (\xi) \times \notag \\
&\quad \times
\mbox{ exp} \left \{
L_b \, g^{LL}_1 \left( a_S (\mu) L_b \right) + 
g^{LL}_2 \left( a_S (\mu) L_b \right)
\right \}
\label{eq:LL_sud}
\end{align}
where $L_b = \log{{\mu}/{\mu_b}}$, $a_S = {\alpha_S}/{4\pi}$ and:
\begin{align}
&g^{LL}_1 (x) = \frac{\gamma_K^{[1]}}{4 \beta_0} + 
\frac{\gamma_K^{[1]}}{8 \, x \, \beta_0^2}
\, \log{(1 - 2 \beta_0 x)}, \\
&g^{LL}_2 (x) = 
\frac{1}{8 \beta_0} \, \gamma_K^{[1]} \, 
\log{\frac{\zeta}{\mu^2}} 
\, \log{(1 - 2 \beta_0 x)}.
\label{eq:LL_functions}
\end{align}
Notice that the function $g^{LL}_2$ contributes to the LL estimate even if it typically appears at NLL. This is due to the presence of three scales instead of two.
In fact, if $\sqrt{\zeta}$ equals either $\mu$ or $\mu_b$, only $g^{LL}_1$ contributes to LL.

\bigskip

As a consequence of the previous arguments, integrated TMDs are bare quantities when approached perturbatively. Formally:
\begin{align}
\int d^{D-2} \vec{k}_T \, C_{j,\,H}(\xi,\, k_{T}; \, \mu, \,\zeta) =
c_{j,\,H}^{(0)}(\xi,\,\mu) = 
\begin{cases}
f_{j/H}^{(0)}(x,\,\mu )
&\mbox{ initial state;} \\[10pt]
z^{-2+2\epsilon}
d_{H/j}^{(0)}(z,\,\mu)
&\mbox{ final state.} 
\end{cases}
\label{eq:bare_int} 
\end{align}
The bare integrated TMDs in the equation above acquire their dependence on $\mu$ through the renormalized fields used to compute them.
Real bare quantities are defined through bare fields and are obtained by multiplying by $Z_2$ as in Eq.~\eqref{eq:fact_defTMDs}.
Notice that integration makes the soft-collinear subtractions trivial, because the $\soft{2}$ appearing in the factorization definition is unity when integrated over all soft transverse momentum.
The required UV counterterm depends on the plus component of the momentum of the reference parton, i.e. on the collinear momentum fraction $\xi$. 
Hence, the renormalized quantities are not simple products of the bare quantities with the UV counterterm, like in Eq.~\eqref{eq:fact_defTMDs}, but rather convolutions
\begin{align}
c_{j,\,H}(\xi,\,\mu) = 
\left( 
\left(Z_{\mbox{\tiny int}}\right)_{j}^{\,k} (\alpha_S(\mu)) 
\otimes
c_{k,\,H}^{(0)} 
\right) (\xi),
\label{eq:renorm_int} 
\end{align}
where now $c_{k,\,H}^{(0)}$ denotes a bare quantity computed with bare fields. With this definition, we can interpret the renormalized integrated TMDs as the usual PDFs and FFs used in collinear factorized cross sections as in Eq.~\eqref{eq:coll_fact}.

The factorization procedure applied to the TMD at small $b_T$ does not give   Eq.~\eqref{eq:ope_tmds} directly. Instead, it expresses the final result as a convolution between a collinear part, represented by the unrenormalized integrated TMDs, and a hard factor $\mathbb{H}$ which has to be properly subtracted in order to cancel the double counting due to the overlapping between the hard and the collinear momentum region.
This subtraction mechanism is completely anologous to that used in the definition of the subtracted collinear part in Eq.~\eqref{eq:sub_coll}.
Roughly speaking, the UV part of the bare integrated TMDs is (minus) $Z_{\mbox{\tiny int}}$, then the subtracted hard part acquires the divergence induced by the counterterm.
Despite this, we can still define a finite hard part by interpreting $\mathbb{H}^{\mbox{\small sub}}$ as a bare quantity as well, with its renormalized finite counterpart represented by the Wilson Coefficients in the OPE.
As a consequence, the required counterterm will be exactly $Z^{-1}_{\mbox{\tiny int}}$.
Then, a straightforward application of the convolution property shows that:
\begin{align}
&\widetilde{C}_{j,\,H}(b_{T}; \, \mu, \,\zeta)
\overset{\mbox{low }b_T}{\sim} \hspace{.1cm}
\left(\mathbb{H}^{\mbox{\small sub}}\right)_j^{\,k}
(b_{T}; \, \mu, \,\zeta) 
\otimes c_{k,\,H}^{(0)}
= \notag \\
&\quad= \left[ 
\left(\mathbb{H}^{\mbox{\small sub}}\right)_j^{\,k}
(b_{T}; \, \mu, \,\zeta) 
\otimes 
\left(Z^{-1}_{\mbox{\tiny int}}\right)_{k}^{\,l} (\alpha_S(\mu))
\right] 
\otimes
\left[ 
\left(Z_{\mbox{\tiny int}}\right)_{l}^{\,m} (\alpha_S(\mu))
\otimes 
c_{m,\,H}^{(0)} 
\right] = \notag \\[8pt]
&\quad= \widetilde{\mathcal{C}}_j^{\,k} (b_{T}; \, \mu, \,\zeta) 
\otimes c_{k,\,H} (\mu) .
\label{eq:ope_tmds_renorm}
\end{align}
Therefore, the functions $c_{j,\,H}$ appearing in the OPE are the renormalized integrated TMDs.
Notice that the same procedure is used in the cross sections where the usual PDFs and FFs appear.

Different renormalizations of the integral over $\vec{k}_T$ are allowed.
A common procedure, for instance, is to introduce a cut-off as we did for the $2$-h soft factor in~\ref{subapp:smallbT_2h_S}, 
by introducing a new parameter $b_{\mbox{\tiny MIN}} \neq 0$ that provides a minimum value for $b_T$,
for istance as in Eq.~\eqref{eq:mod_bstar}.
Then, the integrated TMD is given by the unintegrated TMD evaluated in $b_T^\star\left(b_c (0)\right) = b_{\mbox{\tiny MIN}}$:
\begin{align}
&\int d^{D-2} \vec{k}_T \, C_{f,\,H}(\xi,\, k_{T}; \, \mu, \,\zeta) = 
\widetilde{C}_{f,\,H}(\xi,\, b_{\mbox{\tiny MIN}}; \, \mu, \,\zeta) 
\sim \hspace{.1cm} \notag \\ 
&\quad \sim 
\left( 
\widetilde{\mathcal{C}}_j^{\,k} (b_{\mbox{\tiny MIN}}; \, \mu, \,\zeta) 
\otimes c_{k,\,H} (\mu)
\right) (\xi),
\end{align}
where in the last step we used the OPE expansion valid at small $b_T$.
In general, this result does not coincide with $c_{f,\,H}(\xi,\,\mu)$, but it will do if the Wilson Coefficients can be well approximated by their lowest order. 
If $\mu$ can be considered a large energy scale (e.g. if it can be set equal to the hard energy scale $Q$ of the process) 
then we can set $b_{\mbox{\tiny MIN}} \propto {1}/{\mu}$.
Then all the logs inside the Wilson Coefficients are heavily suppressed and the lowest order approximation is reliable.
Therefore, if $\mu$ is large enough, the cut-off approach gives the same result of the renormalization through the UV counterterm $Z_{\mbox{\tiny int}}$.
Thanks to $b_{\mbox{\tiny MIN}}$, the subtraction mechanism implemented in the factorization procedure applied to the TMD at small $b_T$ is now applied to the collinear parts instead of the hard factor.
Therefore, we do not have to worry about subtracting the hard part.
However, the final result coincides with that of Eq.~\eqref{eq:ope_tmds} because, trivially, $\mathbb{H}^{\mbox{\small sub}} \otimes \coll^{\mbox{\small unsub}} = \mathbb{H}^{\mbox{\small unsub}} \otimes \coll^{\mbox{\small sub}}$.

\bigskip

The integration over $\vec{k}_T$ of the TMD, actually gives the area under the curve designed by the TMD in $k_T$-space.
Even with the introduction of an explicit $b_{\mbox{\tiny MIN}}$, the value of such integral is very small. 
Since in momentum space, at small $k_T$, the TMD is positive (e.g. Gaussian behavior), the small value of the integrand implies that the TMD has to change sign at a certain $k_T$.
This is equivalent to say that the TMD loses its physical meaning when $k_T$ becomes too large.
In fact, the power counting imposes $k_T \sim \lambda$, where $\lambda$ is some small IR energy scale.

\bigskip


\section{Kinematics \label{app:kin}}


\bigskip

As stressed in Section \ref{subsec:coll_vs_tmd}, kinematics play a crucial role in 
factorization, as it determines whether we need to apply a TMD or a collinear 
factorization scheme.
The study of kinematics is strictly connected to the choice of the frame.
In the case of $\epm \to H\,X$, three four-vectors underlay the kinematical 
configuration:
\begin{itemize}
\item The momentum $k$ of the fragmenting parton. 
\item The momentum $P$ of the outgoing detected hadron $H$ of mass $M$, $P^2 = 
M^2$.
\item The momentum $q$ of the highly virtual time-like photon that makes the 
partonic state. 
Its squared momentum gives the square of the center of mass energy $Q >> M$, 
$q^2 = Q^2$.
\end{itemize}
Clearly, the choice of the frame is completely arbitrary since the cross 
section will be Lorentz invariant.
Three main frames are useful in deriving the final form of the factorized cross 
section: in this appendix we will provide a short description of all of them.
\begin{enumerate}
\item  \textbf{Hadron frame}, labeled by $h$. 
This is the frame where the outgoing hadron $H$ has no transverse components 
and it moves very fast along the (positive) 
$z_h$-direction:
\begin{equation} \label{eq:hframe_cond}
\vec{P}_{T,\,h} = \vec{0}_T.
\end{equation}
Furthermore, since $H$ is strongly boosted in the plus direction its plus 
component is very large, of order $\sim Q$.
As a consequence, its minus component has to be very small in order to satisfy 
the on-shell condition $P^2 = 2 P_h^+ \,P_h^- = M^2$. 
Therefore, in this frame, the full four-momentum $P$ can be written  as:
\begin{equation} \label{eq:P_h}
P =  \left( P_h^+,\,\dfrac{M^2}{2 P_h^+}, \vec{0}_T \right)_h \sim Q 
\left(1,\,\dfrac{M^2}{Q^2}, \,0 \right) .
\end{equation}
The fragmenting parton belongs by definition to the same collinear group of the 
outgoing hadron, hence it is \emph{almost} collinear to it: it has a very large 
plus component, a low transverse momentum and an even lower minus component.
It is almost on-shell, with a very low virtuality.
Power counting (see Chapter 5 in Ref.~\cite{Collins:2011zzd}) allows us to 
quantify the sizes of these quantities by introducing a small infrared scale 
$\lambda << Q$.
Then $k^2 = \lambda^2$, which means $k_h^+ \sim Q$, $k_h^- \sim {\lambda^2}/{Q}$ 
and $k_{h,\,T} \sim \lambda$.
Neglecting all the suppressed components, $k$ and $P$ become exactly collinear, i.e. $k 
\propto P$. This can be made explicit by setting:
\begin{equation} \label{eq:z_coll}
k_h^+ = \dfrac{1}{\widehat{z}} P_h^+,
\end{equation}
Therefore $P \sim \widehat{z} k$, and
\begin{equation} \label{eq:k_h}
k =  \left( \frac{ P_h^+}{\widehat{z}},\,k_h^-, \vec{k}_{T,\,h} \right)_h \sim 
Q 
\left(1,\,\dfrac{\lambda^2}{Q^2}, \, \frac{\lambda}{Q} \right) .
\end{equation}
Since power counting rules are defined in the hadron frame, this is the 
most appropriate frame where to implement factorization.
We can interpret $\widehat{z}$ as the collinear momentum fraction that the 
outgoing 
hadron takes off the fragmenting parton.
Clearly $\widehat{z}$ has kinematics boundaries, due to the requirement that all 
the 
particles crossing the final state cut are physical, i.e. they have 
positive energy.
With the help of Fig.~\ref{fig:z_bound} and by applying the power counting 
rules, we obtain the following constraints:
\begin{figure}[t]
\centering
\includegraphics[width=7.5cm]{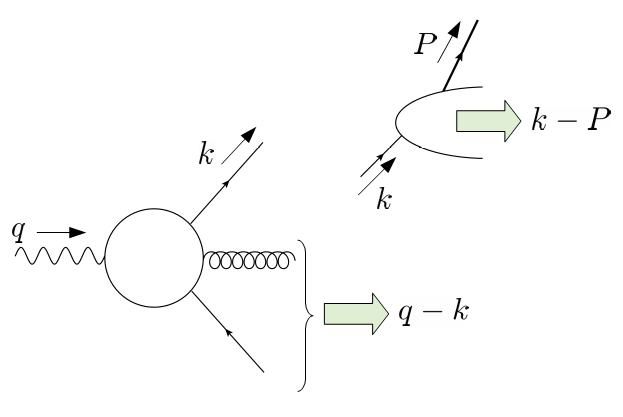}
\caption{Momentum flow that determines the kinematics boundaries on 
$\widehat{z}$.}
\label{fig:z_bound}
\end{figure}
\begin{itemize}
\item Positive energy for the final state of the jet
\begin{equation} \label{eq:upper_limit}
(k - P)_h^0 \sim k_h^+ - P_h^+ = P_h^+ \left( \frac{1}{\widehat{z}} - 1 \right) 
\geq 
0,
\end{equation}
which gives $\widehat{z} \leq 1$.
\item Positive energy in the hard part of the process (given that $q_h^- >0$)
\begin{equation} \label{eq:lower_limit}
(q - k)_h^0 \geq 0 \rightarrow q_h^+ - k_h^+ = \frac{Q}{\sqrt{2}} 
\left( 1 - \frac{\sqrt{2} P_h^+}{Q} \,  \frac{1}{\widehat{z}}\right) \geq 0.
\end{equation}
The \textbf{fractional energy} $z$ is defined as
\begin{equation} \label{eq:z_def}
z = 2 \, \frac{P \cdot q}{Q^2} = 2 \frac{E_{\mbox{\tiny CM}}}{Q} \sim 
\frac{\sqrt{2} P_h^+}{Q},
\end{equation}
where $E_{\mbox{\tiny CM}}$ is the energy of the detected hadron in the center 
of mass frame. 
Then Eq.~\eqref{eq:lower_limit} gives the kinematics boundary: $\widehat{z} \geq z$, 
with  
$z \leq 1$.
\end{itemize}
The scaling of the components of the four-momentum $q$ is obtained from the 
momentum conservation relation:
\begin{equation} \label{eq:mom_cons}
q = k + \sum_{\alpha} k_\alpha ,
\end{equation}
where $k_\alpha$ is the momentum of a generic real emission.
As explained in Section~\ref{subsec:coll_vs_tmd}, since the process $\epm \to H X$ belongs to 
the \hclass{1}, there is always at least one real emission (in this case the 
anti-quark leg that does not fragment) with a hard momentum, i.e. with all  
components very large, at least of order $Q$.
As a consequence, the only component of $k$ that survives in 
Eq.~\eqref{eq:mom_cons} is $k_h^+$, while all the others are strongly 
suppressed by the large momenta $k_\alpha$.
\item \textbf{c.m.    frame}, labeled by $\gamma$.
In this frame the spatial momentum of $q$ is zero
\begin{equation} \label{eq:CMframe_cond}
\vec{q}_{\gamma} = \vec{0}\,,
\end{equation}
which means
\begin{equation} \label{eq:q_CM}
q =  \left( Q, \, \vec{0} \right)_\gamma =  \left( 
\frac{Q}{\sqrt{2}},\,\frac{Q}{\sqrt{2}}, \vec{0}_{T} \right)_\gamma.
\end{equation}
Since rotations send null spatial vectors into null spatial vectors, the 
condition in Eq.~\eqref{eq:CMframe_cond} is defined modulo a rotation in space.
Therefore, if we set the $z$-axis of this frame to be the direction of the 
outgoing hadron, we can identify the hadron frame with the c.m.  frame 
and apply power counting and the whole factorization procedure directly 
in this frame.
This is a big advantage, since usually the calculation of the hard part of the 
cross section is much easier in the c.m.  frame but in general it 
does not coincide with the hadron frame, which on the other hand makes simpler 
the application of the factorization procedure\footnote{For example, this is the 
case of $\epm \to H_A\,H_B\,X$, with the two hadrons almost back-to-back. In 
this 
case, the hadron frame is defined as the frame in which both hadrons have 
zero transverse momentum, i.e. where they are \emph{exactly} back-to-back. 
However, a spatial rotation can fix only one hadron and the c.m.  frame 
cannot be identified with the $h$-frame. 
The two frames are actually connected by a light boost in the transverse 
direction, where the boost parameter is (proportional to) $q_{T,\,h}$.
As a consequence, we need boost-dependent projectors connecting the collinear 
and the hard parts of the cross section.
In principle, we can use a boost also in the case of the production of a single 
hadron, however the boost will depend on $q_{T,\,h}$ which, in this case, is not 
observed.}.
Then we can write the components of $q$ in the $h$-frame as in 
Eq.~\eqref{eq:q_CM}.
From Eq.~\eqref{eq:mom_cons} it follows that the total transverse momentum of 
the real emissions exactly cancels the contribution of $\vec{k}_{T,\,h}$, hence 
$|\sum_\alpha \vec{k}_{\alpha,\,T,\,h}| \sim \lambda$.

Notice that the LAB frame, in which the $z$-axis coincide with the beam axis, 
is a valid c.m.  frame but it is not the hadron frame, as they 
differ by a spatial rotation, 
as shown in Fig.~\eqref{fig:CM_h_frames}.
\begin{figure}[t]
\centering
\includegraphics[width=7.5cm]{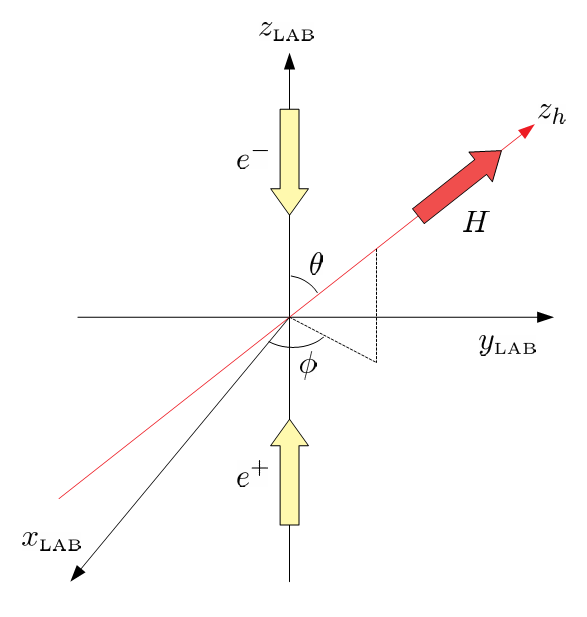}
\caption{The LAB frame and the $h$-frame are both c.m.  frames, but 
differ by a spatial rotation.}
\label{fig:CM_h_frames}
\end{figure}
The lepton pair is back-to-back in both the frames, but the direction of their 
spatial momenta is different.
\item \textbf{Parton frame}, labeled by $p$.
As explained in Ref.~\cite{Collins:2011zzd}, in order to properly define a 
fragmentation function we need a frame in which the fragmenting parton has zero 
transverse momentum. This is the parton frame, defined by requiring
\begin{equation} \label{eq:pframe_cond}
\vec{k}_{T,\,p} = \vec{0}_T.
\end{equation}
In principle we have two Lorentz transformations available that we can use to 
reach the parton frame from the hadron frame: a rotation of the (small) angle 
between the fragmenting parton and the outgoing hadron and a (light) transverse boost in 
the $\vec{k}_{T,\,h}$ direction.
By defining $\vec{k} = {\vec{k}_{T,\,h}}/{k_h^+}$, the angle of the rotation is $\alpha = -\sqrt{2} k$, while the parameter of the boost is $\vec{\beta} =\sqrt{2} \vec{k}$.
The two choices give the same result:
\begin{align}
&k =  \left( k_h^+, \,k_h^{-} - \dfrac{k_{h,\,T}^2}{2 k_h^+}, \vec{0}_T 
\right)_p + 
\mathcal{O} \left( \dfrac{\lambda^2}{Q^2} \right) 
\left(1,\,\dfrac{\lambda^2}{Q^2}, \,1 \right) ;  \label{eq:k_p} \\
&P =  \left( \widehat{z} \, k_h^+, \, \dfrac{M^2 + \widehat{z}^2 
\,k_{h,\,T}^2}{2 \widehat{z} \, 
k_h^+}, \,-\widehat{z} \, \vec{k}_{T,\,h} \right)_p + 
\mathcal{O} \left( \dfrac{M^2,\,\lambda^2}{Q^2} \right) 
\left(1,\,\dfrac{M^2,\,\lambda^2}{Q^2}, \,1 \right). \label{eq:P_p}
\end{align}
Notice that the plus components remain the same in the two frames (apart from 
power suppressed corrections).
In this frame we can identify the $z_p$-axis as the axis of the experimental jet 
of hadrons in which $H$ is detected.
In fact, all the (almost) collinear particles in the jet have been generated by 
the same fragmenting parton and hence the sum of their spatial momenta has to be 
equal to $\vec{k}_p = |\vec{k}| \, \widehat{z}_p$, that lies on the (positive) 
$z$ direction in this frame.
Therefore, measuring $P_{p,\,T}$ gives the transverse momentum of the 
outgoing hadron with respect the jet axis.
By definition, this axis coincides with the \textbf{partonic thrust axis} 
$\widehat{n}_p$, which is the direction that maximizes the partonic thrust $T_p$ 
defined as
\begin{equation} \label{eq:parton_thrust}
T_p = \dfrac{\sum_i |\vec{k}_{h,\,i} \cdot \widehat{n}_p|}{\sum_i 
|\vec{k}_{h,\,i}|},
\end{equation}
where the sum runs over all the partons produced in the hard scattering, and 
$\vec{k}_{h,\,i}$ is the spatial momentum in the c.m.  frame of the 
$i$-th outgoing hard parton.
For example, in the case of two (back-to-back) partons $T_p = 1$ and 
$\widehat{n}_p$ is the axis of the parton pair, while for three partons $T_p = 
\mbox{max} \{x_1, \,x_2,\,x_3 \} \geq {2}/{3}$, with $x_i = 2 
{|\vec{k}_{h,\,i}|}/{Q}$, and $\widehat{n}_p$ is the direction of the $i$-th 
parton.
Since $P_{p,\,T}$ is strictly connected to $k_{h,\,T}$, as shown in 
Eq.~\eqref{eq:P_p}, its measurement offers  powerful information on the 
partonic variables.
However, the experimental measurement is on the transverse momentum of the 
outgoing hadron with respect to the \textbf{hadron thrust axis} $\widehat{n}_h$, 
which is the direction that maximizes the hadronic thrust $T_h$ defined as
\begin{equation} \label{eq:hadron_thrust}
T_h = \dfrac{\sum_i |\vec{P}_{\mbox{\tiny CM},\,i} \cdot \widehat{n}_h|}{\sum_i 
|\vec{P}_{\mbox{\tiny CM},\,i}|},
\end{equation}
where now the sum runs over all the detected particles in the center of mass 
frame (e.g. the LAB frame).
Its value is close to its partonic counterpart, but they are not the same.
As shown in Ref.~\cite{Catani:1991kz}, the observed distribution of hadronic 
thrust is related to the distribution with respect to the partonic thrust (which can be  
computed in perturbation theory) by a correlation function $C(T_h,\,T_p)$ that
is sharply peaked around $T_h \sim T_p$.
Therefore, roughly speaking, we can set $C(T_h,\,T_p) \sim \delta(T_h - T_p)$ 
and the direction which maximizes the hadronic thrust is approximately the same 
axis that maximizes the partonic thrust, i.e. $\widehat{n}_p \sim 
\widehat{n}_h$. The estimate of how much they differ can be made more 
quantitative in the simple case of a $2$-jet configuration.
In fact, in this case we have $T_p = 1$ and $T_h \sim 1 - {(M_1^2 + 
M_2^2)}/{Q^2}$ (see Ref.~\cite{Catani:1991kz}), where $M_{1,\,2}$ is the 
invariant mass of the hadronic jets, hence $T_p - T_h \sim 
\mathcal{O}({M^2}/{Q^2})$.
In this paper, we consider $P_{p,\,T}$ as a valid estimate 
of the transverse momentum of the outgoing hadron with respect to the hadronic 
thrust axis.
\end{enumerate}

 \providecommand{\href}[2]{#2}\begingroup\raggedright\endgroup

\end{document}